\documentclass[11pt,a4paper,oneside]{article}
\pdfoutput=1
\usepackage{jheppub}
\usepackage{hyperref}
\usepackage{setspace}
\usepackage{xspace}
\usepackage{color}
\usepackage[table]{xcolor}
\usepackage{soul}
\usepackage{amsmath, amsfonts, amssymb, amsthm, upgreek, amstext, bm, mathtools, mathrsfs, stmaryrd}
\usepackage[font=footnotesize,labelfont=bf]{caption}
\usepackage{slashed}
\usepackage{subfig}
\usepackage{array} 
\usepackage{bbold}
\usepackage{physics}
\usepackage{booktabs}
\usepackage{fontawesome}
\usepackage[english]{babel}
\usepackage[utf8]{inputenc}
\usepackage{multicol, multirow}
\usepackage{comment}
\usepackage{pdfcomment}           

\usepackage{graphicx} 
\usepackage{pdflscape}

\newcommand{\citewithtip}[2]{%
  \cite{#1}%
  \,\pdftooltip{\textsuperscript{\faInfoCircle}}{#2}%
}

\definecolor{rosso}{cmyk}{0,1,1,0.4}
\definecolor{rossos}{cmyk}{0,1,1,0.55}
\definecolor{rossoc}{cmyk}{0,1,1,0.2}
\definecolor{blu}{cmyk}{1,1,0,0.3}
\definecolor{blus}{cmyk}{1,1,0,0.6}
\definecolor{bluc}{cmyk}{1,1,0,0.1}
\definecolor{verde}{cmyk}{0.92,0,0.59,0.25}
\definecolor{verdec}{cmyk}{0.92,0,0.59,0.15}
\definecolor{verdes}{cmyk}{0.92,0,0.59,0.4}

\newcommand{\be}{\begin{equation}}
\newcommand{\ee}{\end{equation}} 
\newcommand{\bry}{\begin{array}}
\newcommand{\ery}{\end{array}} 
\newcommand{\dst}{\displaystyle} 
\newcommand{\bit}{\begin{itemize}} 
\newcommand{\eit}{\end{itemize}} 
\newcommand{\ben}{\begin{enumerate}} 
\newcommand{\een}{\end{enumerate}}

\RequirePackage[normalem]{ulem} 
\RequirePackage{color}\definecolor{RED}{rgb}{1,0,0}\definecolor{BLUE}{rgb}{0,0,1} 

\begin{document}

\title{Isolating chirality-breaking SMEFT operators with Drell-Yan angular analysis}

\author[a,b]{Samuele Grossi,}
\author[a]{Xu Li,}
\author[a,b]{Lorenzo Rolla,}
\author[a]{Riccardo Torre,}

\emailAdd{riccardo.torre@ge.infn.it}
\emailAdd{samuele.grossi@ge.infn.it}
\emailAdd{xu.li@ge.infn.it}
\emailAdd{lorenzo.rolla@ge.infn.it}

\affiliation[a]{INFN, Sezione di Genova, Via Dodecaneso 33, I-16146 Genova, Italy}
\affiliation[b]{Department of Physics, University of Genova, Via Dodecaneso 33, I-16146 Genova, Italy}

\abstract{
    We present a comprehensive strategy to isolate the effect of a class of chirality-breaking interactions in the Standard Model Effective Field Theory (SMEFT) by exploiting Drell-Yan angular analysis and the violation of the Lam-Tung relation. Unlike most SMEFT interpretation of Drell-Yan measurements, dominated by growing-with-energy effects generated by the interference of SMEFT-induced and SM amplitudes, this method isolates operators that contribute only quadratically in the Wilson coefficients, allowing for an independent probe of non-interfering directions in the EFT parameter space. Denoting with $v$ the electroweak vev, with $\sqrt{s}$ the center-of-mass energy, and with $\Lambda$ the scale of new physics, depending on the nature of the SMEFT operators, the non-interfering  contributions to the amplitude generated by the chirality-breaking operators can be proportional to $v\sqrt{s}/\Lambda^{2}$ or $s/\Lambda^{2}$. We argue that these two classes can be further distinguished by analyzing the angular observables of the lepton pair in the transverse momentum and in the invariant mass distribution of the lepton pair. We therefore present an analysis of the lepton-pair angular observables in both the transverse momentum and invariant mass distributions. Based on a precise estimate of the Standard Model contribution to the relevant observables for the $pp\to l^{+}l^{-}+X$ process up to $\mathcal{O}(\alpha_{S}^{2})$, we present realistic projections for the sensitivity of the LHC with $300$ fb$^{-1}$ and for the HL-LHC with $3$ ab$^{-1}$ to chirality-breaking interactions, demonstrating that angular observables provide an independent and clean handle on SMEFT effects, especially in regions where the Standard Model contribution is naturally suppressed thanks to the Lam-Tung relation. This analysis becomes crucial to go beyond single parameter global fits, since it helps breaking degeneracies with chirality preserving operators and to disentangle overlapping directions in the EFT parameter space.
}

\maketitle

\tableofcontents

\section{Introduction}\label{sec::Intro}
The Standard Model Effective Field Theory (SMEFT) is the proper tool to look for new physics (NP) in the absence of direct evidence of new particles. It offers a consistent framework, both in terms of symmetries and, when empowered by essential assumptions on the ultraviolet (UV) structure of NP, of power counting, to parametrize the effects of heavy NP at energies accessible to existing and future colliders.

Even though, already at dimension six, the number of operators appearing in the SMEFT is extremely large (especially if no assumptions on the flavor structure are made), not all such operators contribute to all processes and observables. Instead, there are usually specific classes of operators, including just a few of them, that contribute to specific classes of observables in specific kinematic regions. This consideration, together with the observation that the more differential observables are, the more information they carry about the underlying physics, often allows one to disentangle the effects of different operators and remove degeneracies and flat directions in the SMEFT parameter space.

One of the most studied processes, with all its related observables, in the context of testing the electroweak sector of the SMEFT at colliders, is the inclusive Drell-Yan (DY) process, $pp\to \ell^+\ell^-+X$, where $\ell$ is a lepton, typically an electron or a muon, and $X$ is any additional particle over which the cross-section is integrated. The reason for this interest in the DY process is multifold: first, this is the cleanest experimental channel that can be studied at hadron colliders; second, it offers a very large statistics at the LHC, with order bilions of events collected at the end of the LHC and tens of bilions at the HL-LHC; third, it is one of the best-known processes from the point of view of theoretical predictions, with $\mathcal{O}(\alpha_\text{S}^3)$  \cite{Duhr:2020seh,Chen:2021vtu,Duhr:2020sdp,Duhr:2021vwj} and $\mathcal{O}(\alpha \alpha_{\text{S}})$ \cite{Armadillo:2022bgm,Bonciani:2021zzf,Armadillo:2024ncf} calculations available in the literature; finally, as we argue in the present paper, even being a simple two-body final state, DY can offer a wide spectrum of observables, in different kinematic regimes, that allow one to disentangle the effects of different SMEFT operators. This is the reason that made the DY process the prototype process in the context of the high energy precision program at hadron colliders \cite{Farina:2016rws}. In this paper, we aim to further develop this program in a direction that has only recently gained attention \cite{Li:2024iyj}: the possibility of constraining classes of dimension-6 operators that do not interfere with the SM by analyzing regions of phase space where the SM contribution is suppressed. We refer to this emerging direction as the ``high-energy precision beyond interference" program.

We can identify a relatively large class of dimension-six operators that contributes to the DY process, through contact interactions involving two quarks and two leptons, through modified vertices of leptons and/or quarks with neutral electroweak gauge bosons and the Higgs, and through modified gauge bosons and Higgs propagators.

Following the classification outlined in Ref.~\cite{Grzadkowski:2010es}, the so-called Warsaw basis, the full set of such operators is summarized in Table \ref{tab:Warsaw}. Here the notation is the following: $\varphi$ is the Higgs doublet, $D_{\mu}$ is the covariant derivative acting on $\varphi$, $\ell$ and $e$ are the left-handed lepton doublet and right-handed lepton singlet, respectively, $q$ and $u$ ($d$) are the left-handed quark doublet and right-handed up (down) quark singlet, respectively, $\tilde{\varphi} = i \sigma_2 \varphi^*$ is the conjugate Higgs doublet, $W_{\mu\nu}^I$ and $B_{\mu\nu}$ are the field strength tensors of the SU(2)$_L$ and U(1)$_Y$ gauge groups, respectively, and $\tau^{I}$ are the SU(2)$_L$ generators in the fundamental representation. Furthermore, the indices $p,r,s,t$ run over the different flavors of leptons and quarks, the lowercase indices $j,k$ and the capital index $I$ run over the fundamental and adjoint representations of SU(2)$_L$, respecitvely, and $\epsilon_{jk}$ is the Levi-Civita symbol in two dimensions. Finally, we clarify that we refer to chirality-breaking operators as those higher dimensional interaction terms that are not invariant under a $SU(N_f)_L\times SU(N_f)_R$ transformation, with $N_f$ the number of flavors of the SM and $SU(N_f)_L$ ($SU(N_f)_R$) acting non trivially only on left-handed (right-handed) spinors. 

\begin{table}[t!] 
\centering
\begin{tabular}{|c|c||c|c|}
\hline
\hline
\multicolumn{2}{|c||}{\textbf{1.\,$ \varphi^4 D^2 $}} &
\multicolumn{2}{c|}{\textbf{2.\,$ \psi^2 \varphi^3 $}} \\
\hline
$ Q_{\varphi\square} $ & $ (\varphi^\dagger \varphi)\square(\varphi^\dagger \varphi) $ &
$ Q_{e\varphi} $  & $ (\varphi^\dagger \varphi)(\bar{\ell}_p e_r \varphi) $ \\
$ Q_{\varphi D} $ & $ (\varphi^\dagger D^\mu \varphi)^* (\varphi^\dagger D_\mu \varphi) $ &
$ Q_{u\varphi} $  & $ (\varphi^\dagger \varphi)(\bar{q}_p u_r \tilde{\varphi}) $ \\
&&
$ Q_{d\varphi} $  & $ (\varphi^\dagger \varphi)(\bar{q}_p d_r \varphi) $ \\
\hline
\hline
\multicolumn{2}{|c||}{\textbf{3.\,$ \psi^2 X \varphi $}} &
\multicolumn{2}{c|}{\textbf{4.\,$ \psi^2 \varphi^2 D $}} \\
\hline
$ Q_{eW} $             & $ (\bar{\ell}_p \sigma^{\mu\nu} e_r) \tau^I \varphi W_{\mu\nu}^I $ &
$ Q_{\varphi l}^{(1)} $ & $ (\varphi^\dagger i\!\!\stackrel{\leftrightarrow}{D}_\mu \varphi)(\bar{\ell}_p \gamma^\mu \ell_r) $ \\
$ Q_{eB} $             & $ (\bar{\ell}_p \sigma^{\mu\nu} e_r) \varphi B_{\mu\nu} $ &
$ Q_{\varphi l}^{(3)} $ & $ (\varphi^\dagger i\!\!\stackrel{\leftrightarrow}{D}_\mu^I \varphi)(\bar{\ell}_p \tau^I \gamma^\mu \ell_r) $ \\
$ Q_{uW} $             & $ (\bar{q}_p \sigma^{\mu\nu} u_r)\, \tau^I \tilde{\varphi}\, W_{\mu\nu}^I $ &
$ Q_{\varphi e} $      & $ (\varphi^\dagger i\!\!\stackrel{\leftrightarrow}{D}_\mu \varphi)(\bar{e}_p \gamma^\mu e_r) $ \\
$ Q_{uB} $             & $ (\bar{q}_p \sigma^{\mu\nu} u_r)\, \tilde{\varphi}\, B_{\mu\nu} $ &
$ Q_{\varphi q}^{(1)} $ & $ (\varphi^\dagger i\!\!\stackrel{\leftrightarrow}{D}_\mu \varphi)(\bar{q}_p \gamma^\mu q_r) $ \\
$ Q_{dW} $             & $ (\bar{q}_p \sigma^{\mu\nu} d_r)\, \tau^I \varphi\, W_{\mu\nu}^I $ &
$ Q_{\varphi q}^{(3)} $ & $ (\varphi^\dagger i\!\!\stackrel{\leftrightarrow}{D}_\mu^I \varphi)(\bar{q}_p \tau^I \gamma^\mu q_r) $ \\
$ Q_{dB} $             & $ (\bar{q}_p \sigma^{\mu\nu} d_r)\, \varphi\, B_{\mu\nu} $ &
$ Q_{\varphi u} $      & $ (\varphi^\dagger i\!\!\stackrel{\leftrightarrow}{D}_\mu \varphi)(\bar{u}_p \gamma^\mu u_r) $ \\
&&
$ Q_{\varphi d} $      & $ (\varphi^\dagger i\!\!\stackrel{\leftrightarrow}{D}_\mu \varphi)(\bar{d}_p \gamma^\mu d_r) $ \\
&&
$ Q_{\varphi ud} $     & $ i(\tilde{\varphi}^\dagger D_\mu \varphi)(\bar{u}_p \gamma^\mu d_r) $ \\
\hline
\hline
\multicolumn{2}{|c||}{\textbf{5.\,$(\bar{L}L)(\bar{L}L)$}} &
\multicolumn{2}{c|}{\textbf{6.\,$(\bar{R}R)(\bar{R}R)$}} \\
\hline
$ Q_{lq}^{(1)} $ & $ (\bar{\ell}_p \gamma_\mu \ell_r)(\bar{q}_s \gamma^\mu q_t) $ &
$ Q_{eu} $ & $ (\bar{e}_p \gamma_\mu e_r)(\bar{u}_s \gamma^\mu u_t) $ \\
$ Q_{lq}^{(3)} $ & $ (\bar{\ell}_p \gamma_\mu \tau^I \ell_r)(\bar{q}_s \gamma^\mu \tau^I q_t) $ &
$ Q_{ed} $ & $ (\bar{e}_p \gamma_\mu e_r)(\bar{d}_s \gamma^\mu d_t) $ \\
\hline
\hline
\multicolumn{2}{|c||}{\textbf{7.\,$(\bar{L}L)(\bar{R}R)$}} &
\multicolumn{2}{|c|}{\textbf{8.\,$(\bar{L}R)(\bar{R}L)$ and $(\bar{L}R)(\bar{L}R)$}} \\
\hline
$ Q_{lu} $ & $ (\bar{\ell}_p \gamma_\mu \ell_r)(\bar{u}_s \gamma^\mu u_t) $ &
$ Q_{ledq} $ & $ (\bar{\ell}^j_p e_r)(\bar{d}_s q^j_t) $ \\
$ Q_{ld} $ & $ (\bar{\ell}_p \gamma_\mu \ell_r)(\bar{d}_s \gamma^\mu d_t) $ &
$ Q_{lequ}^{(1)} $ & $ (\bar{\ell}^j_p e_r) \epsilon_{jk} (\bar{q}^k_s u_t) $ \\
$ Q_{qe} $ & $ (\bar{q}_p \gamma_\mu q_r)(\bar{e}_s \gamma^\mu e_t) $ &
$ Q_{lequ}^{(3)} $ & $ (\bar{\ell}^j_p \sigma_{\mu\nu} e_r) \epsilon_{jk} (\bar{q}^k_s \sigma^{\mu\nu} u_t) $ \\
\hline
\end{tabular}
\caption{Operators contributing to the Drell-Yan process in the Warsaw basis. The notation is the same as in Ref.~\cite{Grzadkowski:2010es}. The operators are grouped according to their field content and the first column indicates the operator's name.}
\label{tab:Warsaw}
\end{table}

Despite, as anticipated, the number of operators that may contribute to the DY process at hadron colliders is rather large, their behavior can be very different, depending on the kinematic region. It is useful to classify them according to two main features: the first is whether they contribute at the linear or at the quadratic level (\textit{i.e.}~whether they interfere with the Standard Model (SM) amplitude or not), the second is their behavior with respect to the energy of the process, \textit{i.e.}~whether they lead to amplitudes that grow with energy or not. Indeed, we generally expect the following different contributions the amplitude: 
\begin{equation}\label{eq:Behaviors}
    \frac{v^{2}}{\Lambda^{2}}, \quad \frac{v\sqrt{s}}{\Lambda^{2}}, \quad \frac{s}{\Lambda^{2}}.
\end{equation}
Let us comment the different operator classes in turn.

\begin{itemize}
    \item \textbf{1.\,$ \varphi^4 D^2 $}\\
    The operators in this class modify the Higgs and gauge boson propagators on-shell, generating corrections to their kinetic terms suppressed by $v^{2}/\Lambda^{2}$. The Drell-Yan amplitude involving these operators is proportional to $v^{2}/\Lambda^{2}$, leading to an interference with the SM amplitude proportional to $v^{2}/\Lambda^{2}$, and to a contribution to the squared amplitude proportional to $v^{4}/\Lambda^{4}$. These contributions do not grow with energy, and are therefore better constrained in very precise experiments at the threshold for on-shell production of SM bosons (such as LEP or future lepton colliders).
    \item \textbf{2.\,$ \psi^2 \varphi^3 $}\\
    These operators are $v^{2}/\Lambda^{2}$ corrections to the Yukawa couplings of the SM. 
    Obviously, their flavor structure may determine stronger or weaker bounds from flavor physics. However, their contribution to the DY amplitude is proportional to $v^{2}/\Lambda^{2}$ and not enhanced with the energy. We can therefore neglect the contribution of these operators to the DY process.
    \item \textbf{3.\,$ \psi^2 X \varphi $} \label{sect: dipoles}\\ 
    This class of operators contains dipole interactions between fermions and gauge bosons. Some combinations of these operators, such as those leading to electron dipole moments, are very strongly constrained by low-energy experiments. However, as we will see, other combinations can effectively be constrained from DY measurements. Indeed, the contribution of these chirality-breaking operators to the DY amplitude is proportional to $v\sqrt{s}/\Lambda^{2}$. Due to the chirality structure of the fermionic tensor, the DY amplitude generated by these operators does not interfere with the SM amplitude. Therefore, their contribution to the squared amplitude scales as $v^{2}s/\Lambda^{4}$, growing linearly with $s$.
    \item \textbf{4.\,$ \psi^2 \varphi^2 D $}\\
    These operators have the form of product of a fermion and a Higgs current. They lead to a $v^{2}/\Lambda^{2}$ corrections to the interactions of fermions with gauge bosons and the Higgs. Their contribution to the DY amplitude does interfere with the SM ones, but does not grow with energy, generating terms proportional to $v^{2}/\Lambda^{2}$ and $v^{4}/\Lambda^{4}$ in the squared amplitude. As for the operators of class 1, these contributions are better constrained in precise experiments at the threshold for on-shell production of SM bosons.
    \item \textbf{5.\,$(\bar{L}L)(\bar{L}L)$, 6.\,$(\bar{R}R)(\bar{R}R)$, 7.\,$(\bar{L}L)(\bar{R}R)$}
    The seven operators in these classes correspond to the product of two fermion currents. They are the only operators that lead to contributions to the DY amplitude that are proportional to $s/\Lambda^{2}$, and, at the same time, interfere with the SM amplitudes. Therefore, they generate contributions to the squared amplitude proportional to $s/\Lambda^{2}$ and $s^{2}/\Lambda^{4}$. These operators have been extensively studied in the context of the DY process at the LHC \cite{Farina:2016rws,Torre:2020aiz,Panico:2021vav,Grossi:2024tou,Corbett:2025oqk}.
    \item \textbf{8.\,$(\bar{L}R)(\bar{R}L)$ and $(\bar{L}R)(\bar{L}R)$}\label{sect: cb4f}\\
    The last class of operators contains chirality-breaking interactions of two scalar or two tensor fermion bilinears. We will refer to them as ``scalar'' and ``tensor'' four-fermion operators, respectively. Their contribution to the DY amplitude has the same energy behavior of the current-current four-fermion operators.\footnote{The growth with energy of the four-fermion interactions is entirely determined by the spinor polarization of the external fermions, which carry a power of $\sqrt{E}$, and not by the Lorentz structure of the operator, so that scalar, vector, and tensor four-fermion operators all lead to amplitudes that grow with the square of the energy $s$.}
    However, due to the chirality structure of the fermionic tensors, they do not interfere with the SM amplitude, leading to contributions to the squared amplitude that are proportional to $s^{2}/\Lambda^{4}$. These operators, together with those in class 3, can be constrained by DY measurements, and will constitute the main focus of the present paper. In the following, we denote them as $\psi^4_{(8)}$.
\end{itemize}

Summarizing, operators in classes 1, 2, and 4 lead to contributions to the DY amplitude that do not grow with energy and are therefore better constrained in experiments at the SM boson production threshold, such as LEP or future lepton colliders, or by flavor measurements. The operators in classes 5, 6, and 7 lead to contributions to the DY amplitude that grow with energy and that interfere with SM amplituedes, therefore generating $s/\Lambda^{2}$ contributions in the squared amplitude. The effect of these operators in the DY process have been extensively studied in the literature. Finally, operators in classes 3 and 8 lead to contributions to the DY amplitude that grow with energy, but do not interfere with the SM amplitude, generating contributions to the squared amplitude that are proportional to $v^{2}s/\Lambda^{4}$ (for class 3) and $s^{2}/\Lambda^{4}$ (for class 8), respectively. Considering only the first two generations, which are those relevant for the present paper, we rewrite these operators, as
\begin{equation}\label{eq:lag}
    \begin{array}{lll}
        \dst\mathcal{L}_{\psi^2 X\varphi}
        &\hspace{-3mm}=&\hspace{-3mm}\dst \frac{1}{\Lambda^2}\sum_{p= e,\mu}\Big[(\overline{\ell}_{p}\sigma^{\mu\nu}e_{p})\!\left(c_{e_{p}B}B_{\mu\nu}+c_{e_{p}W}\tau^{I}W_{\mu\nu}^{I}\right)\varphi + (\overline{q}_{p}\sigma^{\mu\nu}d_{p})\!\left(c_{d_{p}B}B_{\mu\nu}+c_{d_{p}W}\tau^{I}W_{\mu\nu}^{I}\right)\varphi \vspace{2mm}\\
        && \dst\quad+(\overline{q}_{p}\sigma^{\mu\nu}u_{p})\!\left(c_{u_{p}B}B_{\mu\nu}+c_{u_{p}W}\tau^{I}W_{\mu\nu}^{I}\right)\tilde{\varphi} + \text{h.c.}\Big], \vspace{4mm}\\
        \dst\mathcal{L}_{\psi^4}
        &\hspace{-3mm}=&\hspace{-3mm} \dst\frac{1}{\Lambda^2} \sum_{\scriptsize\begin{array}{c} p = e,\mu\\ r = u,d,s,c \end{array}} \Big[c_{\ell_{p} e_{p}dq}\,(\bar{\ell}^{j}_{p} e_{p})(\bar{d}_{r} q^{j}_{r}) + c^{(1)}_{\ell_{p} e_{p}qu}(\bar{\ell}^{j}_{p} e_{p}) \epsilon_{jk} (\bar{q}^{k}_{r} u_{r}) \vspace{2mm}\\
        && \dst + c^{(3)}_{\ell_{p} e_{p}qu}(\bar{\ell}^{j}_{p} \sigma_{\mu\nu} e_{p}) \epsilon_{jk} (\bar{q}^{k}_{r} \sigma^{\mu\nu} u_{r}) + \text{h.c.}\Big],
    \end{array}
\end{equation}
where we have introduced the dimensionless Wilson coefficients $c_{i}$ and we have made explicit our assumption of a diagonal flavor structure in the lepton sector and of flavor universality in the quark sector (the Wilson coefficients of the four-fermion operators do not depend on the quark flavor index $r$). The UV origin of these operators can be diverse and a detailed discussion of explicit models is beyond the phenomenological scope of this paper. However, we present in Appendix \ref{app::UVmodels} a brief overview of possible UV completions that can generate some of these operators, possibly with a sizable coefficient.

As it will be clear in the following, the main idea of this paper is on the one hand to profit of the energy growth of the chirality-breaking operators discussed above, and on the other hand to isolate their contribution from the SM and the SMEFT operators that interfere with the SM by considering observables where the contribution of the latter is suppressed. To do so, we focus on the angular distribution of the lepton pair in the DY process, which is sensitive to the chirality structure of the operators, and on the violation of the Lam-Tung relation~\cite{Lam:1978pu,Lam:1978zr}, which guarantees a suppression of the SM contribution, and of the NP contribution with the same chirality structure of the SM, to the relevant observables. On top of considering observables that are suppressed for the SM due to the Lam-Tung relation, we also exploit the growing-with-energy behavior of the chirality-breaking operators to enhance their contribution with respect to the SM. 

The paper is organized as follows. In Section \ref{sec::LamTung} we describe the angular decomposition of the Drell-Yan differential cross-section, the related angular coefficients, and the Lam-Tung relation, which is a key ingredient in our analysis. In Section \ref{sec::ExpStatus} we review the experimental status of the measurements of the angular coefficients in DY, while in Section \ref{sec::SMPredictions} we present our precise estimates of the cross-section and of the angular observables in the SM, both in the transverse momentum and invariant mass distributions of the lepton pair, for the LHC and HL-LHC. In Section \ref{sec::SMEFTpredictions} we discuss the Lam-Tung relation and compute the contribution of the chirality-breaking SMEFT operators to the relevant observables. In Section \ref{sec::SMEFTanalysis} we present our analysis of the LHC and HL-LHC sensitivity on the relevant SMEFT operators, based on a likelihood fit to the aforementioned observables. In Section \ref{sec::Results} we summarize the results, while in Section \ref{sec::Conclusions} we draw our conclusions. The paper is complemented by a number of appendices reporting our choices on the definition of the observables and some analytic results relevant for our analysis.

\section{Angular coefficients and the $A_{0}-A_{2}$ observable}\label{sec::LamTung}
The fully-differential inclusive cross-section of the neutral DY process $pp\to \ell^+\ell^-$, where $\ell$ is a lepton\footnote{Throughout this paper, we use $\ell$ to denote either an electron or a muon. Although electrons and muons differ from an experimental point of view, they have been treated equivalently in our theoretical discussion.}, can be expressed in terms of the invariant mass $m_{\ell\ell}$, the transverse momentum $p_{T}^{\ell\ell}$, the rapidity $y_{\ell\ell}$ of the lepton pair, and the polar and azimuthal angles $\theta$ and $\phi$ of the negative charged lepton in the Collins-Soper (CS) frame~\cite{Collins:1977iv} (see Appendix \ref{app:CSframe} for details), as
\begin{equation}\label{eq:CrossSection}
\begin{array}{lll}
    \dst \frac{d \sigma}{dm^2_{\ell\ell}dp_{T}^{\ell\ell}dy_{\ell\ell}d\cos{\theta}d\phi} & = & \dst \frac{3}{16\pi}\frac{d \sigma}{dm^2_{\ell\ell}dp_{T}^{\ell\ell}dy_{\ell\ell}} \vspace{2mm}\\
     && \dst \left\{(1 + \cos^2\theta) + \frac{1}{2}A_0(1 - 3\cos^2\theta) + A_1\sin{2\theta}\cos{\phi}\right. \vspace{2mm}\\
     && \dst + \frac{1}{2}A_2\sin^2\theta\cos{2\phi} + A_3\sin{\theta}\cos{\phi} + A_4\cos{\theta} \vspace{2mm}\\
     && \dst \left.+ A_5\sin^2\theta\sin{2\phi} + A_6\sin{2\theta}\sin{\phi} + A_7\sin{\theta}\sin{\phi} \right\}\,.
\end{array}  
\end{equation}
The coefficients $A_{l}$ are dimensionless and can be extracted from experimental data by making a fit to the angular distribution of the lepton pair. This expression is completely general and also holds in the presence of QCD and electroweak corrections.

A crucial observable for this process is $A_{0} - A_{2}$, which, according to the so-called Lam-Tung relation \cite{Lam:1978pu}, is expected to vanish in the Standard Model (SM) up to corrections of order $\mathcal{O}(\alpha_\text{S}) $. As a result, the SM contribution to this observable is suppressed by QCD effects at order $\mathcal{O}(\alpha_\text{S}^2) $, making it an excellent probe for potential new physics (NP) effects. Moreover, as noted in Ref.~\cite{Li:2024iyj}, the only dimension-six operators that can violate the Lam-Tung relation are the chirality-breaking ones, specifically those belonging to classes 3 and 8 discussed in the previous section. For these reasons, $ A_0 - A_2 $ stands out as a distinctive observable where the contribution of such operators can be effectively isolated. This makes it a particularly suitable candidate for constraining them, especially when compared to the differential cross-section, where their effect is subleading relative to operators that do interfere with the SM. Accordingly, we propose a systematic study of the sensitivity of the LHC and HL-LHC to chirality-breaking SMEFT operators via the $A_{0} - A_{2}$ observable.

The $A_{l}$ coefficients are defined through expectation values of suitable combinations of spherical harmonics in the $\theta$ and $\phi$ angles, normalized to the cross-section integrated over such angles.\footnote{see the latest discussion in Refs.~\cite{Li:2025fom,Gauld:2024glt,Lyubovitskij:2025oig,Bandeira:2025log,Petriello:2025lur}.} In particular, for each bin of $m_{\ell\ell}$, $p_{T,\ell \ell}$, and $y_{\ell\ell}$, one can define
\begin{equation}\label{eq:Pl}
    S_{l} = \langle P_{l}(\cos{\theta},\phi)\rangle=\frac{\dst\int d \sigma(\cos{\theta},\phi,m_{\ell\ell},p_{T}^{\ell\ell},y_{\ell\ell}) d\cos{\theta}\,d\phi\, P_{l}(\cos{\theta},\phi)}{\dst\int d \sigma(\cos{\theta},\phi,m_{\ell\ell},p_{T}^{\ell\ell},y_{\ell\ell}) d\cos{\theta}\,d\phi}\,.
\end{equation}
It can be verified that the $P_{l}(\cos{\theta},\phi)$ choice that projects the amplitude onto the $A_{l}$ coefficients is given by
\begin{equation}\label{eq:Sl}
    \begin{array}{l}
        \dst S_{0} = \langle P_{0}(\cos{\theta},\phi) \rangle = \langle \frac{1}{2}\left(1-3\cos^2{\theta}\right) \rangle = \frac{3}{20}\left(A_{0}-\frac{2}{3}\right), \vspace{1mm}\\
        \dst S_{1} = \langle P_{1}(\cos{\theta},\phi) \rangle = \langle \sin{2\theta}\cos{\phi} \rangle = \frac{1}{5}A_{1}, \vspace{1mm}\\
        \dst S_{2} = \langle P_{2}(\cos{\theta},\phi) \rangle = \langle \sin^2{\theta}\cos{2\phi} \rangle = \frac{1}{10}A_{2}, \vspace{1mm}\\
        \dst S_{3} = \langle P_{3}(\cos{\theta},\phi) \rangle = \langle \sin{\theta}\cos{\phi} \rangle = \frac{1}{4}A_{3}, \vspace{1mm}\\
        \dst S_{4} = \langle P_{4}(\cos{\theta},\phi) \rangle = \langle \cos{\theta} \rangle = \frac{1}{4}A_{4}, \vspace{1mm}\\
        \dst S_{5} = \langle P_{5}(\cos{\theta},\phi) \rangle = \langle \sin^2{\theta}\sin{2\phi} \rangle = \frac{1}{5}A_{5}, \vspace{1mm}\\
        \dst S_{6} = \langle P_{6}(\cos{\theta},\phi) \rangle = \langle \sin{2\theta}\sin{\phi} \rangle = \frac{1}{5}A_{6}, \vspace{1mm}\\
        \dst S_{7} = \langle P_{7}(\cos{\theta},\phi) \rangle = \langle \sin{\theta}\sin{\phi} \rangle = \frac{1}{4}A_{7},
    \end{array}
\end{equation}
so that the $A_{l}$ coefficients can be extracted from the $S_{l}$ ones as
\begin{equation}\label{eq:Al}
    \begin{array}{ll}
        \dst A_{0} = \frac{20}{3} S_{0} + \frac{2}{3}, \qquad\qquad\qquad   & A_{1} = 5 S_{1}, \vspace{1mm}\\
        \dst A_{2} = 10 S_{2},                                              & \dst A_{3} = 4 S_{3}, \vspace{1mm}\\
        \dst A_{4} = 4 S_{4},                                               & \dst A_{5} = 5 S_{5}, \vspace{1mm}\\
        \dst A_{6} = 5 S_{6},                                               & \dst A_{7} = 4 S_{7}.
    \end{array}
\end{equation}
The exact relation between the $P_{l}(\cos{\theta},\phi)$ functions and the ordinary spherical harmonics is given in Appendix \ref{app::SphArm}.
The SM contribution to the $A_{l}$ observables can be explitly written as a function of
\begin{equation}
\label{eq:Sl_SM}
        S_{l}^{\text{SM}}
        = \frac{\dst\int d\sigma^{\text{SM}}(\cos{\theta},\phi,m_{\ell\ell},p_{T}^{\ell\ell},y_{\ell\ell}) d\cos{\theta}\,d\phi\, P_{l}(\cos{\theta},\phi)}{\dst\int d \sigma^{\text{SM}}(\cos{\theta},\phi,m_{\ell\ell},p_{T}^{\ell\ell},y_{\ell\ell}) d\cos{\theta}\,d\phi}\,.
\end{equation}
As already mentioned, in this paper we do not make use of the lepton system rapidity $y_{\ell\ell}$ distribution, and we only consider the dependence of the $A_{0}-A_{2}$ observables either on the transverse momentum $p_{T}^{\ell\ell}$ or on the invariant mass $m_{\ell\ell}$ distribution of the lepton pair.

When any of the dipole, scalar, or tensor four-fermion operators, corresponding to the Wilson coefficient $c_{\text{NP}}$ is switched on, we can rewrite the differential cross-section, integrated over $y_{\ell\ell}$, as
\begin{equation}
\label{eq:differential_cross_section}
\begin{split}
    \displaystyle d\sigma(\cos{\theta},\phi,m_{\ell\ell},p_{T}^{\ell\ell}) 
    &= d\sigma^{\text{SM}}(\cos{\theta},\phi,m_{\ell\ell},p_{T}^{\ell\ell}) + \left(\frac{c_{\text{NP}}}{\Lambda^2}\right)^2 d\tilde{\sigma}^{\text{NP}}(\cos{\theta},\phi,m_{\ell\ell},p_{T}^{\ell\ell})\,,
\end{split}
\end{equation}
where $d\tilde{\sigma}^{\text{NP}}$ is, up to the NP coefficient $\left(c_{\text{NP}}/\Lambda^2\right)^2$, the differential cross-section generated by the NP operator, which does not interfere with the SM. 
Substituting Eq.~\eqref{eq:differential_cross_section} into Eq.~\eqref{eq:Pl}, we obtain the $S_{l}$ observable in the presence of NP contributions:
\begin{equation}
\label{eq:Sl_SMEFT}
\begin{array}{lll}
    \displaystyle S_{l}
    &=& \dst \frac{\dst \int d\sigma^{\text{SM}}(\cos{\theta},\phi,m_{\ell\ell},p_{T}^{\ell\ell}) d\cos{\theta}\,d\phi\,P_{l}(\cos{\theta},\phi)}{\dst d\sigma^{\text{SM}}(m_{\ell\ell},p_{T}^{\ell\ell})+\left(\frac{c_{\text{NP}}}{\Lambda^2}\right)^2 d\tilde{\sigma}^{\text{NP}}(m_{\ell\ell},p_{T}^{\ell\ell})}\vspace{2mm}\\
    &&\dst +\left(\frac{c_{\text{NP}}}{\Lambda^2}\right)^2\frac{\dst \int d\tilde{\sigma}^{\text{NP}}(\cos{\theta},\phi,m_{\ell\ell},p_{T}^{\ell\ell}) d\cos{\theta}\,d\phi\,P_{l}(\cos{\theta},\phi)}{\dst d\sigma^{\text{SM}}(m_{\ell\ell},p_{T}^{\ell\ell})+\left(\frac{c_{\text{NP}}}{\Lambda^2}\right)^2 d\tilde{\sigma}^{\text{NP}}(m_{\ell\ell},p_{T}^{\ell\ell})},
\end{array}
\end{equation}
where we made explicit the integration of the denominator over $\cos\theta$ and $\phi$ by omitting the corresponding variables in the argument of the differential cross-sections. The $A_{l}$ observables in the presence of NP contributions can then be obtained from Eq.~\eqref{eq:Sl_SMEFT} by using Eq.~\eqref{eq:Al}.

In the limit $(c_{\text{NP}}/\Lambda^2)\to 0$, this expression reduces to the SM one in Eq.~\eqref{eq:Sl_SM}. Moreover, we know that, in this limit, $A_{0}-A_{2}$ is non-zero only starting from $\mathcal{O}(\alpha_\text{S}^{2})$, due to the Lam-Tung relation. This means that, to correctly account for the SM contribution to the $A_0-A_2$ observable, we need to evaluate the SM differential cross-section at order $\mathcal{O}(\alpha_\text{S}^{2})$. Equation \eqref{eq:Sl_SMEFT} also shows that generally, even though the SM and the SMEFT contributions do not interfere, their contributions to the $S_{l}$ observables, and therefore also to the $A_{l}$ and $A_{0}-A_{2}$ combinations, can not be simply disentangled, since the denominator contains both the SM and the NP contributions. In other words, the behavior with respect to the Wilson coefficients $c_{\text{NP}}$ of the $A_{l}$ observables can be approximated as quadratic only in the limit where the NP contribution to the differential cross-section is subdominant with respect to the SM one. This may not be the case in the regions of the phase space where the SMEFT contribution is enhanced, for example at large invariant mass or transverse momentum of the lepton pair, which turn out to be the most sensitive regions to constrain the chirality-breaking operators. Therefore, in our analysis we present results obtained both with the exact relation of Eq.~\eqref{eq:Sl_SMEFT}, which would of course be subject to corrections from operators of dimension higher than six that we do not consider, and with its quadratic approximation in $c_{\text{NP}}/\Lambda^{2}$, which corresponds to a fixed order truncation in the SMEFT expansion. This allows one to assess the possible impact of higher-order terms in the SMEFT expansion on the bounds that we derive.

In order to compute the $p_{T}^{\ell\ell}$ and $m_{\ell\ell}$ dependence of the $A_{l}$ observables we proceeded as follows. We estimated the SM differential cross-section for the process $pp\to \ell^+\ell^-$ at $\mathcal{O}(\alpha_\text{S}^{2})$ by generating events with the \texttt{MiNNLO}$_{\text{\texttt{PS}}}$ \cite{Monni:2019whf,Monni:2020nks} Monte Carlo generator, implemented in the \texttt{POWHEG} framework \cite{Nason:2004rx,Frixione:2007vw,Alioli:2010xd}. We have checked that the inclusion of showering effects, available in the \texttt{MiNNLO}$_{\text{\texttt{PS}}}$ framework through a matching with \texttt{PYTHIA8} \cite{Sjostrand:2007gs,Sjostrand:2014zea,Bierlich:2022pfr} showering, does not affect the results of our analysis, and we therefore proceeded computing parton level events with a lepton--anti-lepton pair and up to two jets in the final state. Notice that, since we consider the differential $p_{T}^{\ell\ell}$ distribution for a finite $p_{T}^{\ell\ell}\geq 10$ GeV cut, we could have in principle computed the NLO QCD correction to the process $pp\to \ell^+\ell^- + j$. However, we decided to use the \texttt{MiNNLO}$_{\text{\texttt{PS}}}$ framework which allowed us to directly generate weighted events in Les Houches (lhe) format \cite{Alwall:2006yp} and to properly check the effect of the parton showering.

The generated SM events were used to define the differential SM cross-sections $d\sigma^{\text{SM}}/dp_{T}^{\ell\ell}$ within a window $80\text{ GeV}<m_{\ell\ell}<100$ GeV and $d\sigma^{\text{SM}}/dm^2_{\ell\ell}$ in the region $p_{T}^{\ell\ell}>10$ GeV, by binning the events in the relevant kinematic variable.

In order for the differential cross-section to be defined at finite $p_{T}^{\ell\ell}$, the SMEFT contribution was computed analytically  by considering the process $pp\to \ell^+\ell^- + j$ at the leading order in QCD. The partonic cross-sections for all sub-processes have been calculated using the \texttt{FeynCalc}~\cite{Shtabovenko:2020gxv} and \texttt{FeynArts}~\cite{Hahn:2000kx} packages in \texttt{Mathematica}~\cite{Mathematica}, after truncating the squared amplitude at $\mathcal{O}(1/\Lambda^4)$, which is equivalent to considering a single NP insertion. The hadronic cross-section was then derived by convoluting these results with the PDFs. For this purpose, we employed the \texttt{ManeParse} package~\cite{Clark:2016jgm} for \texttt{Mathematica}, which provides all the necessary tools for PDF integration. We adopted the PDF set \texttt{NNPDF31\_nnlo\_as\_0118} (ID 303600 in the LHAPDF~\cite{Buckley:2014ana} set), neglecting contributions from the $b$ and $t$ quarks. Details about the analytical expression of the integral can be found in appendix~\ref{app::PDFintegration}. This procedure allowed us to compute the differential cross-sections $d\tilde{\sigma}^{\text{NP}}/dp_{T}^{\ell\ell}$ and $d\tilde{\sigma}^{\text{NP}}/dm^2_{\ell\ell}$ in the same kinematic region as the SM ones, and to use them in Eq.~\eqref{eq:Sl_SMEFT} to compute the $S_{l}$ and $A_{l}$ observables in the presence of NP contributions. The evaluation of Eq.~\eqref{eq:Sl_SMEFT} also required the calculation of the projections of the cross-section onto the corresponding angular polynomials $P_{l}(\cos{\theta},\phi)$. This was done by Monte Carlo integration for the SM, using the generated events, and by numerical integration of the analytic expression for the NP contributions.

\subsection{Monte Carlo simulation and uncertainty estimation}\label{subsec:uncestimate}
The angular observables defined in the previous section can be predicted theoretically by Monte Carlo estimation of the integrals. Given a sample of weighted events $i$ in bin $I$ with weights $w_{i}$, the cross-section in that bin is given by
\begin{equation}
    \sigma_{I} = \sum_{i\in I} w_{i}\,.
\end{equation}
The angular observables in bin $I$ are then defined from the projections of the cross-section in that bin on the corresponding polynomial functions $P_{l}(\cos{\theta},\phi)$, which appear in the numerator of Eq.~\eqref{eq:Pl}. We can compute this by multiplying the MC weights $w_{i}$ by the value of the relevant polynomial functions $P_{l}^{i}(\cos{\theta},\phi)$, given in Eq.~\eqref{eq:Al}, computed for the $i$-th event kinematic, and by defining a new set of angular weights
\begin{equation}\label{eq:angweights}
    w_{i}^{(l)} = w_{i} P_{l}^{i}(\cos{\theta},\phi)\,.
\end{equation}
Then, the differential cross-section projected over the polynomial functions $P_{l}(\cos{\theta},\phi)$, that we simply denote by $\sigma_{I}^{(l)}$, is given by
\begin{equation}
    \sigma_{I}^{(l)} = \sum_{i\in I} w_{i}^{(l)}\,.
\end{equation}
This projection can be used to estimate the value of the $S_{l,I}$ observables of Eq.~\eqref{eq:Sl} in each bin $I$
\begin{equation}
    S_{l,I} = \frac{\sigma_{I}^{(l)}}{\sigma_{I}}
\end{equation}
and, in turn, to compute the $A_{l,I}$ observables through Eq.~\eqref{eq:Al}.

Extracting a meaningful bound on the new physics contribution from the differential angular observables $A_{l}$, and in particular from the $A_0-A_2$ combination, requires a careful estimate of the expected experimentally measured values and uncertainties of the observables, especially for what concerns the SM contribution. Given the very large statistics available at the LHC for the DY process, at relatively low $p_{T}^{\ell\ell}$ and for di-lepton invariant masses $m_{\ell\ell}$ not far above the $Z$ boson mass, the uncertainty is dominated by experimental systematics. Since no recent analysis of the multi-differential DY cross-section is available, and based on the existing experimental results, we decided to make the simplifying assumption of a flat, uncorrelated, $3\%$ uncertainty on the differential cross-section in all bins. 

Concerning the statistical uncertainty, which becomes dominant in corner regions of the phase space, such as very large $p_{T}^{\ell\ell}$ and/or $m_{\ell\ell}$ much above the $Z$ boson mass, we generated enough statistics to cover the expected experimental statistics at the LHC with $\mathcal{L}=300\text{ fb}^{-1}$,\footnote{This required the generation of about a billion events for the $p_{T}^{\ell\ell}$ distribution, which is at the edge of what we could afford with our computing resources. The number of events generated for the $m_{\ell\ell}$ distribution was substantially lower, thanks to the ability of the \texttt{MiNNLO}$_{\text{\texttt{PS}}}$ generator to cut on the di-lepton invariant mass.} which is the target integrated luminosity of the LHC and extrapolated with a factor of square root of ten to the HL-LHC with $\mathcal{L}=3\text{ ab}^{-1}$. The central values were kept unchanged in this extrapolation.

In order to estimate the central values and the statistical uncertainties in each bin of the two distributions, we have proceeded as follows. We denoted with $n_{I}^{\text{MC}}$ the available number of MC-generated events in each bin $I$ and with
\begin{equation}
    \sigma_{I}^{\text{MC}} = \sum_{i\in I}w_{i}
\end{equation}
the MC-estimated cross-section in that bin. We then fixed an initial value for the number of events corresponding to the experimental observation with a given integrated luminosity $\mathcal{L}$
\begin{equation}
    n_{I}^{\text{exp}} = \sigma_{I}^{\text{MC}}\times \mathcal{L}\,.
\end{equation}
To estimate the expected value and the standard deviation of the differential cross-section and of the angular observables in each bin (of the relevant kinematic distribution), we simulated pseudo-experiments, each corresponding to the statistics expected at the LHC with $\mathcal{L}=300\text{ fb}^{-1}$.\footnote{The result for $\mathcal{L}=3\text{ ab}^{-1}$ has been obtained assuming that the variance scales with the ration of the number of events and therefore scaling the error with a square root of ten.} Pseudo experiments were drawn from the available MC sample with replacement.\footnote{When the available MC sample is smaller or roughly equal in size than the expected statistics, such as in low $p_{T}^{\ell\ell}$ bins, our procedure may underestimate the standard deviation. On the other hand these are the bins where the expected statistical uncertainty is much smaller than the expected systematic uncertainty, so that the effect of such underestimation becomes negligible.} In order to fix the correct value of the cross-section, for each pseudo-experiment we rescaled the value of the corresponding weights. In order to also introduce a fluctuation in the normalization, and not only in the shape, we also considered fluctuations in the normalization, computed from the original MC sample. Obviously, fluctuations on the normalization only affect the cross-section estimate and not the angular coefficients, which are built from ratios. The detailed procedure is outlined in the following:
\begin{itemize}
\item For each pseudo-experiment $p$ and bin $I$, we consider a Poisson fluctuation of the number of MC events 
\begin{equation}
    n_{I}^{\text{MC},(p)} = \text{Poisson}(\mu = n_{I}^{\text{MC}})\,,
\end{equation}
and draw, with replacement, $n_{I}^{\text{MC},(p)}$ events from the available MC events. We use this sample to estimate a fluctuation of the cross-section in that bin, that we denote by $\sigma_{I}^{\text{MC},(p)}$, and that we use as fluctuating normalization of the cross-section in that bin.
\item For each pseudo-experiment $p$ and bin $I$, we consider a Poisson fluctuation of the number of experimental events
\begin{equation}
    n_{I}^{\text{exp},(p)} = \text{Poisson}(\mu = n_{I}^{\text{exp}})\,,
\end{equation}
and draw, with replacement, $n_{I}^{\text{exp},(p)}$ from the available MC events. 
\item For each pseudo-experiment $p$ and bin $I$, we use the corresponding $n_{I}^{\text{exp},(p)}$ events to define a new set of normalized weights
\begin{equation}\label{eq:weightsnorm}
w_{i}^{\text{exp},(p)} = \kappa_{I}^{(p)} w_{i}^{(p)}\quad\quad \forall i\in I\,,
\end{equation}
with normalization
\begin{equation}
    \kappa_{I}^{(p)} = \frac{\sigma_{I}^{\text{MC},(p)}}{\sum_{i\in I}w_{i}^{(p)}}\,.
\end{equation}
Notice that this definition does automatically set the corresponding cross-section to its correct fluctuating normalization:
\begin{equation}\label{eq:normcrosssection}
    \sigma_{I}^{\text{exp},(p)} \equiv \sum_{i\in I}w_{i}^{\text{exp},(p)} = \kappa_{I}^{(p)}\sum_{i\in I}w_{i}^{(p)} = \sigma_{I}^{\text{MC},(p)}\,.
\end{equation}
\item For each pseudo-experiment $p$ and bin $I$, we use the corresponding $n_{I}^{\text{exp},(p)}$ events to define a new set of normalized angular weights, analog to those in Eq.~\eqref{eq:angweights},
\begin{equation}\label{eq:angweightsnorm}
w_{i}^{\text{exp},(l),(p)} = \kappa_{I}^{(p)} w_{i}^{(p)} P_{l}^{i,(p)}(\cos{\theta},\phi)\quad\quad \forall i\in I\,.
\end{equation}
\item For each pseudo-experiment $p$ and bin $I$, we use the new set of normalized angular weights to estimate the cross-section angular projections as
\begin{equation}\label{eq:normangularproj}
    \sigma_{I}^{\text{exp},(l),(p)} = \sum_{i\in I} w_{i}^{\text{exp},(l),(p)} = \kappa_{I}^{(p)} \sum_{i\in I} w_{i}^{(p)} P_{l}^{i,(p)}(\cos{\theta},\phi)\,.
\end{equation}
\item For each pseudo-experiment $p$ and bin $I$, we use Eqs. \eqref{eq:normcrosssection} and \eqref{eq:angweightsnorm} to compute the $S_{l}$'s obsevables as
\begin{equation}
    S_{l,I}^{\text{exp},(p)} = \frac{\sigma_{I}^{\text{exp},(l),(p)}}{\sigma_{I}^{\text{exp},(p)}} = \frac{\sum_{i\in I} w_{i}^{(p)} P_{l}^{i,(p)}(\cos{\theta},\phi)}{\sum_{i\in I} w_{i}^{(p)}}\,,
\end{equation}
which is, as expected, independent on the cross-section normalization $\kappa_{I}^{(p)}$.
\item We estimate the central values and (one dimensiona) covariance matrix of the differential cross-section as
\begin{equation}
    \mu_{\sigma_{I}} = \text{Mean}_{p\in \mathcal{P}} \left(\sigma_{I}^{\text{MC},(p)}\right)\,,\qquad (\Sigma_{\sigma_{I},\sigma_{I}})^{\text{stat}} = \text{Var}_{p\in \mathcal{P}} \left(\sigma_{I}^{\text{MC},(p)}\right)\,,
\end{equation}
where we denoted with $\text{Mean}_{p\in \mathcal{P}}$ and $\text{Var}_{p\in \mathcal{P}}$ the mean and variance computed over a set of 100 pseudo-experiments.
\item We estimate the central values and covariance matrix of the $S_{l}$'s observables as
\begin{equation}
    \mu_{S_{l,I}} = \text{Mean}_{p\in \mathcal{P}} \left(S_{l,I}^{\text{exp},(p)}\right)\,,\qquad (\Sigma_{S_{l,I},S_{m,I}})^{\text{stat}} = \text{Cov}_{p\in \mathcal{P}} \left(S_{l,I}^{\text{exp},(p)},S_{m,I}^{\text{exp},(p)}\right)\,,
\end{equation}
where we denoted with $\text{Mean}_{p\in \mathcal{P}}$ and $\text{Cov}_{p\in \mathcal{P}}$ the mean and covariance matrix computed over a set of 100 pseudo-experiments.
\item Analogously, we estimate the central values and covariance matrix of the $A_{l}$ observables through Eq.~\eqref{eq:Al}.
\end{itemize}

Once the central values $\mu_{\mathcal{O}}$ and covariance matrix $(\Sigma_{\mathcal{O},\mathcal{O}'})^{\text{stat}}$ of the observables $\mathcal{O}$ are computed, we combine the uncertainties with the corresponding quantities parametrizing the systematic uncertainty $(\Sigma_{\mathcal{O},\mathcal{O}'})^{\text{syst}}$ and get the final estimate of the observables as
\begin{equation}
    \mu_{\mathcal{O}}\pm \sqrt{\Sigma_{\mathcal{O},\mathcal{O}}} 
    \qquad\text{with}\qquad \Sigma_{\mathcal{O},\mathcal{O}'} = \Sigma_{\mathcal{O},\mathcal{O}'}^{\text{stat}}+\Sigma_{\mathcal{O},\mathcal{O}'}^{\text{syst}}\,.
\end{equation}

\section{Experimental status}\label{sec::ExpStatus}
The measurements relevant for the present paper fall in the realm of precision measurements for new physics searches.\footnote{We omitted LHCb measurements of the inclusive cross-section since they generally have less statistics. Obviously, in the case in which the forward region becomes more relevant the LHCb measurements become important.} Such precision measurements usually take a long term experimental effort and this is why not many such measurements are already available from LHC Run 2 and Run 3. Here we briefly summarize the most relevant existing measurements in the Drell-Yan channel, and use the information we can gather from them to motivate our assumptions on the uncertainties discussed in the previous section. Table~\ref{tab:ExpStatus} lists the measurements of differential cross-sections and angular coefficients in di-lepton final states performed over the past decade.

\begin{table}[t!]
    \centering
    \small
    \begin{tabular}{c|c|c|c|c|c}
        \toprule
        Collab. & Year & Energy & Luminosity & Observable & Ref. \\ 
        \midrule
            \multirow{1}{*}{ATLAS}
            & \multirow{1}{*}{2024}
            & \multirow{1}{*}{$5.02,13$ TeV}
            & \multirow{1}{*}{$255,338$ pb$^{-1}$}
            & \multirow{1}{*}{$d\sigma_{\ell\ell}/d p_{T,ll}$}
            & \multirow{1}{*}{\citewithtip{ATLAS:2024nrd}{ATLAS Collaboration, G. Aad et al., “Precise measurements of W- and Z-boson transverse momentum spectra with the ATLAS detector using pp collisions at $\sqrt{s} = 5.02$ TeV and 13 TeV”, Eur. Phys. J. C 84 (2024) 1126, arXiv:2404.06204.}} \\
            \multirow{1}{*}{ATLAS}
            & \multirow{1}{*}{2024}
            & \multirow{1}{*}{$8$ TeV}
            & \multirow{1}{*}{$20.2$ fb$^{-1}$}
            & \multirow{1}{*}{$d\sigma_{\ell\ell}/d p_{T,ll} dy_{\ell\ell}$}
            & \multirow{1}{*}{\citewithtip{ATLAS:2023lsr}{ATLAS Collaboration, G. Aad et al., “A precise measurement of the Z-boson double-differential transverse momentum and rapidity distributions in the full phase space of the decay leptons with the ATLAS experiment at $\sqrt{s} = 8$ TeV”, Eur. Phys. J. C 84 (2024) 315, arXiv:2309.09318.}} \\
            \multirow{1}{*}{CMS}
            & \multirow{1}{*}{2023}
            & \multirow{1}{*}{$13$ TeV}
            & \multirow{1}{*}{$36.3$ fb$^{-1}$}
            & \multirow{1}{*}{$d\sigma_{\ell\ell}/d p_{T,ll} dm^2_{\ell\ell}$, $d\sigma_{\ell\ell}/d \phi^{*}_{\eta} dm^2_{\ell\ell}$}
            & \multirow{1}{*}{\citewithtip{CMS:2022ubq}{CMS Collaboration, A. Tumasyan et al., “Measurement of the mass dependence of the transverse momentum of lepton pairs in Drell-Yan production in proton-proton collisions at $\sqrt{s} = 13$ TeV”, Eur. Phys. J. C 83 (2023) 628, arXiv:2205.04897.}} \\
            \multirow{1}{*}{LHCb}
            & \multirow{1}{*}{2022}
            & \multirow{1}{*}{$13$ TeV}
            & \multirow{1}{*}{$5.1$ fb$^{-1}$}
            & \multirow{1}{*}{$A_{0},\ldots,A_{7}$ as functions of $p_{T,ll}$}
            & \multirow{1}{*}{\citewithtip{LHCb:2022tbc}{LHCb Collaboration, R. Aaij et al., “First Measurement of the $Z\to\mu^{+}\mu^{-}$ Angular Coefficients in the Forward Region of $pp$ Collisions at $\sqrt{s}=13$ TeV”, Phys. Rev. Lett. 129 (2022) 091801, arXiv:2203.01602.}} \\
            \multirow{1}{*}{ATLAS}
            & \multirow{1}{*}{2020}
            & \multirow{1}{*}{$13$ TeV}
            & \multirow{1}{*}{$36.1$ fb$^{-1}$}
            & \multirow{1}{*}{$d\sigma_{\ell\ell}/d p_{T,ll}$, $d\sigma_{\ell\ell}/d \phi^{*}_{\eta}$}
            & \multirow{1}{*}{\citewithtip{ATLAS:2019zci}{ATLAS Collaboration, G. Aad et al., “Measurement of the transverse momentum distribution of Drell–Yan lepton pairs in proton–proton collisions at $\sqrt{s} = 13$ TeV with the ATLAS detector”, Eur. Phys. J. C 80 (2020) 616, arXiv:1912.02844.}} \\
            \multirow{1}{*}{CMS}
            & \multirow{1}{*}{2019}
            & \multirow{1}{*}{$13$ TeV}
            & \multirow{1}{*}{$35.9$ fb$^{-1}$}
            & \multirow{1}{*}{$d\sigma_{\ell\ell}/d p_{T,ll}$, $d\sigma_{\ell\ell}/d \lvert y_{\ell\ell} \rvert $, $d\sigma_{\ell\ell}/d \phi^{*}_{\eta}$}
            & \multirow{1}{*}{\citewithtip{CMS:2019raw}{CMS Collaboration, A. M. Sirunyan et al., “Measurements of differential Z boson production cross-sections in proton-proton collisions at $\sqrt{s} = 13$ TeV”, JHEP 12 (2019) 061, arXiv:1909.04133.}} \\
            \multirow{1}{*}{CMS}
            & \multirow{1}{*}{2019}
            & \multirow{1}{*}{$13$ TeV}
            & \multirow{1}{*}{$2.8^{(\mu)}/2.3^{(e)}$ fb$^{-1}$}
            & \multirow{1}{*}{$d\sigma_{\ell\ell}/dm^2_{\ell\ell}$}
            & \multirow{1}{*}{\citewithtip{CMS:2018mdl}{CMS Collaboration, A. M. Sirunyan et al., “Measurement of the differential Drell-Yan cross-section in proton-proton collisions at $\sqrt{s} = 13$ TeV”, JHEP 12 (2019) 059, arXiv:1812.10529.}} \\
            \multirow{1}{*}{CMS}
            & \multirow{1}{*}{2018}
            & \multirow{1}{*}{$8$ TeV}
            & \multirow{1}{*}{$19.7$ fb$^{-1}$}
            & \multirow{1}{*}{$d\sigma_{\ell\ell}/d \phi^{*}_{\eta}$, $d\sigma_{\ell\ell}/d \phi^{*}_{\eta}d \lvert y_{\ell\ell} \rvert$}
            & \multirow{1}{*}{\citewithtip{CMS:2017lvz}{CMS Collaboration, A. M. Sirunyan et al., “Measurement of differential cross-sections in the kinematic angular variable $\phi^*$ for inclusive Z boson production in pp collisions at $\sqrt{s} = 8$ TeV”, JHEP 03 (2018) 172, arXiv:1710.07955.}} \\
            \multirow{1}{*}{ATLAS}
            & \multirow{1}{*}{2017}
            & \multirow{1}{*}{$8$ TeV}
            & \multirow{1}{*}{$20.2$ fb$^{-1}$}
            & \multirow{1}{*}{$d\sigma_{\ell\ell}/d m^2_{\ell\ell} d \lvert y_{\ell\ell} \rvert d\cos\theta^{*}$}
            & \multirow{1}{*}{\citewithtip{ATLAS:2017rue}{ATLAS Collaboration, M. Aaboud et al., “Measurement of the Drell-Yan triple-differential cross-section in pp collisions at $\sqrt{s} = 8$ TeV”, JHEP 12 (2017) 059, arXiv:1710.05167.}} \\
            \multirow{1}{*}{CMS}
            & \multirow{1}{*}{2017}
            & \multirow{1}{*}{$8$ TeV}
            & \multirow{1}{*}{$18.4$ fb$^{-1}$}
            & \multirow{1}{*}{$d\sigma_{\ell\ell}/d p_{T,ll}$}
            & \multirow{1}{*}{\citewithtip{CMS:2016mwa}{CMS Collaboration, V. Khachatryan et al., “Measurement of the transverse momentum spectra of weak vector bosons produced in proton-proton collisions at $\sqrt{s} = 8$ TeV”, JHEP 02 (2017) 096, arXiv:1606.05864.}} \\
            \multirow{1}{*}{ATLAS}
            & \multirow{1}{*}{2016}
            & \multirow{1}{*}{$8$ TeV}
            & \multirow{1}{*}{$20.3$ fb$^{-1}$}
            & \multirow{1}{*}{$d\sigma_{\ell\ell}/d m^2_{\ell\ell} d\lvert y_{\ell\ell} \rvert$, $d\sigma_{\ell\ell}/d m^2_{\ell\ell} d\lvert \Delta \eta_{\ell\ell} \rvert$}
            & \multirow{1}{*}{\citewithtip{ATLAS:2016gic}{ATLAS Collaboration, G. Aad et al., “Measurement of the double-differential high-mass Drell-Yan cross-section in pp collisions at $\sqrt{s} = 8$ TeV with the ATLAS detector”, JHEP 08 (2016) 009, arXiv:1606.01736.}} \\
            \multirow{1}{*}{ATLAS}
            & \multirow{1}{*}{2016}
            & \multirow{1}{*}{$8$ TeV}
            & \multirow{1}{*}{$20.3$ fb$^{-1}$}
            & \multirow{1}{*}{$A_{0},\ldots,A_{7}$ as functions of $p_{T,ll}$}
            & \multirow{1}{*}{\citewithtip{ATLAS:2016rnf}{ATLAS Collaboration, G. Aad et al., “Measurement of the angular coefficients in Z-boson events using electron and muon pairs from data taken at $\sqrt{s} = 8$ TeV with the ATLAS detector”, JHEP 08 (2016) 159, arXiv:1606.00689.}} \\
            \multirow{1}{*}{ATLAS}
            & \multirow{1}{*}{2016}
            & \multirow{1}{*}{$8$ TeV}
            & \multirow{1}{*}{$20.3$ fb$^{-1}$}
            & \multirow{1}{*}{$d\sigma_{\ell\ell}/d p_{T,ll}$,  $d\sigma_{\ell\ell}/d \phi^{*}_{\eta}$}
            & \multirow{1}{*}{\citewithtip{ATLAS:2015iiu}{ATLAS Collaboration, G. Aad et al., “Measurement of the transverse momentum and $\phi^{*}_{η}$ distributions of Drell–Yan lepton pairs in proton–proton collisions at $\sqrt{s} = 8$ TeV with the ATLAS detector”, Eur. Phys. J. C 76 (2016) 291, arXiv:1512.02192.}} \\
            \multirow{1}{*}{CMS}
            & \multirow{1}{*}{2015}
            & \multirow{1}{*}{$8$ TeV}
            & \multirow{1}{*}{$19.7$ fb$^{-1}$}
            & \multirow{1}{*}{$A_{0},\ldots,A_{7}$ as functions of $p_{T,ll}$}
            & \multirow{1}{*}{\citewithtip{CMS:2015cyj}{CMS Collaboration, V. Khachatryan et al., “Angular coefficients of Z bosons produced in pp collisions at $\sqrt{s} = 8$ TeV and decaying to $\mu^+ \mu^-$ as a function of transverse momentum and rapidity”, Phys. Lett. B 750 (2015) 154, arXiv:1504.03512.}} \\
            \multirow{1}{*}{CMS}
            & \multirow{1}{*}{2015}
            & \multirow{1}{*}{$8$ TeV}
            & \multirow{1}{*}{$19.7$ fb$^{-1}$}
            & \multirow{1}{*}{$d\sigma_{\ell\ell}/d p_{T,ll}$, $d\sigma_{\ell\ell}/d \lvert y_{\ell\ell} \rvert $}
            & \multirow{1}{*}{\citewithtip{CMS:2015hyl}{CMS Collaboration, V. Khachatryan et al., “Measurement of the Z boson differential cross-section in transverse momentum and rapidity in proton–proton collisions at 8 TeV”, Phys. Lett. B 749 (2015) 187, arXiv:1504.03511.}} \\
        \bottomrule
    \end{tabular}\normalsize
    \caption{Summary of the possibly relevant experimental measurements of Drell-Yan observable over the last ten years. For ease of reading, hovering over the info icon shows the bibliography item.}
    \label{tab:ExpStatus}
\end{table}

As it can immediately be seen from the table, only a single measurement of the angular coefficients from each experiment ATLAS, CMS, and LHCb is available. The first two measurements are at 8 TeV and an integrated luminosity of around 20 fb$^{-1}$, while the LHCb measurement in the forward region is the only measurement of the angular coefficient to date at an energy of $13$ TeV, with an integrated luminosity of 5.1 fb$^{-1}$. No updated measurements of the angular coefficients have yet been performed at 13 TeV, where the LHC has collected a much larger integrated luminosity. The measurements of the differential cross-section are more abundant, with different differential distributions measured at different energies and integrated luminosities. However, even the most recent measurements at 13 TeV are performed with a limited integrated luminosity of approximately 36 fb$^{-1}$, which is only a small fraction of the total integrated luminosity collected at 13 TeV during Run 2 (around 160 fb$^{-1}$).

Both the CMS \cite{CMS:2015cyj} and ATLAS \cite{ATLAS:2016rnf} measurements of the angular coefficients at 8 TeV are performed in the $Z$-boson mass peak region, and, to take into account the effect of the finite acceptance on the leptons, which affects the angular distributions in Eq.~\eqref{eq:CrossSection}, are performed through matching with Monte Carlo templates of the different angular observables in the $\theta$-$\phi$ plane. Entering in the details of the experimental measuremets is beyond the scope of this paper, and we refer the reader to the original experimental papers for more details. Here, we are only interested in understanding the expectation about the measurement uncertainties, which we need to consider for a realistic projection of the sensitivity to new physics in the angular coefficients.

From Tables 11 to 14 of Ref.~\cite{ATLAS:2016rnf}, we can see that the statistical and systematic uncertainties on the $p_{T}^{\ell\ell}$ distribution of the $A_{0}$, $A_{2}$, and $A_{0}-A_{2}$ observables are comparable over the whole spectrum, but in the last few bins, where, obviously, the statistical uncertainty dominates. Even though most of the systematic uncertainties, such as Monte Carlo statistics, have large margins for improvement, it is reasonable to expect that, with the full LHC and HL-LHC integrated luminosity, the systematic uncertainties will be dominant in the whole spectrum, but the last few bins, where the statistical uncertainty will still be dominant. We have already discussed in the previous section how we model the statistical uncertainty through pseudo-experiments simulation. Concerning the systematic uncertainty, assuming a projected uncertainty directly on the $A_{l}$ coefficients from the aforementioned angular analyses is not possible, since we can not guess how this will improve in the future. However, we can take the expected systematic uncertainties on the differential cross-section measurements from recent $13$ TeV analyses as a guideline. We consider in particular the ATLAS analysis of Ref.~\cite{ATLAS:2019zci} as reference analysis of the $p_{T}^{\ell\ell}$ spectrum and the CMS analysis of Ref.~\cite{CMS:2018mdl} as reference analysis of the $m_{\ell\ell}$ spectrum. We adopt the same binning for the two distributions and use the experience of those analyses to motivate our assumption on the systematic uncertainties on the differential cross-sections. Starting from such uncertainties, and assuming the same uncertainties for the cross-sections projections on the different angular polynomials, we propagate the uncertainty to the $A_{l}$ coefficients. The result is consistent with a systematic uncertainty that dominates in the low $p_{T}^{\ell\ell}$ and $m_{\ell\ell}$ region (above the $Z$ boson mass) and becomes subleading in the tails of the distributions.

In the next section, we show projections for the expected measurements of the $p_{T}^{\ell\ell}$ and $m_{\ell\ell}$ distributions, as well as the angular coefficients.

\section{Standard Model predictions}\label{sec::SMPredictions}
In this section, we present SM predictions for the $p_{T}^{\ell\ell}$ and $m_{\ell\ell}$ distributions of the cross-section and angular observables in the Drell-Yan process. Such predictions will be combined with predictions of the effect of new physics and used to derive projected limits on the chirality-breaking operator of Eq.~\eqref{eq:lag} in the next section.

\subsection{Transverse momentum distribution}\label{sec::pTdist}

\begin{figure}[t!]
    \centering
    \begin{minipage}{0.49\textwidth}
        \centering
        \includegraphics[width=\textwidth]{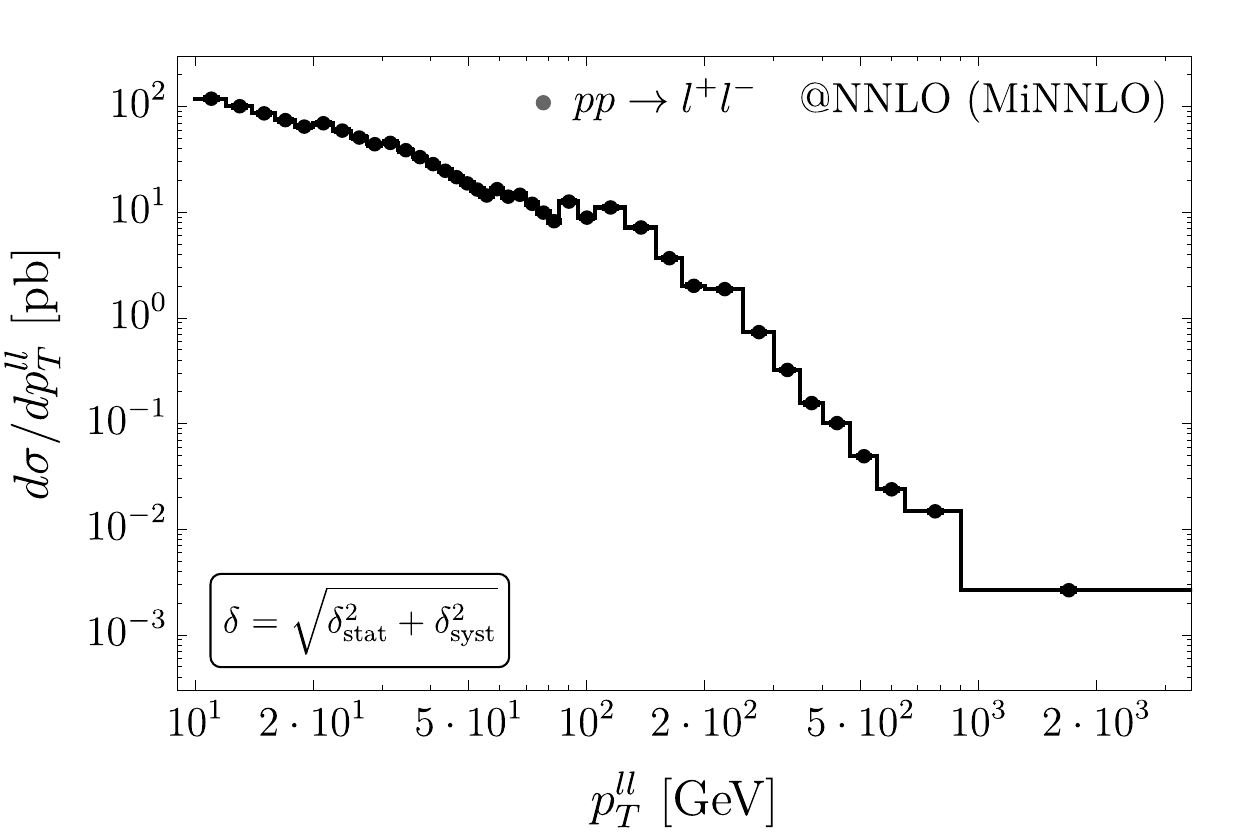}
    \end{minipage}
    \begin{minipage}{0.49\textwidth}
        \centering
        \includegraphics[width=\textwidth]{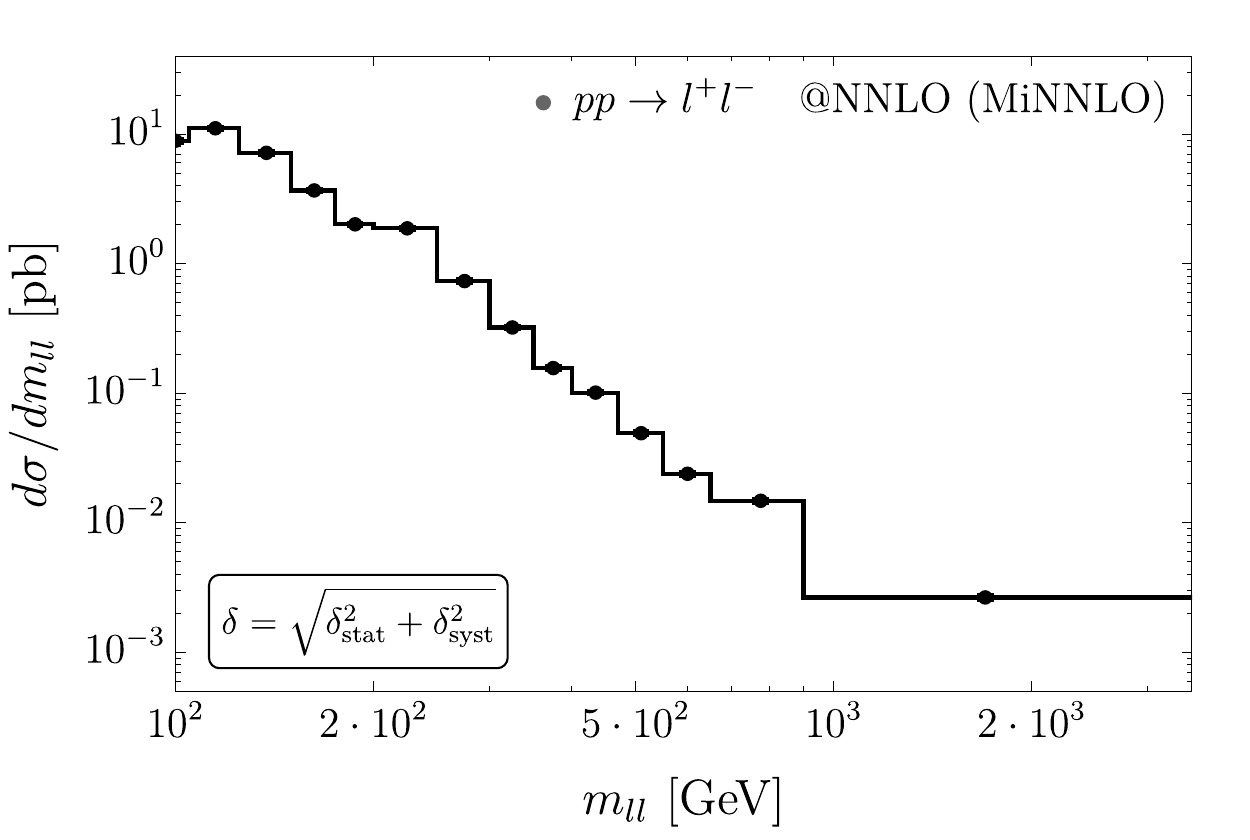}
    \end{minipage}
    \caption{Standard Model prediction for the $p_{T}^{\ell\ell}$ (left panel) and $m_{\ell\ell}$ (right panel) spectra in the DY process at the LHC at $13$ TeV with $300$ fb$^{-1}$. Uncertainties are too small to be visible, so the result is visually the same for both $300$ fb$^{-1}$ and $3$ ab$^{-1}$ of integrated luminosity. The exact numerical values (both central values and uncertainties) corresponding to $300$ fb$^{-1}$ and $3$ ab$^{-1}$ of integrated luminosity are given in Tables \ref{tab:ptxsdistributions} and \ref{tab:mllxsdistributions} of Appendix \ref{app::Tables}.}
    \label{fig:xsdistributions}
\end{figure}

The left panel of Figure \ref{fig:xsdistributions} shows the SM prediction for the di-lepton $p_{T}^{\ell\ell}$ spectrum in the DY process at the LHC at $13$ TeV, assuming a flat, uncorrelated systematic uncertainty of $3\%$, corresponding, for instance, to a $2\%$ systematic from the integrated luminosity measurement and about $2\%$ additional systematic uncertainty, including missing higher orders (scale variation), PDFs, Monte Carlo, and experimental uncertainties. These numbers for the systematic uncertainties are consistent with those reported in the aforementioned ATLAS analysis of Ref.~\cite{ATLAS:2019zci}, at least for the low and intermediate $p_{T}^{\ell\ell}$ region. The high $p_{T}^{\ell\ell}$ region has much larger systematic uncertainties, which are clearly statistically dominated. We do not have a clear prescription to project this uncertainty to the future measurements, and therefore we assume the flat $3\%$ uncertainty also for the high $p_{T}^{\ell\ell}$ region. Statistical and systematic uncertainties are added in quadrature, and the statistical uncertainty is derived through pseudo-experiments as explained in the previous section. Uncertainties are too small to be visible in the plot, so the result is visually the same for both $300$ fb$^{-1}$ and $3$ ab$^{-1}$ of integrated luminosity. The exact numerical values (both central values and uncertainties) corresponding to $300$ fb$^{-1}$ and $3$ ab$^{-1}$ of integrated luminosity are given in Table \ref{tab:ptxsdistributions} of Appendix \ref{app::Tables}.

Figure \ref{fig:ptangcoefficients} shows the SM prediction for the angular coefficients $A_{0}$, $A_{2}$, and $A_{0}-A_{2}$ as functions of $p_{T}^{\ell\ell}$ in the DY process at the LHC at $13$ TeV with two assumptions for the statistical uncertainty: the larger error bars represent the combined statistical and systematic uncertainty with $300$ fb$^{-1}$ of integrated luminosity, while the smaller error bars represent the combined statistical and systematic uncertainty with $3$ ab$^{-1}$ of integrated luminosity. In both cases, we assume a flat, uncorrelated systematic uncertainty of $3\%$ on the measurement of the cross-section and of its angular projections (denoted as $\sigma_{I}^{(l)}$ in Section \ref{sec::LamTung}) and propagate it to the angular coefficients as explained in the previous section. As before, statistical and systematic uncertainties are added in quadrature, and the statistical uncertainty is derived through pseudo-experiments as explained in the previous section.

Notice that the systematic uncertainty on the $A_{0}$ observable is much smaller than that on the $A_{2}$ observable, so that, even including correlation among them, the combined systematic uncertainty on $A_{0}-A_{2}$ is not largely affected, and remains dominated by the uncertainty on $A_{2}$. For this reason we only show the result under the assumption of no correlation among the systematic uncertainties of the different angular coefficients.

\begin{figure}[t!]
    \centering
    \begin{minipage}{0.49\textwidth}
        \centering
        \includegraphics[width=\textwidth]{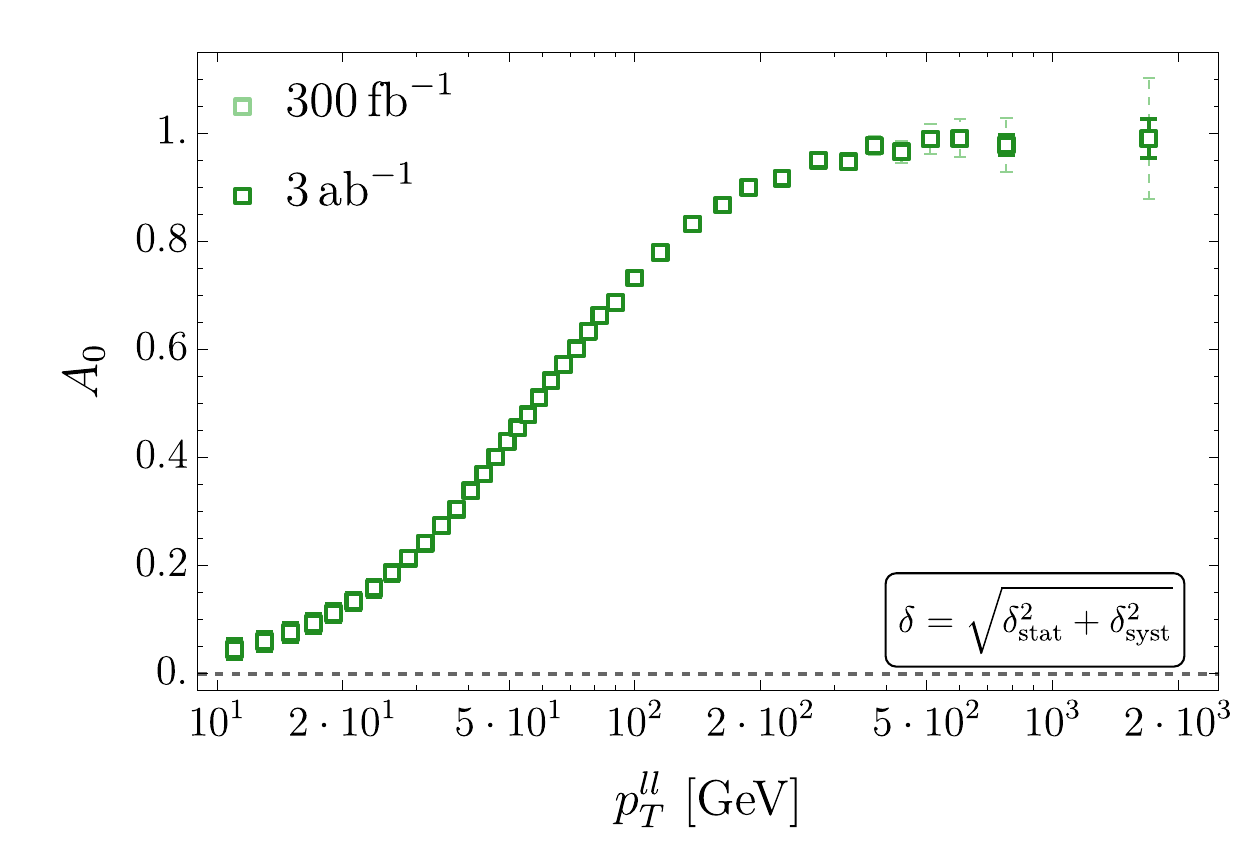}
    \end{minipage}
    \begin{minipage}{0.49\textwidth}
        \centering
        \includegraphics[width=\textwidth]{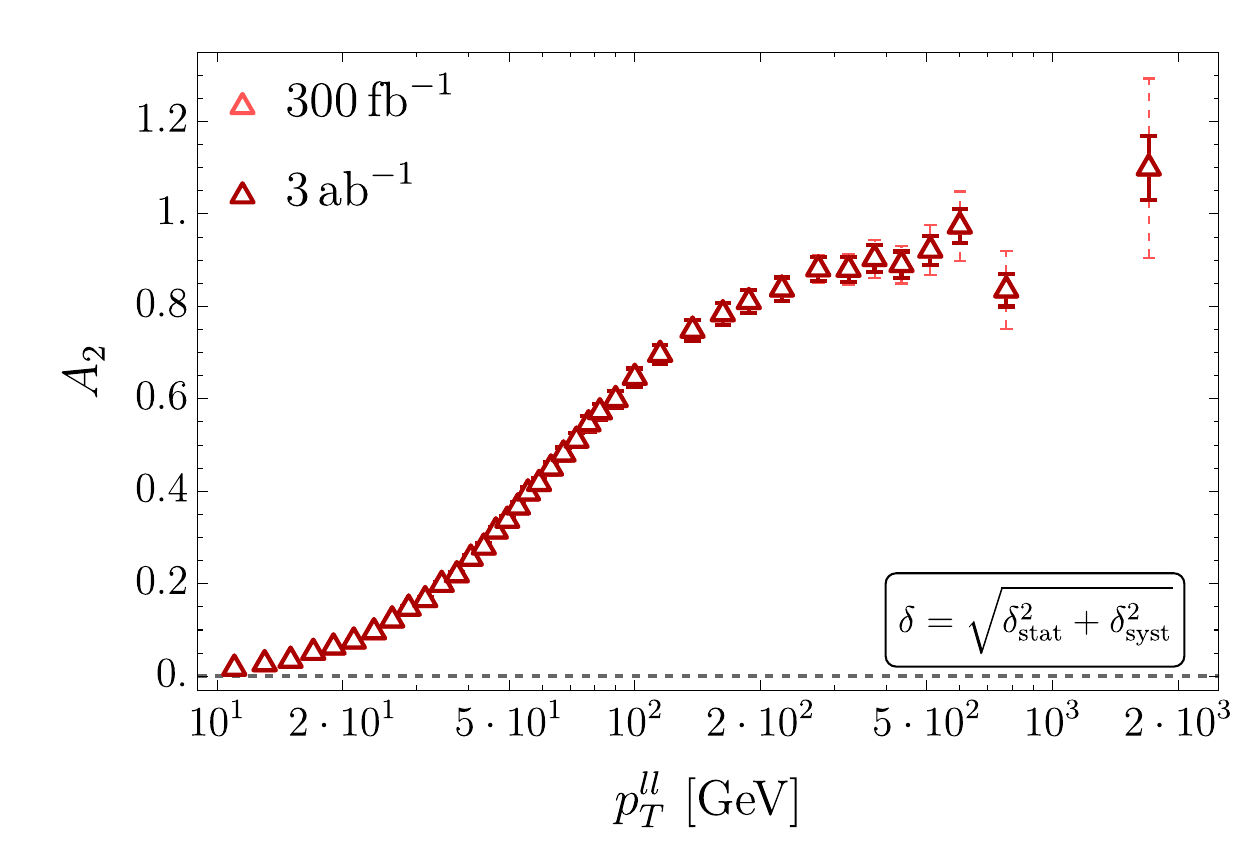}
    \end{minipage}
    \begin{minipage}{0.49\textwidth}
        \centering
        \includegraphics[width=\textwidth]{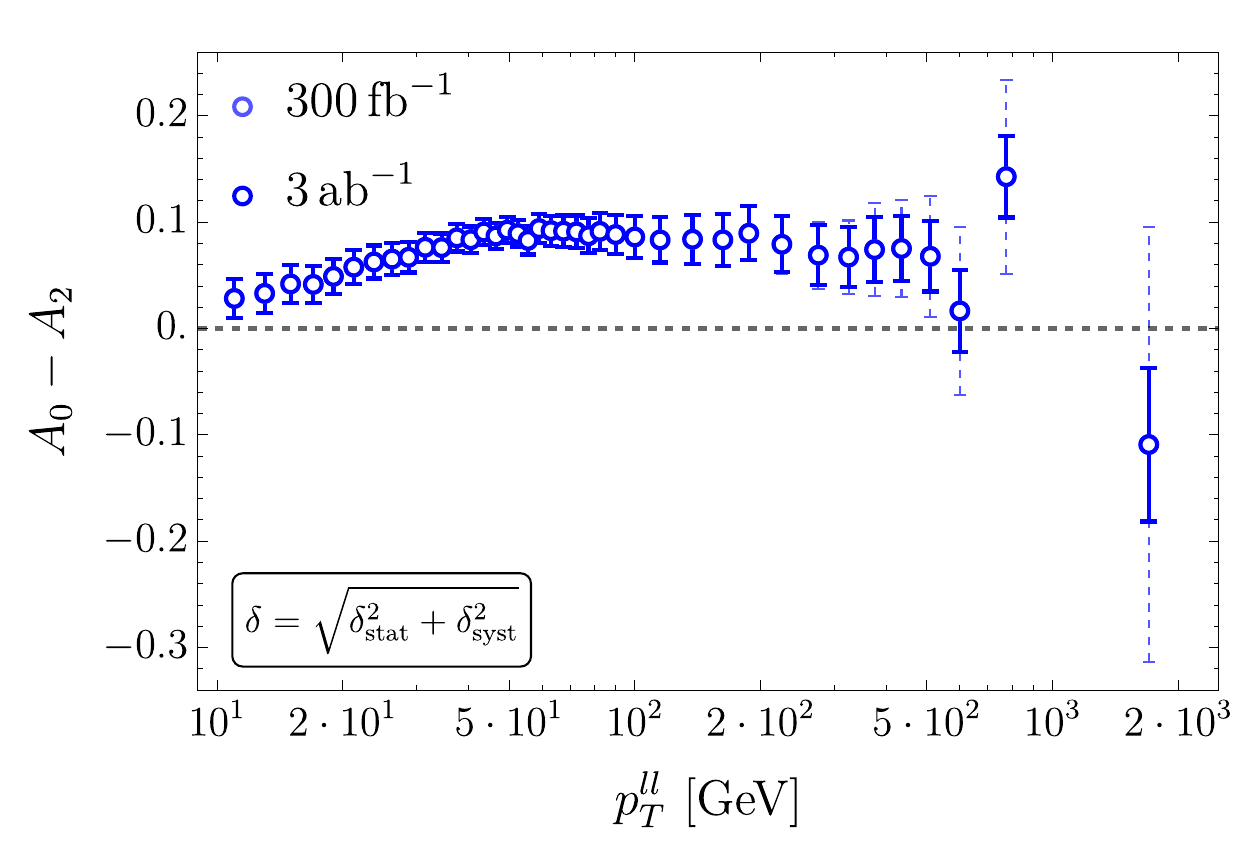}
    \end{minipage}
    \begin{minipage}{0.49\textwidth}
        \centering
        \includegraphics[width=\textwidth]{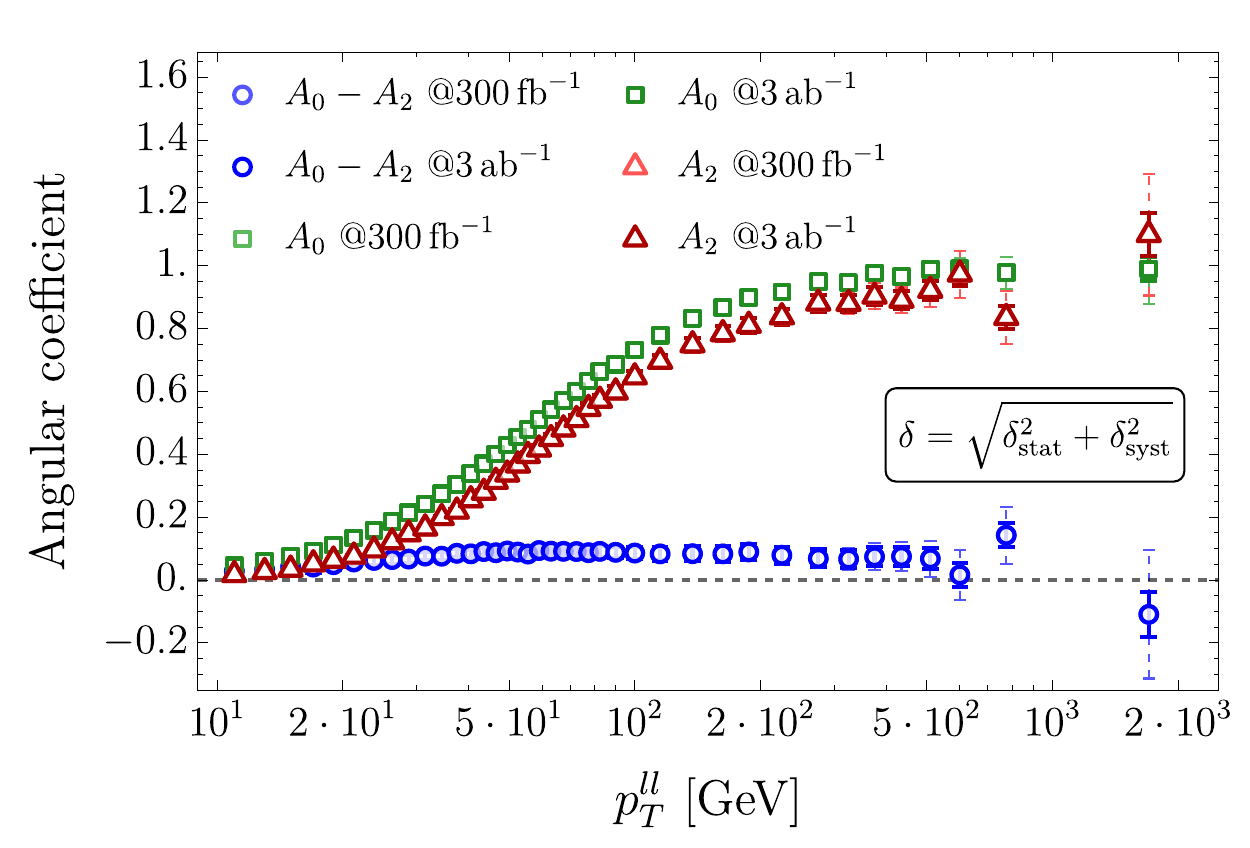}
    \end{minipage}
    \caption{Standard Model prediction for the $p_{T}^{\ell\ell}$ dependence of the angular observables $A_{0}$ (upper left), $A_{2}$ (upper right), and $A_{0}-A_{2}$ (lower left) in the DY process at the LHC at $13$ TeV with $300$ fb$^{-1}$ and $3$ ab$^{-1}$ both inclusive in rapidity. The lower right panel gives a combined view of the result. The numerical values of the observables appearing in the plots are given in Tables \ref{tab:ptangcoefficientsA0}, \ref{tab:ptangcoefficientsA2}, and \ref{tab:ptangcoefficientsA0mA2} of Appendix \ref{app::Tables}.}
    \label{fig:ptangcoefficients}
\end{figure}

All the numbers of the cross-sections and angular coefficients predictions shown in Figures \ref{fig:xsdistributions} (left) and \ref{fig:ptangcoefficients} are summarized in Tables \ref{tab:ptxsdistributions}, \ref{tab:ptangcoefficientsA0}, \ref{tab:ptangcoefficientsA2}, and \ref{tab:ptangcoefficientsA0mA2} of Appendix \ref{app::Tables}.

\subsection{Invariant mass distribution}\label{sec::mllDist}

\begin{figure}[t!]
    \centering
    \begin{minipage}{0.49\textwidth}
        \centering
        \includegraphics[width=\textwidth]{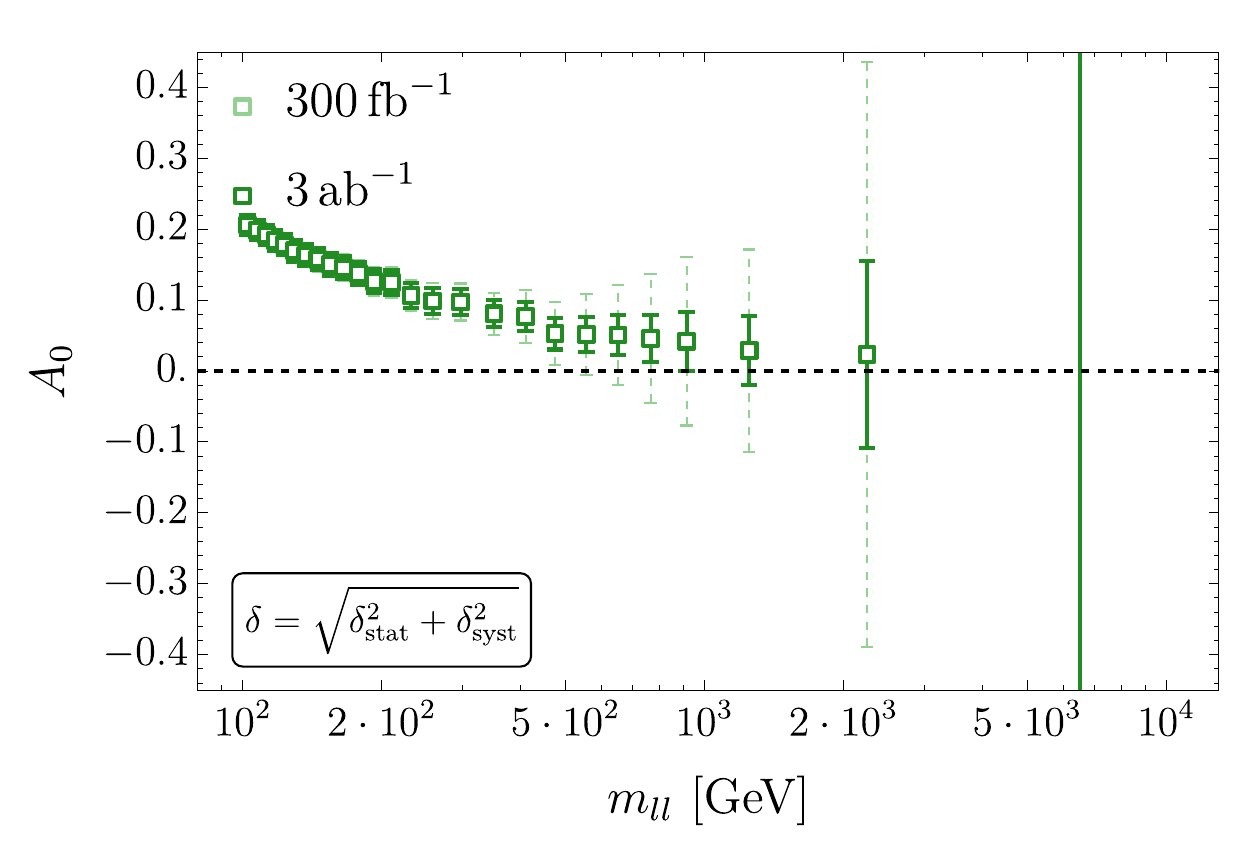}
    \end{minipage}
    \begin{minipage}{0.49\textwidth}
        \centering
        \includegraphics[width=\textwidth]{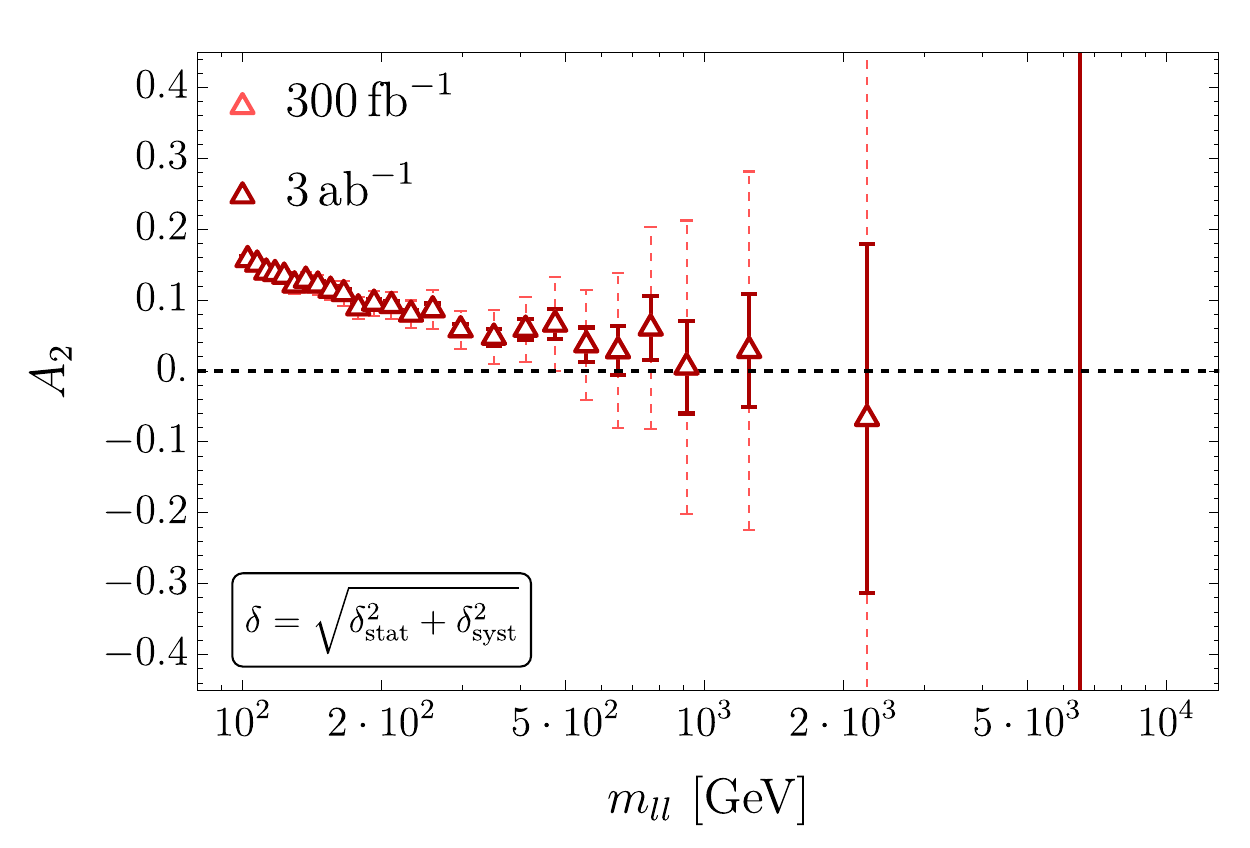}
    \end{minipage}
    \begin{minipage}{0.49\textwidth}
        \centering
        \includegraphics[width=\textwidth]{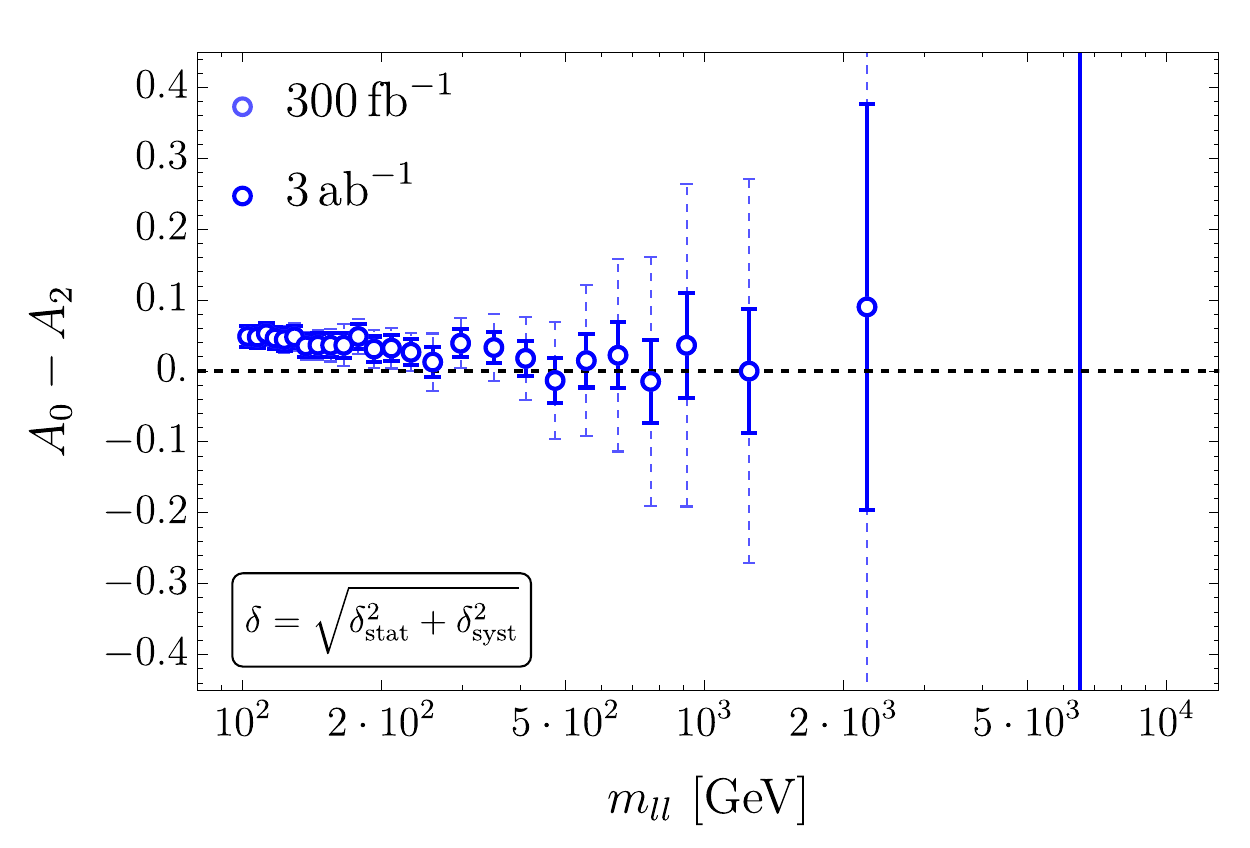}
    \end{minipage}
    \begin{minipage}{0.49\textwidth}
        \centering
        \includegraphics[width=\textwidth]{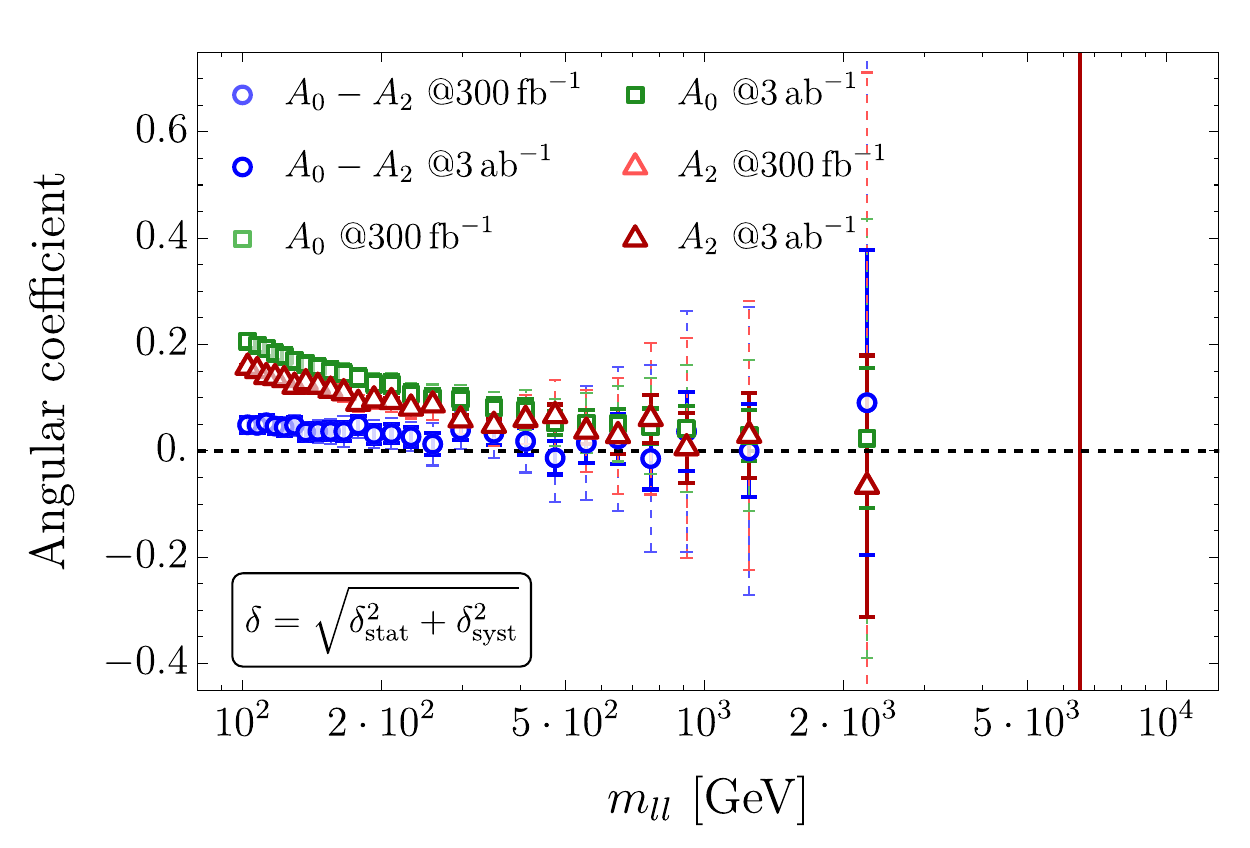}
    \end{minipage}
    \caption{Standard Model prediction for the $m_{\ell\ell}$ dependence of the angular observables $A_{0}$ (upper left), $A_{2}$ (upper right), and $A_{0}-A_{2}$ (lower left) in the DY process at the LHC at $13$ TeV with $300$ fb$^{-1}$ and $3$ ab$^{-1}$ both inclusive in rapidity. The lower right panel gives a combined view of the result. The numerical values of the observables appearing in the plots are given in Tables \ref{tab:mllangcoefficientsA0}, \ref{tab:mllangcoefficientsA2}, and \ref{tab:mllangcoefficientsA0mA2} of Appendix \ref{app::Tables}.}
    \label{fig:mllangcoefficients}
\end{figure}

The right panel of Figure \ref{fig:xsdistributions} shows the SM prediction for the di-lepton $m_{\ell\ell}$ spectrum in the DY process at the LHC at $13$ TeV, assuming, as in the case of the $p_{T}^{\ell\ell}$ distribution, a flat, uncorrelated systematic uncertainty of $3\%$ (see above). These numbers for the systematic uncertainties are consistent with those reported in the aforementioned CMS analysis of Ref.~\cite{CMS:2018mdl}, at least for systematic uncertainties that are not statistically dominated. In the high $m_{\ell\ell}$ region, we do not have a clear procedure to project the systematic uncertainty to the future measurements, and therefore we assume the flat $3\%$ uncertainty also in that region. As before, statistical and systematic uncertainties are added in quadrature, the statistical uncertainty is derived through pseudo-experiments as explained in the previous section, and uncertainties are too small to be visible in the plot, so the result is visually the same for both $300$ fb$^{-1}$ and $3$ ab$^{-1}$ of integrated luminosity. The exact numerical values (both central values and uncertainties) corresponding to $300$ fb$^{-1}$ and $3$ ab$^{-1}$ of integrated luminosity are given in Table \ref{tab:mllxsdistributions} of Appendix \ref{app::Tables}.

Figure \ref{fig:mllangcoefficients} shows the SM prediction for the angular coefficients $A_{0}$, $A_{2}$, and $A_{0}-A_{2}$ as functions of $m_{\ell\ell}$ in the DY process at the LHC at $13$ TeV with two assumptions for the statistical uncertainty: the larger error bars represent the combined statistical and systematic uncertainty with $300$ fb$^{-1}$, while the smaller error bars represent the combined statistical and systematic uncertainty with $3$ ab$^{-1}$. In both cases, we assume a flat, uncorrelated systematic uncertainty of $3\%$ on the measurement of the cross-section and of its angular projections (denotes as $\sigma_{I}^{(l)}$ in Section \ref{sec::LamTung}) and propagate it to the angular coefficients as explained in the previous section. As before, statistical and systematic uncertainties are added in quadrature, and the statistical uncertainty is derived through pseudo-experiments as explained in the previous section.

Notice that, also in this case, the systematic uncertainty on the $A_{0}$ observable is much smaller than that on the $A_{2}$ observable, so that, even including correlation among them, the combined systematic uncertainty on $A_{0}-A_{2}$ is not largely affected, and remains dominated by the uncertainty on $A_{2}$. For this reason we only show the result under the assumption of no correlation among the systematic uncertainties of the different angular coefficients.

All the numbers of the cross-sections and angular coefficients projections shown in Figures \ref{fig:xsdistributions} (right) and \ref{fig:mllangcoefficients} are summarized in Tables \ref{tab:mllxsdistributions}, \ref{tab:mllangcoefficientsA0}, \ref{tab:mllangcoefficientsA2}, and \ref{tab:mllangcoefficientsA0mA2} of Appendix \ref{app::Tables}.
\section{SMEFT predictions}\label{sec::SMEFTpredictions}
In this section we analyze the effects of the dimension-6 operators introduced in Section~\ref{sec::Intro} on the $A_0-A_2$ observable. In particular, we argue that the dipoles and scalar/tensor four-fermion operators are the only dimension-6 operators that can break the Lam-Tung relation at order $\mathcal{O}(\alpha_{\text{S}}^0)$, that is when QCD corrections are ignored.
This remains true also at order $\mathcal{O}(\alpha_{\text{S}})$, motivating our focus on just these two classes of operators. From now on, we omit the flavor index in the Wilson coefficients, and it is understood that all predictions and constraints hold separately, and identically, for electron and muons.

\subsection{Lam-Tung relation breaking at $\mathcal{O}(\alpha_\text{S}^0)$}

At the zero-th order in $\alpha_\text{S}$, and under the assumption of negligible intrinsic transverse momentum of the partons and of negligible effects from the possible QED radiation, the transverse momentum of the lepton pair vanishes. As a consequence, in the di-lepton rest frame the parton beams are collinear, and, in turn, the $\hat{z}$ axis of the CS frame (see Appendix \ref{app:CSframe} for its definition) lies in the direction of the beams and can be identified with the beam axis, denoted by $z$. Therefore, the process features azimuthal symmetry, the CS frame can be identified with the center of mass frame (the $\phi$ angle is arbitrary), and the angular distribution of the leptons in such frame only depends on the angle $\theta$, that is the angle between the negatively charged lepton and the $z$ axis. 

The fully differential cross-section of $pp\to \ell^{+}\ell^{-}+X$ in the CS frame in the SM is given by Eq.~\eqref{eq:CrossSection}. This five-differential cross-section reduces, at the zero-th order in $\alpha_\text{S}$, to the triple differential cross-section for the process $pp\to \ell^{+}\ell^{-}$. Integrating over the arbitrary angle $\phi$ we can write
\begin{equation} \label{eq:LOL}
    \int_{0}^{2\pi}d\phi\lim_{\alpha_S\to 0} \left( \frac{d \sigma}{dm^2_{\ell\ell}dp_{T}^{\ell\ell}dy_{\ell\ell}d\cos{\theta}d\phi} \right) = \left(\frac{d \sigma}{dm^2_{\ell\ell}dy_{\ell\ell}d\cos{\theta}} \right)_{\text{LO}}.
\end{equation}
Since, at $\mathcal{O}(\alpha_S^0)$, Eq.~\eqref{eq:CrossSection} only depends on $\theta$ and not on $\phi$, that means that it should hold for arbitrary values of $\phi$, then all the terms proportional to functions of $\phi$ must vanish, implying that only the coefficients $A_0$ and $A_4$ can be non-zero in this limit. Therefore, we can formally write the limit in Eq.~\eqref{eq:LOL} as\footnote{The additional factor $1/(2\pi)$ appearing in Eq.~\eqref{eq:CrossSection} has been absorbed by the integral over $\phi$.}
\begin{equation} \label{eq:LOLAngDist}
    \begin{array}{lll}
    \dst \int_{0}^{2\pi}d\phi\lim_{\alpha_S\to 0} \left( \frac{d \sigma}{dm^2_{\ell\ell}dp_{T}^{\ell\ell}dy_{\ell\ell}d\cos{\theta}d\phi} \right) & = &
    \dst \frac{3}{8}\left(\frac{d \sigma}{dm^2_{\ell\ell}dy_{\ell\ell}} \right)_{\text{LO}} \left[ (1 + \cos^2\theta) \right.\vspace{2mm}\\
    && \dst \left.+ \frac{1}{2}A_0(1 - 3\cos^2\theta) + A_4\cos{\theta}       \right].
    \end{array}
\end{equation}
It is important to notice that Eq.~\eqref{eq:LOLAngDist} does not imply that $A_0$ and $A_4$ are both non-vanishing in the leading order limit. 

Table \ref{tab:angularSMEFT} summarizes the angular dependence of the squared amplitude at leading order in $\alpha_S$ arising from the SM and the different dimension-6 operators that we consider. In particular, we immediately see that the contributions proportional to a linear combination of $(1-\cos\theta)^{2}$ and $(1+\cos\theta)^{2}$ can only arise from Eq.~\eqref{eq:LOLAngDist} if $A_{0}=0$. This is the Lam-Tung relation at order $\mathcal{O}(\alpha_S^0)$:
\begin{equation}
    \lim_{\alpha_S\to 0} (A_{0}- A_{2}) = A_{0} = A_{2} = 0.
\end{equation}
Table \ref{tab:angularSMEFT} also shows that the only operators that can break this relation at order $\mathcal{O}(\alpha_S^0)$ are the dipole operators $\mathcal{O}_{(3)}$ and the chirality-breaking scalar/tensor four-fermion operators $\mathcal{O}_{(8)}$ with angular dependences respectively proportional to $1-\cos^2\theta$ and to a constant term. 

\begin{table}[]
\centering
\begin{tabular}{l|lllll}
                        & SM                      & $\mathcal{O}_{(3)}$                       & $\mathcal{O}_{(1,2,4)}$     & $\mathcal{O}_{(5,6,7)}$  & $\mathcal{O}_{(8)}$ \\ 
    \hline SM               & $(1\pm \cos \theta)^2$    &                                       &                           &                           & \\
    $\mathcal{O}_{(3)}$     & 0                         & \cellcolor{gray!20}$1-\cos^2 \theta$  &                           &                           & \\
    $\mathcal{O}_{(1,2,4)}$ & $(1\pm \cos \theta)^2$    & 0                & $(1\pm \cos \theta)^2$    &                           &\\
    $\mathcal{O}_{(5,6,7)}$ & $(1\pm \cos \theta)^2$    & 0                                     & 0                         & $(1\pm \cos \theta)^2$    & \\
    $\mathcal{O}_{(8)}$     & 0                         & 0                                     & 0                         & 0                         & \cellcolor{gray!20}$1$                    
\end{tabular}
\caption{Dependence on the $\cos\theta$ angular variable of the squared amplitude of the process $pp \to \ell^+\ell^-$ (at leading order in $\alpha_S$) arising from the interference of the SM and the SMEFT operators considered in our analysis. The notation $(1\pm \cos\theta)^{2}$ is used to indicate that the angular dependence is a linear combination of both $(1+\cos\theta)^2$ and $(1-\cos\theta)^2$ terms. The operators are labeled as in Section~\ref{sec::Intro}: $\mathcal{O}_{(1,2)}$ are the operators $\phi^4 D^2$ and $\psi^2 \phi^3$, $\mathcal{O}_{(3)}$ are the dipole operator $\psi^2 X \varphi$, $\mathcal{O}_{(4)}$ are the operators $\psi^2 \varphi^2 D$, $\mathcal{O}_{(5,6,7)}$ are the current-current four-fermion operators, and $\mathcal{O}_{(8)}$ are the scalar/tensor four-fermion operators. .
The table is symmetric and we only filled the lower part. The two shaded entries highlight the cases where the Lam-Tung relation is violated. 
}
\label{tab:angularSMEFT}
\end{table}

\subsection{Lam-Tung relation breaking at $\mathcal{O}(\alpha_S)$}
\begin{figure}[t!]
  \centering
  \begin{minipage}{0.5\textwidth}
    \centering
    \includegraphics[width=\linewidth]{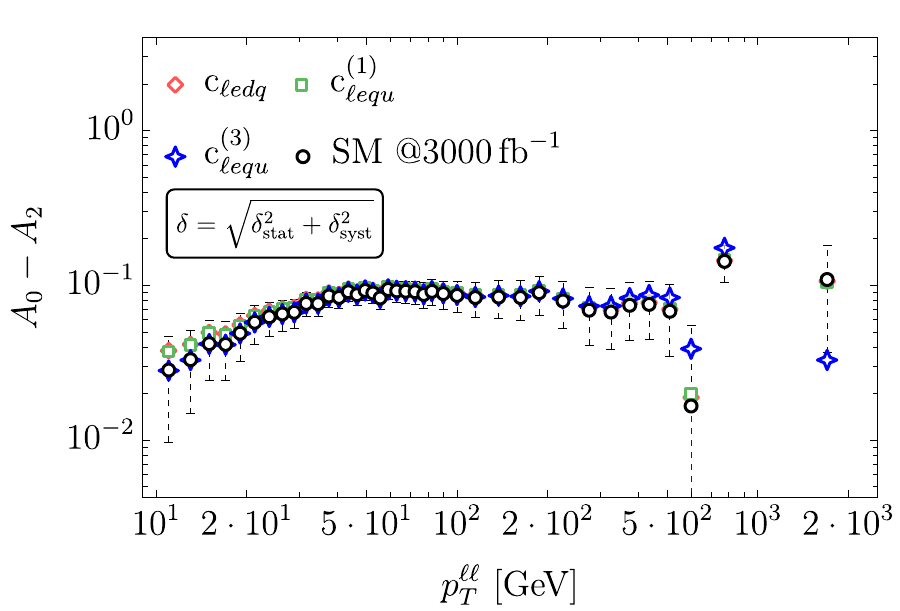}
  \end{minipage}\hfill
  \begin{minipage}{0.5\textwidth}
    \centering
    \includegraphics[width=\linewidth]{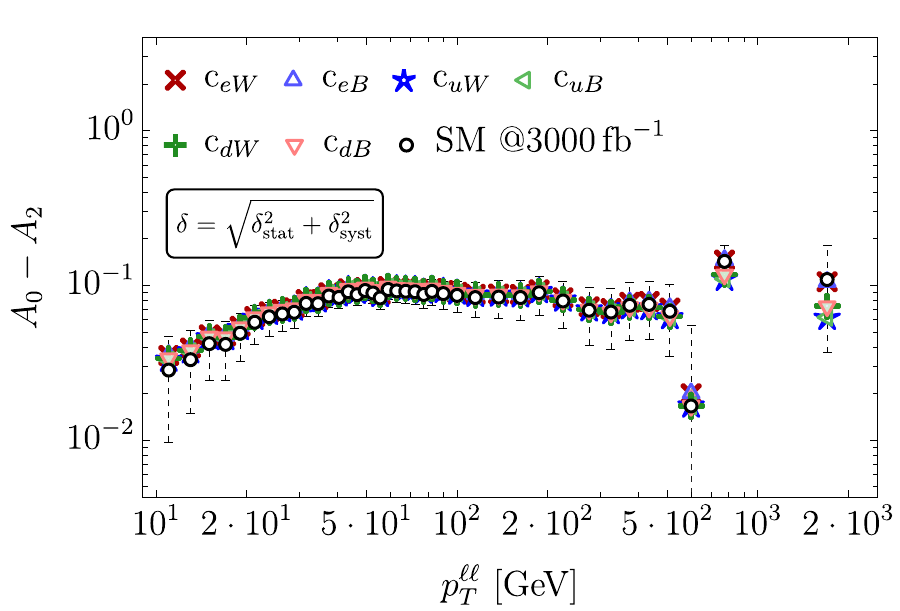}
  \end{minipage}
  \begin{minipage}{0.5\textwidth}
    \centering
    \includegraphics[width=\linewidth]{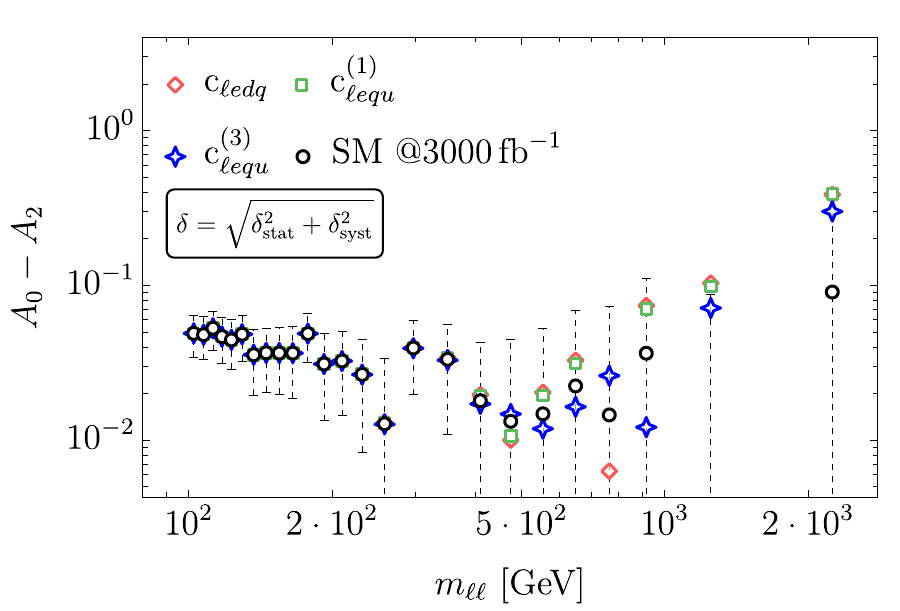}
  \end{minipage}\hfill
  \begin{minipage}{0.5\textwidth}
    \centering
    \includegraphics[width=\linewidth]{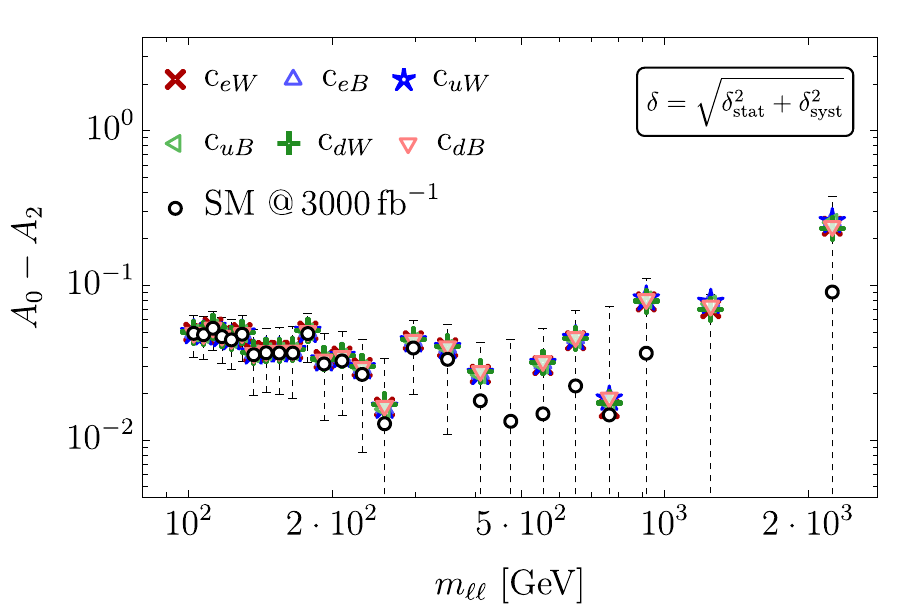}
  \end{minipage}
  \caption{Comparison between the contribution of the different SMEFT operators to the transverse momentum (first row) and invariant mass distribution (second row) of the $A_0-A_2$ observable. The left column shows scalar/tensor four-fermion operators, the right one shows the dipole operators. For each distribution, the Wilson coefficients are fixed equal to the largest positive value consistent with the bounds from the corresponding kinematic distribution at $3000\text{ fb}^{-1}$ (Table~\ref{tab:pT_bounds} for $p_{T}^{\ell \ell}$ and Table~\ref{tab:mll_bounds} for $m_{\ell \ell}$).}
  \label{fig:ComparisonSMEFTvsSM}
\end{figure}

At $\mathcal{O}(\alpha_\text{S})$ the SM does not violate the Lam-Tung relation as a consequence of the fact that gluons couple to the vector quark current, as explained in details in Ref.~\cite{Arteaga-Romero:1983llb}. This implies that four-fermion operators involving the vector quark current (namely $\mathcal{O}_{(5,6,7)}$, following the notation in Table~\ref{tab:angularSMEFT}) can not violate the Lam-Tung relation. As already mentioned in Section \ref{sec::Intro}, the effect of the operators $\mathcal{O}_{(1,2,4)}$ does not grow with energy and can safely be neglected.\footnote{One could expect that contributions to the Drell–Yan process from the $s$-channel Higgs boson exchance could break Lam-Tung already at $\mathcal{O}(\alpha_\text{S}^{0})$. However, even if they did, these contributions are too small to be observed.}
Thus, even at order $\mathcal{O}(\alpha_\text{S})$, the only operators expected to give an observable Lam-Tung breaking effect are the dipole and the scalar/tensor four-fermion operators. The $\mathcal{O}(\alpha_\text{S})$ contributions of these operators to the $A_0-A_2$ observable is compared to the first non-vanishing SM contribution (corresponding to $\mathcal{O}(\alpha_\text{S}^2)$) in Figure~\ref{fig:ComparisonSMEFTvsSM}. The SM curves are the same as those shown in Figures~\ref{fig:ptangcoefficients} and~\ref{fig:mllangcoefficients}, while the SMEFT contributions are computed at order $\mathcal{O}(\alpha_\text{S})$, for Wilson coefficients fixed to the $3000$ fb$^{-1}$ (positive) exclusion bound reported in Tables~\ref{tab:pT_bounds} and~\ref{tab:mll_bounds} (see Section~\ref{sec::SMEFTanalysis} for details on how these bounds are derived).

\section{Analysis and projected limits}\label{sec::SMEFTanalysis}

In this section, we detail the procedure used to extract the projected constraints on the Wilson coefficient appearing in Eq.~\eqref{eq:lag}, and we specify the hypotheses underlying our analysis.

We considered, at the LHC, a center-of-mass energy of 13 TeV and two benchmark integrated luminosities: 300 fb$^{-1}$, corresponding approximately to the dataset expected by the end of Run 3, and 3000 fb$^{-1}$, representative of the High-Luminosity LHC (HL-LHC) scenario. Since measurements of $A_0 - A_2$ at 13 TeV are not yet available, we rely on simulations to obtain pseudo-data, as specified in Section~\ref{sec::LamTung}. These simulations are performed under the SM-only hypothesis for the process $pp \rightarrow \ell^+\ell^-+X$ at next-to-next-to-leading order (NNLO) in QCD, \textit{i.e.}, at $\mathcal{O}(\alpha_\text{S}^2)$, using the \texttt{MiNNLO}$_{\text{\texttt{PS}}}$ framework. The resulting predictions are treated as pseudo-data for the purpose of our projections, with uncertainties estimated as explained in Section~\ref{subsec:uncestimate}. Predictions for the $A_0 - A_2$ observable in the SMEFT framework have been calculated analytically as explained in Section~\ref{sec::LamTung} and are expressed as functions of $c_{\text{NP}}^2/\Lambda^4$, where $c_{\text{NP}}^2$ generically denotes the square of a Wilson coefficient of those appearing in Eq.~\eqref{eq:lag}.

The $A_0 - A_2$ observable in the presence of NP contributions can be expressed as:
\begin{equation}
    \label{eq:A0-A2_SMEFT}
    \begin{array}{lll}
    \dst (A_0-A_2)^{\text{SMEFT}}
    &=&\dst 4 - 10 \Bigg[
        \frac{\dst \int d\sigma^{\text{SM}}(\cos{\theta},\phi,m_{\ell\ell},p_{T}^{\ell\ell}) 
        \, d\cos{\theta}\,d\phi \,(\cos^2{\theta} + \sin^2{\theta}\cos{2\phi})}
        {d\sigma^{\text{SM}}(m_{\ell\ell},p_{T}^{\ell\ell}) 
        + \displaystyle\left(\frac{c_{\text{NP}}}{\Lambda^2}\right)^2 
        d\tilde{\sigma}^{\text{NP}}(m_{\ell\ell},p_{T}^{\ell\ell})} \\[2mm]
    &&\hspace{-5mm}\dst + \left(\frac{\dst c_{\text{NP}}}{\Lambda^2}\right)^2
     \frac{\dst \int d\tilde{\sigma}^{\text{NP}}(\cos{\theta},\phi,m_{\ell\ell},p_{T}^{\ell\ell}) 
           \, d\cos{\theta}\,d\phi \,(\cos^2{\theta} + \sin^2{\theta}\cos{2\phi})}
          {d\sigma^{\text{SM}}(m_{\ell\ell},p_{T}^{\ell\ell}) 
           + \dst \left(\frac{c_{\text{NP}}}{\Lambda^2}\right)^2d\tilde{\sigma}^{\text{NP}}(m_{\ell\ell},p_{T}^{\ell\ell})}
     \Bigg]\,.
\end{array}
\end{equation}
Due to the NP contribution appearing in the denominator, the dependence of $(A_0-A_2)^{\text{SMEFT}}$ on the Wilson coefficient is non-linear, as anticipated above, and also affects the normalization of the distribution. This is expected to potentially break the Gaussian assumption that would allow a straightforward $\chi^2$ analysis. For this reason, we decided to extract the bounds using the log-likelihood-ratio (LLR) test-statistic, which represents a more robust procedure for non-linear $\chi^2$ (non-Gaussian likelihood). The LLR is defined as:
\begin{equation}
    t_{\boldsymbol{c}} = -2\ln\frac{\mathscr{L}_{H_0}}{\mathscr{L}_{H_1}(\boldsymbol{c})}\,,
\end{equation}
where, generically, $H_0$ represents the null hypothesis and $H_1$ the alternative hypothesis, that depends on some parameters $\boldsymbol{c}$. In our specific case, $H_0$ is  the SM-only hypothesis, $H_1$ the SMEFT hypothesis, and the $\boldsymbol{c}$ parameters are the relevant Wilson coefficients. For the two hypotheses, we assumed:
\begin{itemize}
    \item bin-by-bin, the likelihood for the SM hypothesis is considered to be a gaussian distribution, centered around the $A_0-A_2$ central value provided by the simulations, with a width fixed by the estimated uncertainty. The total likelihood is then:
    \begin{equation}
    \label{eq::SM_likelihood}
        \mathscr{L}_{H_0} = \prod_{i=1}^{n_{bins}}N\left((A_0-A_2)^{\text{SM}}_i\,;\,\sigma_i\right)\,.
    \end{equation}
    \item analogously, the likelihood for the SMEFT hypothesis is constructed as the product of gaussian distributions that, bin-by-bin, are centered around the $(A_0-A_2)^{\text{SMEFT}}$ central value, calculated as explained above, with the same variance as in the SM-only hypothesis:\footnote{Here we assume that the uncertainty is not significantly affected by the presence of NP contributions. This is a reasonable assumption, which can possibly be relaxed, if needed, by assuming $\sigma_i = \sigma_i(c_{\text{NP}})$ in Eq.~\eqref{eq::SMEFT_likelihood}.}
    \begin{equation}\label{eq::SMEFT_likelihood}
        \mathscr{L}_{H_1}(c_{\text{NP}}) = \prod_{i=1}^{n_{bins}}N\left((A_0-A_2)^{\text{SMEFT}}_i(c_{\text{NP}})\,;\,\sigma_i\right)\,,
    \end{equation}
    where we have made explicit the $(A_0-A_2)^{\text{SMEFT}}$ dependence on the Wilson coefficients.
\end{itemize}

Each Wilson coefficient is constrained individually by setting all others to zero.\footnote{Obviously, since we have an implementation of the full likelihood in Eq.~\eqref{eq::SMEFT_likelihood}, we could also perform a simultaneous fit of some, or all coefficients, or a combination with other analyses.} The analysis is performed using two kinematic distributions: the transverse momentum of the lepton pair $p_{T}^{\ell\ell}$ for above $p_{T}^{\ell\ell}>10$ GeV and $80\,\text{GeV} < m_{\ell\ell} < 100\,\text{GeV}$, and the di-lepton invariant mass $m_{\ell\ell}$ for above $m_{\ell\ell}>100$ GeV and $p_{T}^{\ell\ell}>10$ GeV. The binning schemes adopted for both cases are provided in Tables~\ref{tab:ptxsdistributions}-\ref{tab:mllangcoefficientsA0mA2} of Appendix \ref{app::Tables}, and are inspired by the existing measurements of Refs.~\cite{ATLAS:2019zci,CMS:2018mdl}.

To constrain each Wilson coefficient, we estimated the distribution of the test-statistic $t_{c_{\text{{NP}}}}$ under the SM hypothesis $H_0$, for a fixed, reasonable, initial value of $c_{\text{NP}}$, by computing its value on $10^{4}$ pseudo-experiments drawn from $\mathscr{L}_{H_0}$. More explicitly, we used the available Monte Carlo to draw, with replacement, pseudo-data and used them to compute $t_{c_{\text{{NP}}}}$ $10^{4}$ times. This yields a distribution of values of $t_{c_{\text{{NP}}}}$. From this distribution we can identify a threshold, denoted by $\overline{t}_{c_{\text{{NP}}}}$, defining the $95\%$ confidence level (CL) threshold for rejecting the null hypothesis $H_0$ when testing against the alternative hypothesis $H_1$ corresponding to the fixed value $c_{\text{NP}}$.

Next, we compute the average value of $t_{c_{\text{NP}}}$ over $10^{3}$ samples drawn from $\mathscr{L}_{H_1}$. This corresponds to generating pseudo-experiments corresponding to the $H_{1}$ hypothesis with fixed $c_{\text{{NP}}}$, computing $t_{c_{\text{NP}}}$ for each of these pseudo-experiments, and taking the average.\footnote{Comparing the average test-statistic under the alternative hypothesis with the distribution of the test-statistic under the null hypothesis is one possible procedure to compute a bound. Another option would be to compute the full distribution under the alternative hypothesis and compare the two distributions in terms of a ``confusion matrix", or to require a given power of the test at fixed CL. These are arbitrary choices of hypothesis testing.} The corresponding average value is denoted by $t^{*}_{c_{\text{NP}}}$. 

Finally, both the threshold value $\overline{t}_{c_{\text{{NP}}}}$ and the average value $t^{*}_{c_{\text{NP}}}$ are iteratively computed adjusting the value of $c_{\text{NP}}$ (using a bisection method) until they are equal, within a fixed threshold
\begin{equation}\label{eq:convergence}
    \delta_{c_{\text{NP}}} = 2 \left| \frac{t^{*}_{c_{\text{NP}}} - \overline{t}_{c_{\text{NP}}}}{t^{*}_{c_{\text{NP}}} + \overline{t}_{c_{\text{NP}}}} \right|\,,\qquad \text{with } \delta_{c_{\text{NP}}}< 5\%.
\end{equation}
This is equivalent to solving the optimization problem:
\begin{equation}
    \tilde{c}_{\text{NP}} = \arg \min_{c_{\text{NP}}} \left| \delta_{c_{\text{NP}}} - 5\% \right|\,.
\end{equation}
where $\tilde{c}_{\text{NP}}$ represents the Wilson coefficient value corresponding to the 95\% CL bound.

This procedure is carried out to obtain both the upper and the lower bounds. Since NP contributions manifest themselves as $c^2_{\text{NP}}/\Lambda^4$ and uncertainties are symmetric, the resulting constraints are expected to be symmetric around zero. Nevertheless, we extract both sides independently.\footnote{In some cases the bounds on certain Wilson coefficients are not exactly symmetric. These small deviations arise from small numerical instabilities in the procedure used to estimate the LLR distribution and give an idea of the uncertainty in our numerical optimization procedure.} Results are presented and discussed in the next section.

\begin{table*}[t!]
    \small
    \centering
    \scalebox{1}{
    \begin{tabular}{c|cc|cc}
        \hline
        \multirow{2}{*}{\shortstack{\textbf{95\%CL}, $\Lambda=4$ TeV \\ $p_{T}^{\ell\ell}$ distribution}}  &         
        \multicolumn{2}{c|}{$\mathcal{L}=300\,\text{fb}^{-1}$} & 
        \multicolumn{2}{c}{$\mathcal{L}=3000\,\text{fb}^{-1}$} \\
        & $A_0-A_2$ & cross-section & $A_0-A_2$ & cross-section\\
        \hline
$c^{(3)}_{\ell e q u}$ & $[-19.1,\,19.1]$ & $[-15.6,\,15.8]$ & $[-12.9,\,13]$ & $[-14.1,\,14.1]$ \\
$c^{(1)}_{\ell e q u}$ & $[-352,\,344]$ & $[-363,\,359]$ & $[-346,\,344]$ & $[-359,\,359]$ \\
$c_{\ell e d q}$ & $[-388,\,375]$ & $[-400,\,400]$ & $[-375,\,375]$ & $[-400,\,400]$ \\
$c_{eW}$ & $[-9.61,\,9.61]$ & $[-18,\,17.6]$ & $[-9.38,\,9.38]$ & $[-17.2,\,17.2]$ \\
$c_{eB}$ & $[-17.5,\,17.5]$ & $[-32.8,\,32]$ & $[-17.5,\,17.5]$ & $[-32,\,32]$ \\
$c_{uW}$ & $[-13.1,\,13.4]$ & $[-3.2,\,3.28]$ & $[-11.4,\,11.6]$ & $[-2.85,\,2.86]$ \\
$c_{uB}$ & $[-23.8,\,23.8]$ & $[-5.63,\,5.78]$ & $[-20.9,\,20.6]$ & $[-5.16,\,5.16]$ \\
$c_{dW}$ & $[-14.4,\,14.4]$ & $[-4.45,\,4.42]$ & $[-13.1,\,13.1]$ & $[-3.98,\,3.98]$ \\
$c_{dB}$ & $[-25.8,\,25.8]$ & $[-7.97,\,7.97]$ & $[-23.4,\,23.8]$ & $[-7.03,\,7.03]$ \\
        \hline
    \end{tabular}}
    \caption{95\% CL individual bounds for the Wilson coefficients in Eq.~\eqref{eq:lag} for integrated luminosity values of 300 $\text{fb}^{-1}$ and 3000 $\text{fb}^{-1}$. For each luminosity value, in the left column are reported the bounds obtained looking at the $A_0-A_2$ observable and in the right column are reported the bounds obtained looking at the cross-section in the $p_{T}^{\ell\ell}$ distribution, fixing the NP scale $\Lambda$ as $\Lambda=4$ TeV.}
    \label{tab:pT_bounds}
\end{table*}

\begin{table*}[t!]
    \small
    \centering
    \scalebox{1}{
    \begin{tabular}{c|cc|cc}
        \hline
        \multirow{2}{*}{\shortstack{\textbf{95\%CL}, $\Lambda=4$ TeV \\ $m_{\ell\ell}$ distribution}}  &         
        \multicolumn{2}{c|}{$\mathcal{L}=300\,\text{fb}^{-1}$} & 
        \multicolumn{2}{c}{$\mathcal{L}=3000\,\text{fb}^{-1}$} \\
        & $A_0-A_2$ & cross-section & $A_0-A_2$ & cross-section\\
        \hline
$c^{(3)}_{\ell e q u}$ & $[-0.297,\,0.289]$ & $[-0.0641,\,0.0645]$ & $[-0.141,\,0.141]$ & $[-0.0645,\,0.0625]$ \\
$c^{(1)}_{\ell e q u}$ & $[-1.64,\,1.62]$ & $[-0.152,\,0.15]$ & $[-0.703,\,0.703]$ & $[-0.149,\,0.148]$ \\
$c_{\ell e d q}$ & $[-2.11,\,2.19]$ & $[-0.205,\,0.202]$ & $[-0.898,\,0.938]$ & $[-0.199,\,0.199]$ \\
$c_{eW}$ & $[-5.63,\,5.63]$ & $[-2.34,\,2.34]$ & $[-3.31,\,3.28]$ & $[-2.34,\,2.29]$ \\
$c_{eB}$ & $[-9.38,\,9.06]$ & $[-3.63,\,3.59]$ & $[-5.31,\,5.42]$ & $[-3.59,\,3.63]$ \\
$c_{uW}$ & $[-7.24,\,7.27]$ & $[-2.87,\,2.86]$ & $[-4.22,\,4.34]$ & $[-2.93,\,2.81]$ \\
$c_{uB}$ & $[-6.09,\,6.09]$ & $[-2.34,\,2.37]$ & $[-3.57,\,3.52]$ & $[-2.34,\,2.34]$ \\
$c_{dW}$ & $[-8.67,\,8.91]$ & $[-3.81,\,3.84]$ & $[-5.33,\,5.33]$ & $[-3.87,\,3.87]$ \\
$c_{dB}$ & $[-7.59,\,7.31]$ & $[-3.13,\,3.16]$ & $[-4.5,\,4.5]$ & $[-3.17,\,3.13]$ \\
        \hline
    \end{tabular}}
    \caption{95\% CL individual bounds for the Wilson coefficients in Eq.~\eqref{eq:lag} for integrated luminosity values of 300 $\text{fb}^{-1}$ and 3000 $\text{fb}^{-1}$. For each luminosity value, in the left column are reported the bounds obtained looking at the $A_0-A_2$ observable and in the right column are reported the bounds obtained looking at the cross-section in the $m_{\ell\ell}$ distribution, fixing the New Physics scale $\Lambda$ as $\Lambda=4$ TeV.}
    \label{tab:mll_bounds}
\end{table*}

\section{Results and discussion}\label{sec::Results}
In Tables~\ref{tab:pT_bounds} and~\ref{tab:mll_bounds} we report the 95\% CL individual bounds for all the Wilson coefficients in Eq.~\eqref{eq:lag}, equally valid for both electrons and muons, fixing the NP scale to $\Lambda=4$ TeV. For the $m_{\ell\ell}$ analysis, the last bin, corresponding to $3-10$ TeV, was not included because the estimated uncertainty was too large, and the region was partly outside of EFT validity range. For each luminosity value, we also report, for comparison, the bounds obtained from the differential cross-sections. The procedure to extract the latter bounds is the same as the one described above, except that the $A_0-A_2$ observable is replaced by the corresponding differential cross-section. Results in Tables~\ref{tab:pT_bounds} and~\ref{tab:mll_bounds} are obtained considering two benchmark values for the integrated luminosity, corresponding to 300 fb$^{-1}$ and 3000 fb$^{-1}$. We have fixed the NP scale to $\Lambda=4$ TeV, and we focused on the $A_0-A_2$ observable and on the $p_{T}^{\ell\ell}$ and $m_{\ell\ell}$ differential cross-sections.

\begin{figure}[t!]
  \centering
  \begin{minipage}{0.45\textwidth}
    \centering
    \includegraphics[width=\linewidth]{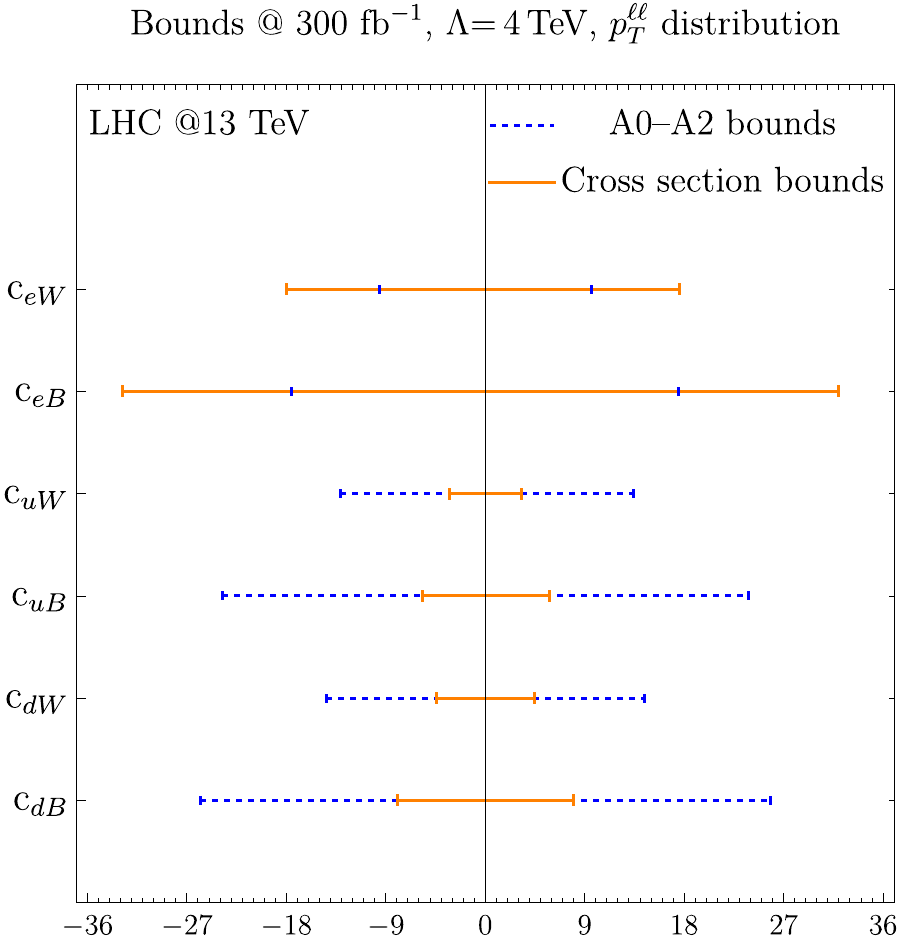}
  \end{minipage}\hfill
  \begin{minipage}{0.45\textwidth}
    \centering
    \includegraphics[width=\linewidth]{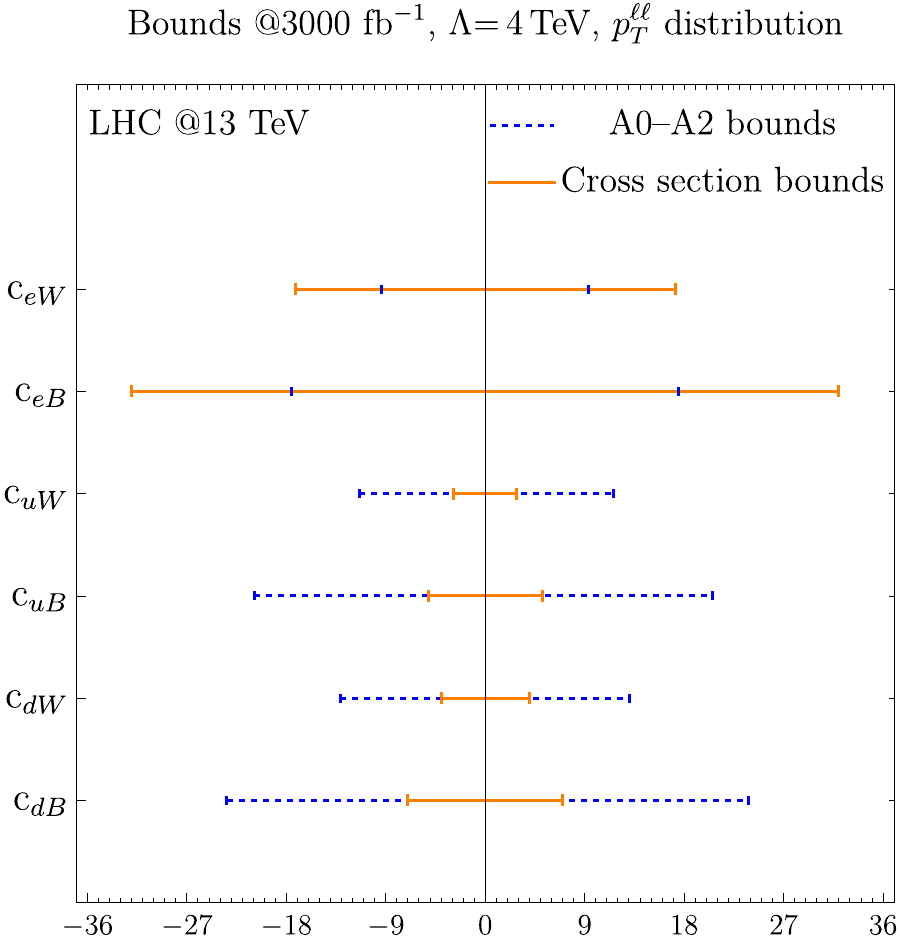}
  \end{minipage}
  \caption{Comparison between the 95\% CL boundaries obtained looking at the $A_0-A_2$ observable (blue) and the cross-section (orange) in the $p_{T}^{\ell\ell}$ distribution. Left: 300 fb$^{-1}$. Right: 3000 fb$^{-1}$ }
  \label{fig:comparisonPT}
\end{figure}

\begin{figure}[t!]
  \centering
  \begin{minipage}{0.45\textwidth}
    \centering
    \includegraphics[width=\linewidth]{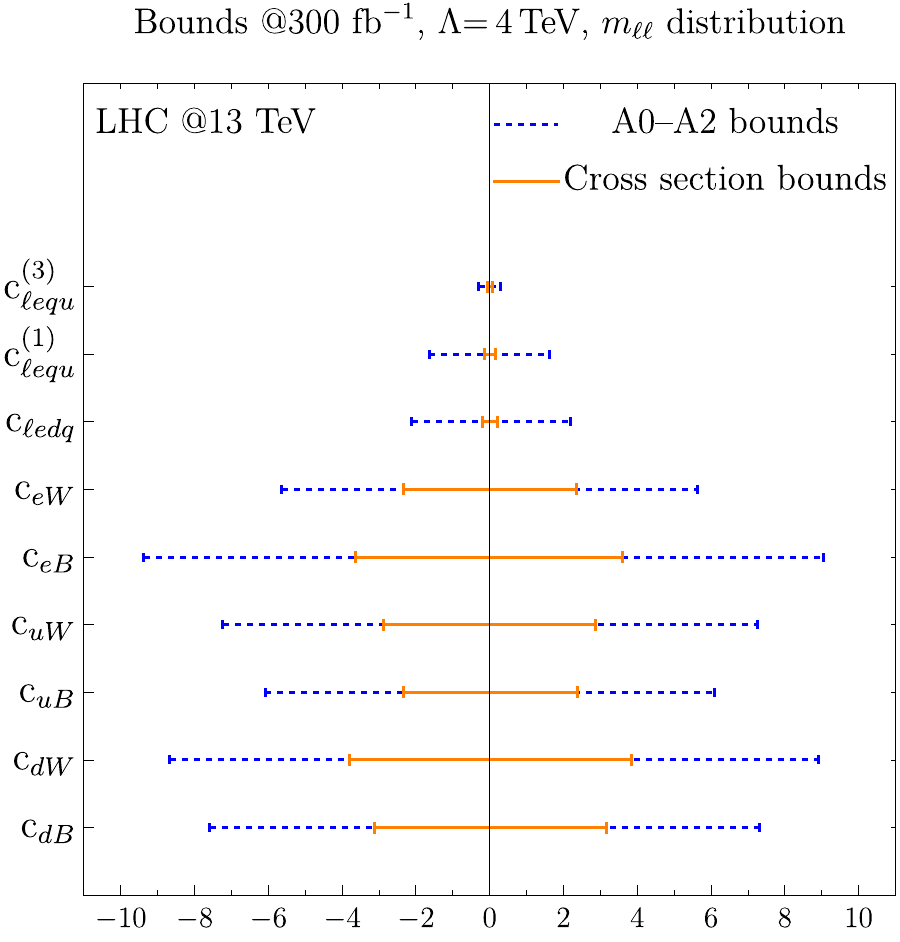}
  \end{minipage}\hfill
  \begin{minipage}{0.45\textwidth}
    \centering
    \includegraphics[width=\linewidth]{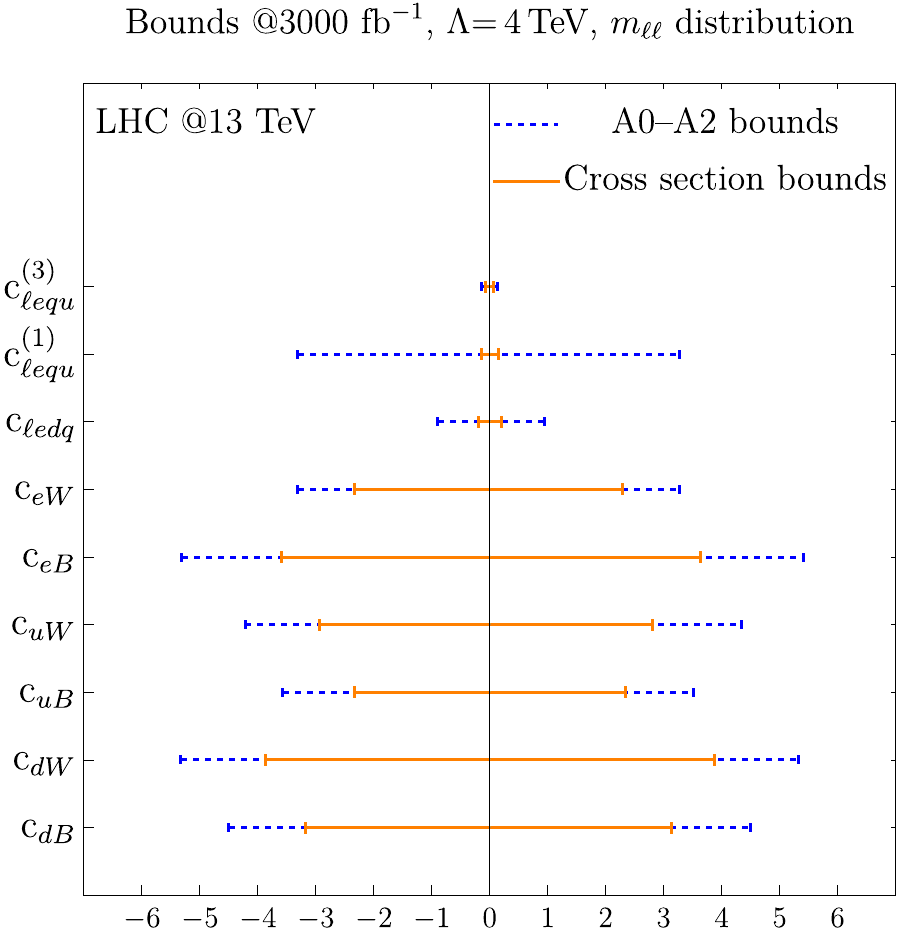}
  \end{minipage}
  \caption{Comparison between the 95\% CL boundaries obtained looking at the $A_0-A_2$ observable (blue) and the cross-section (orange) in the $m_{\ell\ell}$ distribution. Left: 300 fb$^{-1}$. Right: 3000 fb$^{-1}$ }
  \label{fig:comparisonmll}
\end{figure}

For a more immediate visual comparison, the same bounds, except for the four-fermion operators in the $p_{T}^{\ell\ell}$ distribution, are shown graphically in Figures~\ref{fig:comparisonPT} and \ref{fig:comparisonmll}. We have not included the constraints on the Wilson coefficient of the four-fermion operators in Figure \ref{fig:comparisonPT}, since the $p_{T}^{\ell\ell}$ distribution does not provide reasonably stringent bounds on them.\footnote{This is due to the fact that the $p_{T}^{\ell\ell}$ analysis is done in a small window around the $Z$ mass peak ($80\text{ GeV}<m_{\ell\ell}<100 \text{ GeV}$), where the four-fermion operators do not profit of the Breit-Wigner enhancement of contributions involving the $Z$ boson propagator.}

Comparing the bounds shows that, in general, the constraints are tighter when looking at the cross-section rather than $A_0-A_2$ observable (see Appendix~\ref{app::A0A2CScomparison} for an intuitive explanation). However, our analysis neglects the leading operators contributing to the cross-section, namely the curret-current operators that interfere with the SM. In an analysis including all operators, those in Eq.~\eqref{eq:lag} would be sub-leading and thus extremely hard to constrain. The $A_0-A_2$ observable, on the other-hand, offers a complementary channel where the contribution of the current-current operators is highly suppressed, thanks to the interference with the SM and the Lam-Tung relation, and the dipole and scalar and tensor four-fermions operators provide the leading contributions. In the spirit of global fits and of breaking degeneracies between different (classes of) operators, the $A_0-A_2$ observable becomes a very valuable tool, which gives independent and direct access to the chirality-breaking operators. 

From Tables \ref{tab:pT_bounds} and \ref{tab:mll_bounds}, we can also see that the bounds projected for $300~\text{fb}^{-1}$ and for $3000~\text{fb}^{-1}$ are often comparable. This effect is more pronounced in the $p_{T}^{\ell\ell}$ analysis and for the differential cross-section observable. This behavior comes from the fact that the bound on those coefficients is dominated by the systematic uncertainty, which we assume not to improve with luminosity. As we discussed before, systematic uncertainty dominate the low-$p_T$ and low-$m_{\ell\ell}$ regions, so that only operators with a shape significantly different from the SM in the high-$p_T$ or high-$m_{\ell\ell}$ regions will see a reasonable improvement in their bounds when increasing the integrated luminosity. The different shapes shown in Figure \ref{fig:ComparisonSMEFTvsSM} confirm this interpretation and show that the different growth in energy of the dipole and four-fermion operators leads not only to different sensitivities in the two kinematic distributions, but also to a different pattern of improvement of the bounds with the integrated luminosity.

As mentione above, the $m_{\ell\ell}$ distribution, yields tighter bounds than the $p_{T}^{\ell\ell}$ distribution, indicating that higher energies enhance the sensitivity to both classes of operators. This can be seen very neatly in Figure \ref{fig:comparisonmll}, which makes apparent how the four-fermion operators are more tightly constrained than the dipole ones.

Finally, in Tables \ref{tab:pT_bounds_2ndO} and \ref{tab:mll_bounds_2ndO}, we report the bounds obtained using the Taylor expansion of the $A_0-A_2$ observable truncated at order $c_{\text{NP}}^{2}/\Lambda^{4}$. Such bounds do not differ significantly from those obtained using the full expression in Table \ref{tab:pT_bounds} and \ref{tab:mll_bounds}. This is the expected behavior in the region of validity of the EFT expansion, where higher-order terms are sub-leading corrections. In particular, the expansion parameter involves the ratio $d\sigma^{\text{SMEFT}}/d\sigma^{\text{SM}}\times (c_{\text{NP}}/\Lambda^2)^2$, which remains sufficiently small for small enough Wilson coefficients and in most of the allowed phase space. The largest discrepancies between the two approaches appear in the regions of the phase space where the NP contributions are more significant, such as the tails of $p_{T}^{\ell\ell}$ and $m_{\ell\ell}$ distributions. In those regions, for relatively large values of the Wilson coefficients, the expansion parameter may become largish, making the expansion less reliable, and one could expect higher-order terms to become relevant. This is the reason why we believe that showing both sets of bounds is useful, as it gives an idea of the robustness of the results.

\begin{table}[t!]
  \centering
  \begin{tabular}{c|cc}
  \hline
 \shortstack{\textbf{95\% CL}, $\Lambda=4$ TeV \\ $p_{T}^{\ell \ell}$ distribution} & $\mathcal{L}=300\,\text{fb}^{-1}$ & $\mathcal{L}=3000\,\text{fb}^{-1}$ \\
  \hline
$c^{(3)}_{\ell e q u}$ & $[-18.8,\,18.8]$ & $[-12.9,\,12.8]$ \\
$c^{(1)}_{\ell e q u}$ & $[-359,\,348]$ & $[-344,\,344]$ \\
$c_{\ell e d q}$ & $[-388,\,375]$ & $[-375,\,375]$ \\
$c_{eW}$ & $[-9.38,\,9.38]$ & $[-9.38,\,9.3]$ \\
$c_{eB}$ & $[-17.5,\,17.5]$ & $[-17.2,\,16.9]$ \\
$c_{uW}$ & $[-12.6,\,12.5]$ & $[-10.4,\,10.3]$ \\
$c_{uB}$ & $[-23.1,\,23.1]$ & $[-18.9,\,19.2]$ \\
$c_{dW}$ & $[-14.1,\,14.1]$ & $[-12.5,\,12.5]$ \\
$c_{dB}$ & $[-25,\,25.4]$ & $[-22.7,\,22.7]$ \\
 \hline
  \end{tabular}
  \caption{95\% CL individual bounds for the Wilson coefficients in eq.~\ref{eq:lag} for integrated luminosity values of 300 $\text{fb}^{-1}$ and 3000 $\text{fb}^{-1}$. These are obtained looking at the $A_0-A_2$ observable in the $p_T$ distribution expanded at order $c_{\text{NP}}^{2}/ \Lambda^4$, fixing the New Physics scale $\Lambda$ as $\Lambda=4$ TeV.}
  \label{tab:pT_bounds_2ndO}
\end{table}

\begin{table}[t!]
  \centering
  \begin{tabular}{c|cc}
  \hline
 \shortstack{\textbf{95\% CL}, $\Lambda=4$ TeV \\ $m_{\ell \ell}$ distribution} & $\mathcal{L}=300\,\text{fb}^{-1}$ & $\mathcal{L}=3000\,\text{fb}^{-1}$ \\
  \hline
$c^{(3)}_{\ell e q u}$ & $[-0.227,\,0.234]$ & $[-0.133,\,0.129]$ \\
$c^{(1)}_{\ell e q u}$ & $[-1,\,0.988]$ & $[-0.563,\,0.568]$ \\
$c_{\ell e d q}$ & $[-1.31,\,1.31]$ & $[-0.738,\,0.75]$ \\
$c_{eW}$ & $[-5.47,\,5.31]$ & $[-3.32,\,3.28]$ \\
$c_{eB}$ & $[-8.75,\,8.75]$ & $[-5.31,\,5.31]$ \\
$c_{uW}$ & $[-7.03,\,7.03]$ & $[-4.16,\,4.24]$ \\
$c_{uB}$ & $[-6.04,\,5.86]$ & $[-3.52,\,3.52]$ \\
$c_{dW}$ & $[-8.44,\,8.44]$ & $[-5.27,\,5.27]$ \\
$c_{dB}$ & $[-7.03,\,7.23]$ & $[-4.45,\,4.45]$ \\
 \hline
  \end{tabular}
  \caption{95\% CL individual bounds for the Wilson coefficients in Eq.~\ref{eq:lag} for integrated luminosity values of 300 $\text{fb}^{-1}$ and 3000 $\text{fb}^{-1}$. These are obtained looking at the $A_0-A_2$ observable in the $m_{\ell \ell}$ distribution expanded at order $c_{\text{NP}}^{2}/ \Lambda^4$, fixing the New Physics scale $\Lambda$ as $\Lambda=4$ TeV.}
  \label{tab:mll_bounds_2ndO}
\end{table}

\section{Conclusions}\label{sec::Conclusions}

In this work, we analyzed the contributions of chirality-breaking dimension-six operators, specifically dipole and four-fermion operators, to the Drell–Yan (DY) process.

Our focus was on the angular observable $A_{0}-A_{2}$, which vanishes in the Standard Model (SM) up to $\mathcal{O}(\alpha_{\text{S}}^2)$ and does not receive contributions from dimension-six operators that interfere with the SM amplitude. In Ref.~\cite{Li:2024iyj}, bounds on dipole operators were derived at a center-of-mass energy of 8~TeV using the ATLAS analysis of Ref.~\cite{ATLAS:2016rnf}. Here, we extended that study by providing projected bounds at 13~TeV, considering both dipole and four-fermion operators, and using both the $p_T^{\ell\ell}$ and $m_{\ell\ell}$ distributions.

We performed SM Monte Carlo simulations at $\mathcal{O}(\alpha_{\text{S}}^2)$ with the \texttt{MiNNLO}$_{\text{\texttt{PS}}}$ tool, and included the NP contributions analytically to estimate their effect on the $p_T^{\ell\ell}$ and $m_{\ell\ell}$ distributions of the DY cross-section, the angular coefficients $A_\ell$, and the $A_{0}-A_{2}$ observable. A pseudo-analysis was carried out for 300~fb$^{-1}$ of integrated luminosity, corresponding the final LHC dataset, and for 3000~fb$^{-1}$ of integrated luminosity, corresponding to the HL-LHC. We included estimates of both theoretical and experimental systematic uncertainties, based on current measurements and projections for future improvements.

We compared the 95\% CL bounds on Wilson coefficients extracted from $A_{0}-A_{2}$ with those obtained from the differential cross-section. The latter are found to be generally more stringent, which is not surprising given the much larger statistics and smaller experimental uncertainties available for cross-section measurements. However, the key point of this work is that the angular observable $A_{0}-A_{2}$ offers a clean and independent probe of chirality-breaking operators, free from contamination by dimension-six operators that interfere with the SM. This is not the case for the cross-section, where SM contributions are never suppressed and the leading SMEFT effects arise from interference with current–current four-fermion operators. As such, $A_{0}-A_{2}$ becomes particularly valuable in global SMEFT fits, where multiple operators are constrained simultaneously and degeneracies in parameter space must be resolved.

We stress that our study was conducted in a simplified theoretical setup and does not attempt to replicate a full experimental analysis. In particular, we assumed direct access to the angular coefficients and to the $A_{0}-A_{2}$ observable, whereas real analyses typically extract these quantities via template fits. While we strived to provide realistic estimates of theoretical and experimental systematics, a dedicated experimental study will ultimately be required to fully assess the potential of $A_{0}-A_{2}$ for probing chirality-breaking SMEFT operators at the LHC.

\section*{Acknowledgements}
X.L.~is grateful to Bin Yan for early collaboration on the project and for useful discussions. R.T.~thanks Emanuele Re for help with the \texttt{POWHEG} and \texttt{MiNNLO}$_{\text{\texttt{PS}}}$ codes. We thank R.~Rattazzi for comments on the UV origin of the chirality-breaking operators. We also thank the INFN Genova IT department for continuous support with the computing resources.

\newpage
\appendix

\section{The Collins-Soper frame} \label{app:CSframe}
Let the 4-vectors of the two incident beams in the laboratory frame be:
\begin{align}
b_+&=(E_{b_+},\vec{b}_+), \\
b_-&=(E_{b_-},\vec{b}_-).
\end{align}
We call $b'_+$ and $b'_-$ the respective 4-vectors boosted in the di-lepton rest frame. The $z$ axis of the CS frame is defined as the bisector of the unit vectors $\hat{b}'_+$ and $-\hat{b}'_-$, pointing such that its scalar product with the di-lepton 3-momentum, in the laboratory frame, is positive.\\
Another axis, called $q$, is defined as the one laying in the plane defined by $\hat{b}'_+$ and $\hat{b}'_-$, orthogonal to the $z$ axis and pointing in the direction opposite to $\hat{b}'_++\hat{b}'_-$.\\
The angle $\theta$ is defined with respect to the $z$ axis, while the angle $\phi$ is defined with respect to the $q$ axis. \\
This orientation of the CS frame axes can provide, for certain events, angles that are shifted by $\pi$ with respect to those defined in the original paper~\cite{Collins:1977iv}. This can be taken into account by slightly modifying the original definitions, introducing a factor $p^{\ell\ell}_{z}/\mid p^{\ell\ell}_{z}\mid$ that accounts for axis orientation. Therefore, one has:
\begin{align}
    \cos{\theta} &= \frac{p_{z,\ell\ell}}{\mid p_{z,\ell\ell}\mid}\frac{2(p_{z,\ell^+}E_{\ell^-}-p_{z,\ell^-}E_{\ell^+})}{m_{\ell\ell}\sqrt{m_{\ell\ell}^2+p_{T}^{\ell\ell}}}\\
    \tan{\phi} &= \frac{p_{z,\ell\ell}}{\mid p_{z,\ell\ell}\mid}\frac{\sqrt{m^2_{\ell\ell}+p^2_{T,\ell\ell}}}{m_{\ell\ell}}\frac{\Delta p_{T}^{\ell\ell}\cdot \hat{R}_T}{\Delta p_{T}^{\ell\ell}\cdot \hat{p}_{T,\ell\ell}}
\end{align}
where $\hat{p}_{T,\ell\ell}$ is a unit vector in the direction of $p_{T}^{\ell\ell}$ and $\hat{R}_T$ is the normalized cross product between $\hat{b}_+$ and $p_{z,\ell\ell}$.
\begin{figure}
    \centering
   \includegraphics[width=0.5\linewidth]{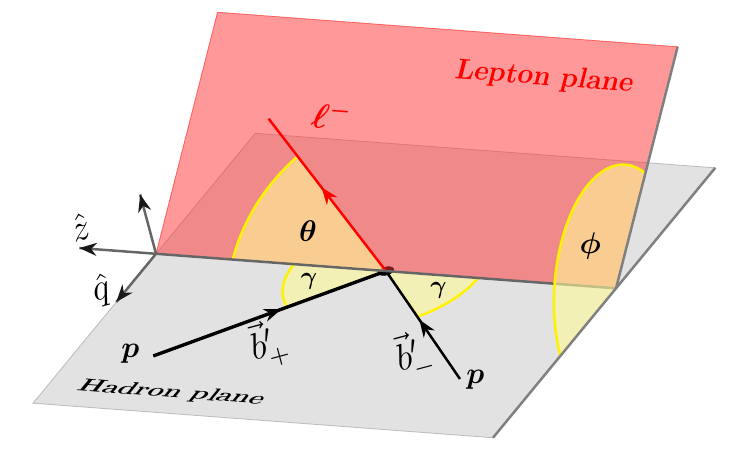}
    \caption{A graphic representation of the Collins-Soper frame.}
    \label{fig:CSframe}
\end{figure}

\section{Angular observables from spherical-harmonics}\label{app::SphArm}
In the following, we report the combination of spherical harmonics that provides the factor multiplying each $A_l$ in Eq.~\eqref{eq:CrossSection}.
\begin{equation*}
    \begin{array}{llll}
    &\dst 1+\cos^2{\theta} &=&\dst \frac{4\sqrt{\pi}}{15}\left(10Y_{0}^{0}+\sqrt{5}Y_{2}^{0}\right)\,,\vspace{2mm}\\
    &\dst \frac{1}{2}(1-3\cos^2{\theta}) &=& \dst -2\sqrt{\frac{\pi}{5}}Y_{2}^{0}\,,\vspace{2mm}\\
    &\dst \sin{2\theta}\cos{\phi} &=& \dst 2\sqrt{\frac{2\pi}{15}}(Y_{2}^{-1}-Y_{2}^{1})\,,\vspace{2mm}\\
    &\dst \frac{1}{2}(\sin^2{\theta}\cos{2\phi}) &=& \dst \sqrt{\frac{2\pi}{15}}(Y_{2}^{-2}+Y_{2}^{2})\,,\vspace{2mm}\\
    &\dst \sin{\theta}\cos{\phi} &=& \dst \sqrt{\frac{2\pi}{3}}(Y_{1}^{-1}-Y_{1}^{1})\,,\vspace{2mm}\\
    &\dst \cos{\theta} &=& \dst 2\sqrt{\frac{\pi}{3}}Y_{1}^{0}\,,\vspace{2mm}\\
    &\dst \sin^2{\theta}\sin{2\phi} &=& \dst -2i\sqrt{\frac{2\pi}{15}}(Y_{2}^{-2}-Y_{2}^{2})\,,\vspace{2mm}\\
    &\dst \sin{2\theta}\sin{\phi} &=& \dst -2i\sqrt{\frac{2\pi}{15}}(Y_{2}^{-1}+Y_{2}^{1})\,,\vspace{2mm}\\
    &\dst \sin{\theta}\sin{\phi} &=& \dst -i\sqrt{\frac{2\pi}{3}}(Y_{1}^{-1}+Y_{1}^{1})\, .\\
    \end{array}
\end{equation*}

One can verify that, apart from the first two, each of these combinations of spherical harmonics is orthogonal to the others. 

\section{Differential cross-section and PDF integration}\label{app::PDFintegration}
In this appendix we derive the expression for the fully differential unpolarized cross-section for the process  $pp \to \ell^+ \ell^- X$ at $\mathcal{O}\left(\alpha_\text{S}\right)$.  This is achieved by splitting it in two subsequent processes: the first is $pp \to ZX$, which on the partonic level gives $q \bar q \to Z g$ or $q g \to Z q $;  the second consists in the $Z$ boson decaying into the lepton pair $Z\to \ell^+\ell^-$. Neglecting the quark masses the kinematic constraint $s+t+u=m_{\ell \ell}^2$ must hold, where $m_{\ell \ell}$ denotes the invariant mass of the lepton pair and $s$, $t$ and $u$ the partonic Mandelstam variables defined for the first subprocess. \\
In the CS frame the $Z$ boson momentum reads $p_Z=\left(E_Z,p_{T}^{\ell\ell},0,p_l\right)$. All the kinematic quantities are expressed in terms of: 
\begin{itemize}
    \item the invariant mass of the lepton pair $m_{\ell \ell}$;
    \item  the transverse momentum of the lepton pair $p_{T}^{\ell\ell}$;
    \item the energy of each hadron in the laboratory frame $E_p$;
    \item the rapidity $y= \log \left( \frac{E_Z+p_l}{E_Z-p_l}\right)$ of the $Z$ boson;
    \item the fraction of hadronic momentum carried by the incoming partons $x_a$.
\end{itemize}
 In terms of these quantities, the energy $E_Z$ and longitudinal momentum $p_l$ become
\begin{equation}
    \begin{aligned}
            E_Z = \sqrt{m_{\ell \ell}^2+(p_{T}^{\ell\ell})^2}\cosh{y}, \quad
            p_{l}  = \sqrt{m_{\ell \ell}^2+(p_{T}^{\ell\ell})^2}\sinh{y},
    \end{aligned}
\end{equation}
and the Mandelstam variables can be rewritten as 
\begin{equation} \label{eq: MandeVar}
s = 4x_a x_b E_p^2, \quad t = m_{\ell\ell}^2 - 2x_a E_p \sqrt{m_{\ell\ell}^2 + (p_{T}^{\ell\ell})^2} e^{-y}, \quad u = m_{\ell\ell}^2 - 2x_b E_p \sqrt{m_{\ell\ell}^2 + (p_{T}^{\ell\ell})^2 }e^{y}.
\end{equation}
The general expression for the hadronic cross-section $\sigma_{pp \to X \ell^+ \ell^-}$ is
\begin{equation} \label{eq: CS}
    \begin{aligned}
        \sigma_{pp \to X \ell^+ \ell^-} & = \sum_{ab}  \int \frac{d^3p_X}{(2\pi)^3 2E_X} \frac{d^3p_{ \ell^+}}{(2\pi)^3 2E_{\ell^+}} \frac{d^3p_{ \ell^-}}{(2\pi)^3 2E_{ \ell^-}} \frac{1}{2s} f_{a/p}(x_a) f_{b/p}(x_b) dx_a dx_b \\
        & \times \langle \left|\mathcal{M}_{ab}\right|^2 \rangle (2\pi)^4 \delta^{(4)}(p_a + p_b - p_X - p_{\ell^+} - p_{\ell^-}),
    \end{aligned}
\end{equation}
with $\mathcal{M}_{ab}$ the amplitudes for the partonic processes; the meaning of the other symbols is understood. Momentum conservation gives
\begin{equation}
     \begin{aligned}
        d\sigma_{pp \to X \ell^+ \ell^-} & = \sum_{ab}  \int d^4 p_Z d\phi^{(2)}_{\ell \ell } \frac{d^3p_X}{(2\pi)^3 2E_X} \frac{1}{2s} f_{a/p}(x_a) f_{b/p}(x_b) dx_a dx_b \\
        & \times \langle |\mathcal{M}_{ab}|^2 \rangle (2\pi)^4 \delta^{(4)}(p_a + p_b - p_X - p_{Z}) , 
    \end{aligned}
\end{equation}
where 
\begin{equation}
    d\phi^{(2)}_{\ell \ell} \equiv \left( 2\pi \right)^4 \delta^{(4)}\left( p_Z - p_{\ell^+}-p_{\ell^-} \right) \frac{d^3p_{ \ell^+}}{(2\pi)^3 2E_{\ell^+}} \frac{d^3p_{ \ell^-}}{(2\pi)^3 2E_{ \ell^-}} 
\end{equation}
is the lepton  pair phase space. \\
The integral on $x_b$ is now performed using the identity $\frac{d^3 p_X }{2 E_X} = d^4 p_X \delta(p_X^2)$: we can substitute in the Dirac $\delta$-function $p_X^2 = s+t+u- m_{\ell \ell}^2$ and then use Eq.~\eqref{eq: MandeVar} to obtain
\begin{equation} \label{eq: delta}
    \delta(p_X^2) = \frac{1}{2 E_p^2 \left| 2x_a - x_T e^y  \right|}\delta \left(x_b - \frac{x_a x_T e^{-y} - \frac{m^2_{\ell \ell}}{2 E_p^2}}{2x_a-x_T e^y} \right) ,
\end{equation}
with $x_T \equiv \sqrt{m_{\ell \ell}^2+(p_{T}^{\ell\ell})^2}/E_p$. Moreover, being  $x_b < 1$, the above expression gives a lower limit on $x_a$ that reads 
\begin{equation}
    x_a > x_{a}^{\text{min}} \equiv \frac{x_T e^y- \frac{m^2_{\ell \ell}}{2E_p^2}}{2-x_T e^{-y}}. 
\end{equation}
The integration over the the lepton pair momenta gives
\begin{equation} \label{eq: cs0}
    \sigma_{pp\to \ell^+\ell^-} = \frac{1}{(2\pi)^6}\frac{1}{64 E_p^4}\int \left[ \sum_{a b} \int_{x_a^{\text{min}}}^1 f_{a/p}(x_a) f_{b/p}(x_b) dx_a\frac{ \langle |\mathcal{M}_{ab}|^2 \rangle}{\left( 2x_a -x_T e^y  \right)x_a x_b^* }\right] dc_\theta d\phi d^4p_Z,
\end{equation}
where the star in $x_b^*$ keeps track of the constraint on $x_b$ given by Eq.~\eqref{eq: delta}. \\
The 4-momentum of the outgoing $Z$ boson can be parametrized as 
\begin{equation}
    p_Z = \left( \sqrt{m_{\ell \ell}^2+(p_{T}^{\ell\ell})^2} \cosh y, p_{T}^{\ell\ell} \cos \alpha , p_{T}^{\ell\ell} \sin \alpha, \sqrt{m_{\ell \ell}^2+(p_{T}^{\ell\ell})^2} \sinh y  \right),
\end{equation}
which gives   
\begin{equation}
    d^4 p_Z = \frac{1}{2}  p_{T}^{\ell\ell} d\alpha dm_{\ell\ell}^2 dp_{T}^{\ell\ell} dy.
\end{equation}
Plugging this into Eq.~\eqref{eq: cs0} and integrating over $\alpha$ we find  
\begin{equation}\label{eq: CSMasterFormula}
    \frac{d\sigma_{pp\to X \ell^+\ell^-}}{dm_{\ell\ell}^2 dp_{T}^{\ell\ell} dc_\theta d\phi} = \frac{1}{(2\pi)^5} \frac{\pi p_{T}^{\ell\ell}}{128 E_p^4} \int_{y_{\text{min}}}^{y_{\text{max}}
} \left[\sum_{ab}\int_{x_a^{\text{min}}}^1 f_{a/p}(x_a) f_{b/p}(x_b^*) dx_a  \frac{ \langle |\mathcal{M}_{ab}|^2 \rangle}{\left( 2x_a -x_T e^y  \right)x_a x_b^* } \right]dy,
\end{equation}
where the upper and lower bounds
\begin{equation}
    \begin{aligned}
        \dst y_{\max} & \dst \equiv \log \left[ \frac{1}{2} \left( \frac{4E_p^2 + m_{\ell\ell}^2}{2E_p^2 x_T} + \sqrt{\left( \frac{4E_p^2 + m_{\ell\ell}^2}{2E_p^2 x_T} \right)^2 - 4} \right) \right], \\
        \dst y_{\min} & \dst \equiv \log \left( \frac{x_T}{2} \right)
    \end{aligned}
\end{equation}
follow from the fact that $x_a^{\text{min}}$ is positive and lesser than $1$. \\ 
The analytic results obtained using formula~\eqref{eq: CSMasterFormula} were found to be in excellent agreement with those obtained with \texttt{Madgraph5} \cite{Alwall:2011uj}.

\section{Cross-section vs $A_0-A_2$ bounds}\label{app::A0A2CScomparison}
\begin{figure}[t!]
  \centering
  \begin{minipage}{0.5\textwidth}
    \centering
    \includegraphics[width=\linewidth]{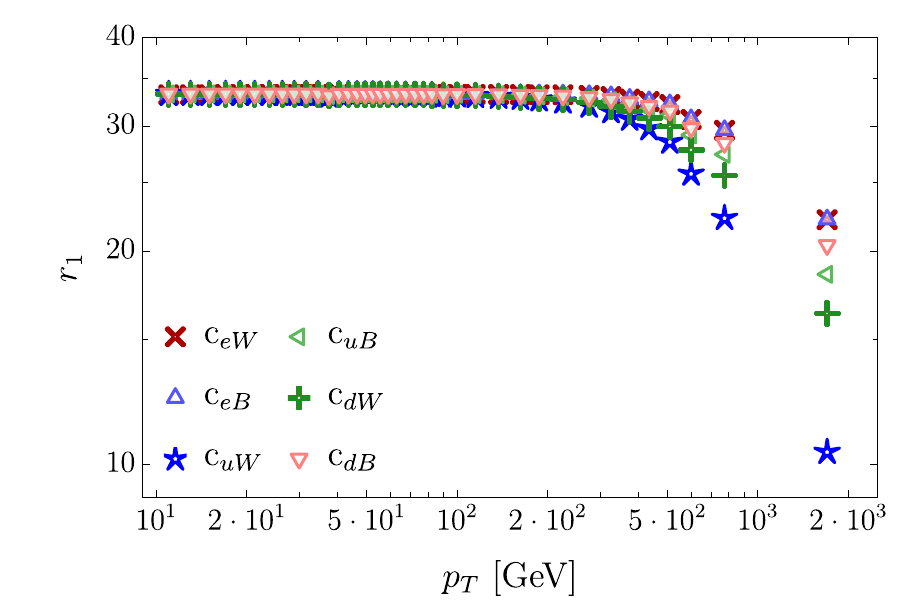}
  \end{minipage}\hfill
  \begin{minipage}{0.5\textwidth}
    \centering
    \includegraphics[width=\linewidth]{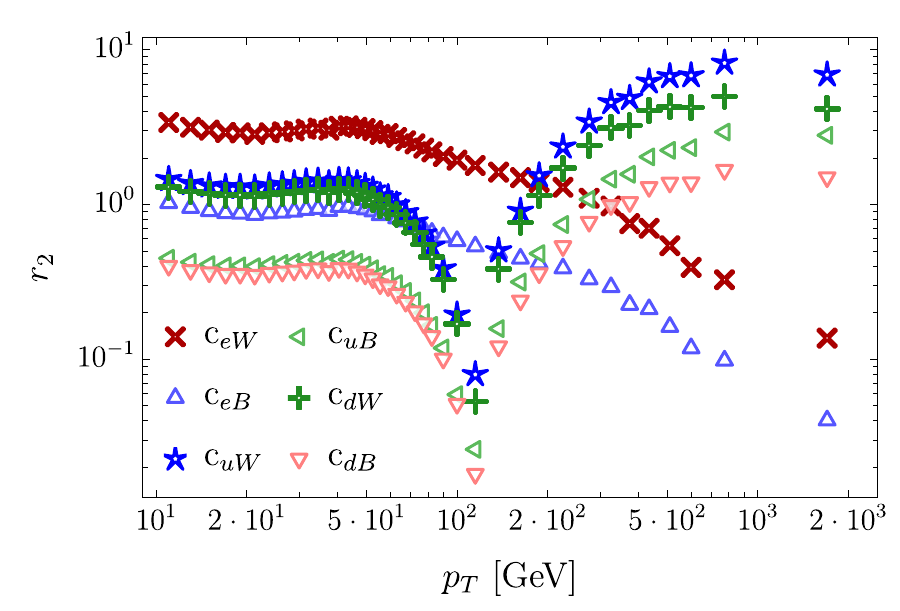}
  \end{minipage}
  \centering
  \begin{minipage}{0.5\textwidth}
    \centering
    \includegraphics[width=\linewidth]{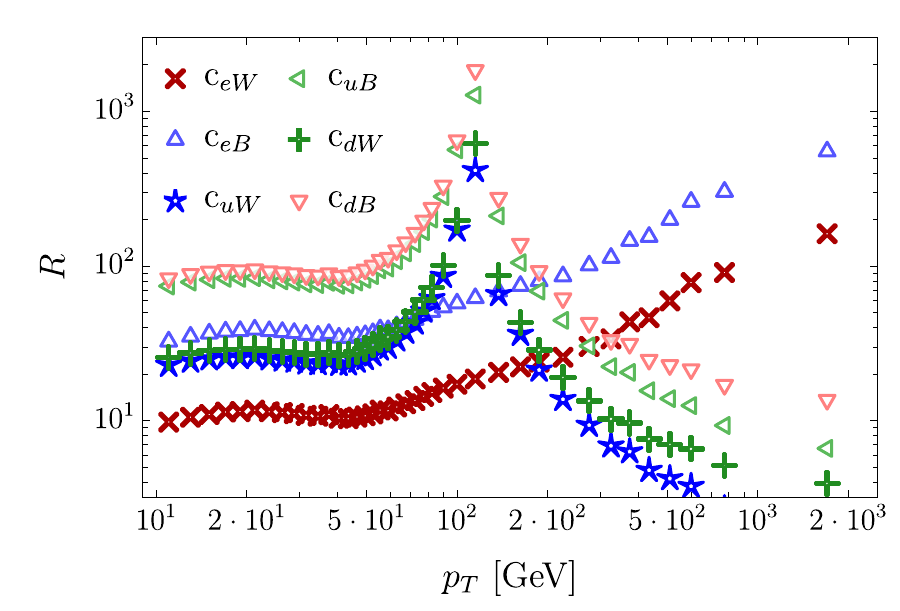}
  \end{minipage}\hfill
  \caption{Plots of the ratios defined in equation \eqref{eq::ratios} assuming $\mathcal{L}=300 \text{ fb}^{-1}$ in the $p_{T}^{\ell\ell}$ distribution. Upper left: $r_1$, upper right: $r_2$, lower panel: $R$. The different behavior of $c_{eW}$ and $c_{eB}$, that are better constrained by the $A_0-A_2$ observable, is evident.}
  \label{fig:IntuitivePlots}
\end{figure}

In this appendix, we provide an intuitive explanation for why the bounds obtained from the cross-section are tighter than those obtained with the $A_0-A_2$ observable. In particular we focus on the ratios:
\begin{align}\label{eq::ratios}
    r_1 &= \left| \frac{\sigma^{\text{SMEFT}} - \sigma^{\text{SM}}}{\delta_{\sigma^{\text{SM}}}} \right|\,, \\
    r_2 &= \left| \frac{(A_0 - A_2)^{\text{SMEFT}} - (A_0 - A_2)^{\text{SM}}}{\delta_{(A_0 - A_2)^{\text{SM}}}} \right|\,, \\
    R   &= \frac{r_1}{r_2}\,,
\end{align}
where $\delta_{\sigma^{\text{SM}}}$ and $\delta_{(A_0 - A_2)^{\text{SM}}}$ indicate the uncertainty on the cross-section and the $A_0-A_2$ observable, respectively. Clearly, at fixed $c_{\text{NP}}$, a larger value of the ratio indicates that the corresponding observable is more sensitive to the considered operator. Figure~\ref{fig:IntuitivePlots} shows the $r1$ (upper left), $r_2$ (upper right) and $R$ (lower panel) ratios for the $p_{T}^{\ell\ell}$ distribution with fixed luminosity of 300 fb$^{-1}$. Each Wilson coefficient $c_{\text{NP}}$ is set to 10 and $\Lambda$ is fixed at 4 TeV. 

It is immediate to notice that $r_1$ is significantly larger that $r_2$, as $R$ ranges from a few units up to more than 1000. This behavior indicates that the cross-section is more sensitive to NP insertion, due to its smaller uncertainties. Furthermore, the $R$ plot shows a peak in the central bins, where $A_0 - A_2$ increases and the statistical uncertainty is small thanks to the high statistics. This pattern does not occur for the $c_{eW}$ and $c_{eB}$ Wilson coefficients, which are, in fact, better constrained through the $A_0 - A_2$ observable. An analogous behavior is observed for the 3000 fb$^{-1}$ integrated luminosity. Similar conclusions hold when considering the $m_{\ell\ell}$ distribution, where $r_1$ remains consistently larger than $r_2$.

\section{Numerical tables}\label{app::Tables}

In this appendix we report the numerical values of the observables appearing in Figures~\ref{fig:xsdistributions}, \ref{fig:ptangcoefficients}, \ref{fig:mllangcoefficients} with explicit separation of statistical and systematic uncertainties.

\begin{table}[h!]
	\centering
	\small
	\begin{tabular}{cccccc}
		\toprule
		$p_{T}^{\ell\ell} \,\,[\text{GeV}]$ & $N_{\text{MC}}$ & $\sigma\,\,[\text{pb}]$ & $\delta\sigma_{\text{stat}}^{300}\,\,[\text{pb}]$ & $\delta\sigma_{\text{stat}}^{3000}\,\,[\text{pb}]$ & $\delta\sigma_{\text{syst}}^{3\%} \,\,[\text{pb}]$\\
		\midrule
        $10.-12.$ & $55355060$ & $118.99$ & $0.0462851$ & $0.0146366$ & $3.5697$ \\
        $12.-14.$ & $44432363$ & $100.94$ & $0.0350784$ & $0.0110928$ & $3.02821$ \\
        $14.-16.$ & $36325175$ & $86.4391$ & $0.0271705$ & $0.00859205$ & $2.59317$ \\
        $16.-18.$ & $30149734$ & $74.566$ & $0.0268787$ & $0.00849979$ & $2.23698$ \\
        $18.-20.$ & $25356520$ & $64.7717$ & $0.0211872$ & $0.00669997$ & $1.94315$ \\
        $20.-22.5$ & $26414219$ & $69.6736$ & $0.0345891$ & $0.010938$ & $2.09021$ \\
        $22.5-25.$ & $21848818$ & $59.3793$ & $0.0205171$ & $0.00648808$ & $1.78138$ \\
        $25.-27.5$ & $18289822$ & $51.0283$ & $0.0207527$ & $0.00656258$ & $1.53085$ \\
        $27.5-30.$ & $15450562$ & $44.1082$ & $0.0153048$ & $0.00483982$ & $1.32325$ \\
        $30.-33.$ & $15561084$ & $45.4264$ & $0.0155019$ & $0.00490215$ & $1.36279$ \\
        $33.-36.$ & $12982145$ & $38.7237$ & $0.0154616$ & $0.00488938$ & $1.16171$ \\
        $36.-39.$ & $10916808$ & $33.2943$ & $0.104793$ & $0.0331385$ & $0.998829$ \\
        $39.-42.$ & $9253016$ & $28.593$ & $0.0125402$ & $0.00396556$ & $0.857789$ \\
        $42.-45.$ & $7892013$ & $24.7582$ & $0.0112726$ & $0.00356472$ & $0.742747$ \\
        $45.-48.$ & $6768494$ & $21.5125$ & $0.0102497$ & $0.00324124$ & $0.645376$ \\
        $48.-51.$ & $5840548$ & $18.7992$ & $0.0102651$ & $0.00324612$ & $0.563975$ \\
        $51.-54.$ & $5053525$ & $16.4331$ & $0.009844$ & $0.00311295$ & $0.492994$ \\
        $54.-57.$ & $4394576$ & $14.4348$ & $0.0106579$ & $0.00337034$ & $0.433044$ \\
        $57.-61.$ & $5002474$ & $16.5707$ & $0.018601$ & $0.00588215$ & $0.497122$ \\
        $61.-65.$ & $4199710$ & $14.0805$ & $0.00805207$ & $0.00254629$ & $0.422415$ \\
        $65.-70.$ & $4330950$ & $14.6711$ & $0.0076631$ & $0.00242328$ & $0.440134$ \\
        $70.-75.$ & $3520638$ & $12.0448$ & $0.00867464$ & $0.00274316$ & $0.361345$ \\
        $75.-80.$ & $2883906$ & $9.93928$ & $0.00625729$ & $0.00197873$ & $0.298178$ \\
        $80.-85.$ & $2374827$ & $8.24452$ & $0.00648999$ & $0.00205232$ & $0.247335$ \\
        $85.-95.$ & $3608604$ & $12.6469$ & $0.021292$ & $0.00673313$ & $0.379406$ \\
        $95.-105.$ & $2524395$ & $8.91138$ & $0.00627343$ & $0.00198383$ & $0.267341$ \\
        $105.-125.$ & $3122264$ & $11.1183$ & $0.00677558$ & $0.00214263$ & $0.33355$ \\
        $125.-150.$ & $1999251$ & $7.18747$ & $0.00551796$ & $0.00174493$ & $0.215624$ \\
        $150.-175.$ & $1017474$ & $3.68138$ & $0.00427167$ & $0.00135082$ & $0.110441$ \\
        $175.-200.$ & $555846$ & $2.01631$ & $0.0030348$ & $0.000959689$ & $0.0604894$ \\
        $200.-250.$ & $518627$ & $1.877$ & $0.00313351$ & $0.000990902$ & $0.05631$ \\
        $250.-300.$ & $203615$ & $0.735178$ & $0.0018908$ & $0.000597924$ & $0.0220553$ \\
        $300.-350.$ & $89323$ & $0.32241$ & $0.00118899$ & $0.000375991$ & $0.00967229$ \\
        $350.-400.$ & $43580$ & $0.156648$ & $0.000841524$ & $0.000266113$ & $0.00469943$ \\
        $400.-470.$ & $28142$ & $0.101251$ & $0.000669394$ & $0.000211681$ & $0.00303754$ \\
        $470.-550.$ & $13814$ & $0.0492041$ & $0.000388475$ & $0.000122847$ & $0.00147612$ \\
        $550.-650.$ & $6735$ & $0.0239324$ & $0.000303223$ & $0.0000958875$ & $0.000717971$ \\
        $650.-900.$ & $4229$ & $0.0148124$ & $0.000228624$ & $0.0000722974$ & $0.000444371$ \\
        $900.-2500.$ & $791$ & $0.00265501$ & $0.0000894172$ & $0.0000282762$ & $0.0000796503$ \\
	\bottomrule
	\end{tabular}
    \caption{Expected results of measurements of the transverse momentum spectrum at the LHC at $13$ TeV with $300$ fb$^{-1}$ and $3$ ab$^{-1}$. The columns show the bin range, the number of Monte Carlo events, the expected cross-section, the statistical uncertainty with $300$ fb$^{-1}$ and $3$ ab$^{-1}$, and a systematic uncertainty of $3\%$.}
    \label{tab:ptxsdistributions}
\end{table}

\begin{table}[h!]
	\centering
	\small
	\begin{tabular}{ccccc}
		\toprule
		$p_{T}^{\ell\ell} \,\,[\text{GeV}]$ & $A_{0}$ & $(\delta A_{0})_{\text{stat}}^{300}$ & $(\delta A_{0})_{\text{stat}}^{3000}$ & $(\delta A_{0})_{\text{syst}}^{3\%}$\\
		\midrule
        $10.-12.$ & $0.0454443$ & $0.00114832$ & $0.000363131$ & $0.0186367$ \\
        $12.-14.$ & $0.0599507$ & $0.00135674$ & $0.000429039$ & $0.0182015$ \\
        $14.-16.$ & $0.0762949$ & $0.0014533$ & $0.000459572$ & $0.0177112$ \\
        $16.-18.$ & $0.0929306$ & $0.00134978$ & $0.000426839$ & $0.0172121$ \\
        $18.-20.$ & $0.112002$ & $0.0013841$ & $0.000437691$ & $0.01664$ \\
        $20.-22.5$ & $0.133622$ & $0.001311$ & $0.000414575$ & $0.0159913$ \\
        $22.5-25.$ & $0.158508$ & $0.0011865$ & $0.000375205$ & $0.0152448$ \\
        $25.-27.5$ & $0.186841$ & $0.00158303$ & $0.000500597$ & $0.0143948$ \\
        $27.5-30.$ & $0.213909$ & $0.00122356$ & $0.000386924$ & $0.0135827$ \\
        $30.-33.$ & $0.24182$ & $0.00124103$ & $0.000392448$ & $0.0127454$ \\
        $33.-36.$ & $0.274575$ & $0.0013978$ & $0.000442023$ & $0.0117627$ \\
        $36.-39.$ & $0.30449$ & $0.00357953$ & $0.00113195$ & $0.0108653$ \\
        $39.-42.$ & $0.338655$ & $0.00157738$ & $0.00049881$ & $0.00984036$ \\
        $42.-45.$ & $0.369891$ & $0.00131081$ & $0.000414515$ & $0.00890328$ \\
        $45.-48.$ & $0.400941$ & $0.00182146$ & $0.000575997$ & $0.00797177$ \\
        $48.-51.$ & $0.429686$ & $0.00183945$ & $0.000581685$ & $0.00710942$ \\
        $51.-54.$ & $0.455141$ & $0.00167453$ & $0.000529532$ & $0.00634577$ \\
        $54.-57.$ & $0.479131$ & $0.00175029$ & $0.00055349$ & $0.00562608$ \\
        $57.-61.$ & $0.510621$ & $0.00232018$ & $0.000733707$ & $0.00468139$ \\
        $61.-65.$ & $0.542085$ & $0.00204103$ & $0.000645431$ & $0.00373744$ \\
        $65.-70.$ & $0.571995$ & $0.00163757$ & $0.000517846$ & $0.00284015$ \\
        $70.-75.$ & $0.601204$ & $0.00168406$ & $0.000532546$ & $0.00196388$ \\
        $75.-80.$ & $0.633106$ & $0.00186528$ & $0.000589854$ & $0.00100681$ \\
        $80.-85.$ & $0.66316$ & $0.00247316$ & $0.000782083$ & $0.0001052$ \\
        $85.-95.$ & $0.686713$ & $0.0060731$ & $0.00192048$ & $0.000601351$ \\
        $95.-105.$ & $0.732296$ & $0.00206278$ & $0.000652307$ & $0.00196888$ \\
        $105.-125.$ & $0.779334$ & $0.00163382$ & $0.000516661$ & $0.00338001$ \\
        $125.-150.$ & $0.832182$ & $0.0024437$ & $0.000772765$ & $0.00496547$ \\
        $150.-175.$ & $0.867126$ & $0.00282857$ & $0.000894471$ & $0.00601376$ \\
        $175.-200.$ & $0.899789$ & $0.0041151$ & $0.00130131$ & $0.00699372$ \\
        $200.-250.$ & $0.916585$ & $0.00481605$ & $0.00152297$ & $0.00749755$ \\
        $250.-300.$ & $0.949971$ & $0.00721618$ & $0.00228196$ & $0.00849908$ \\
        $300.-350.$ & $0.947089$ & $0.0109943$ & $0.00347669$ & $0.00841257$ \\
        $350.-400.$ & $0.977413$ & $0.0146018$ & $0.0046175$ & $0.0093229$ \\
        $400.-470.$ & $0.965547$ & $0.0181728$ & $0.00574673$ & $0.00896602$ \\
        $470.-550.$ & $0.989361$ & $0.0254853$ & $0.00805915$ & $0.00968054$ \\
        $550.-650.$ & $0.9903$ & $0.0337669$ & $0.010678$ & $0.00970749$ \\
        $650.-900.$ & $0.978093$ & $0.0496695$ & $0.0157069$ & $0.00934364$ \\
        $900.-2500.$ & $0.990086$ & $0.111957$ & $0.0354038$ & $0.00968581$ \\
	\bottomrule
	\end{tabular}
    \caption{Expected results of measurements of the $A_{0}$ observable at the LHC at $13$ TeV with $300$ fb$^{-1}$ and $3$ ab$^{-1}$. The columns show the bin range, the expected value of $A_{0}$, the statistical uncertainty with $300$ fb$^{-1}$ and $3$ ab$^{-1}$, and a systematic uncertainty of $3\%$.}
    \label{tab:ptangcoefficientsA0}
\end{table}

\begin{table}[h!]
	\centering
	\small
	\begin{tabular}{ccccc}
		\toprule
		$p_{T}^{\ell\ell} \,\,[\text{GeV}]$ & $A_{2}$ & $(\delta A_{2})_{\text{stat}}^{300}$ & $(\delta A_{2})_{\text{stat}}^{3000}$ & $(\delta A_{2})_{\text{syst}}^{3\%}$\\
		\midrule
        $10.-12.$ & $0.0236693$ & $0.00121024$ & $0.000382711$ & $0.000710079$ \\
		$12.-14.$ & $0.0252714$ & $0.00147315$ & $0.000465851$ & $0.000758143$ \\
		$14.-16.$ & $0.0305507$ & $0.00105898$ & $0.00033488$ & $0.00091652$ \\
		$16.-18.$ & $0.0313687$ & $0.00101226$ & $0.000320105$ & $0.00094106$ \\
		$18.-20.$ & $0.0379034$ & $0.000984646$ & $0.000311372$ & $0.0011371$ \\
		$20.-22.5$ & $0.0410952$ & $0.00108569$ & $0.000343324$ & $0.00123286$ \\
		$22.5-25.$ & $0.045506$ & $0.000973089$ & $0.000307718$ & $0.00136518$ \\
		$25.-27.5$ & $0.0499428$ & $0.000975623$ & $0.000308519$ & $0.00149828$ \\
		$27.5-30.$ & $0.0537455$ & $0.00112234$ & $0.000354915$ & $0.00161236$ \\
		$30.-33.$ & $0.0579562$ & $0.00105402$ & $0.000333309$ & $0.00173869$ \\
		$33.-36.$ & $0.0626439$ & $0.00116594$ & $0.000368703$ & $0.00187932$ \\
		$36.-39.$ & $0.0598799$ & $0.00623964$ & $0.00197315$ & $0.00179644$ \\
		$39.-42.$ & $0.0652713$ & $0.00123154$ & $0.000389449$ & $0.00195814$ \\
		$42.-45.$ & $0.0695346$ & $0.00134794$ & $0.000426257$ & $0.00208604$ \\
		$45.-48.$ & $0.070112$ & $0.00140319$ & $0.000443728$ & $0.00210336$ \\
		$48.-51.$ & $0.0707345$ & $0.00135741$ & $0.000429251$ & $0.00212204$ \\
		$51.-54.$ & $0.0727296$ & $0.00144449$ & $0.000456788$ & $0.00218189$ \\
		$54.-57.$ & $0.0774697$ & $0.00252136$ & $0.000797325$ & $0.00232409$ \\
		$57.-61.$ & $0.0773344$ & $0.00353121$ & $0.00111667$ & $0.00232003$ \\
		$61.-65.$ & $0.076637$ & $0.00150062$ & $0.000474538$ & $0.00229911$ \\
		$65.-70.$ & $0.0759462$ & $0.00163643$ & $0.000517484$ & $0.00227839$ \\
		$70.-75.$ & $0.0734238$ & $0.00154593$ & $0.000488867$ & $0.00220272$ \\
		$75.-80.$ & $0.0749548$ & $0.0019587$ & $0.000619396$ & $0.00224864$ \\
		$80.-85.$ & $0.0728264$ & $0.00160893$ & $0.000508789$ & $0.0021848$ \\
		$85.-95.$ & $0.0776449$ & $0.00432023$ & $0.00136618$ & $0.00232935$ \\
		$95.-105.$ & $0.0722412$ & $0.00178587$ & $0.000564742$ & $0.00216723$ \\
		$105.-125.$ & $0.0682163$ & $0.00171466$ & $0.000542224$ & $0.00204649$ \\
		$125.-150.$ & $0.0606084$ & $0.00219335$ & $0.000693597$ & $0.00181825$ \\
		$150.-175.$ & $0.0549135$ & $0.00265379$ & $0.000839202$ & $0.00164741$ \\
		$175.-200.$ & $0.0500717$ & $0.00341425$ & $0.00107968$ & $0.00150216$ \\
		$200.-250.$ & $0.0438294$ & $0.00375023$ & $0.00118593$ & $0.0013149$ \\
		$250.-300.$ & $0.0255768$ & $0.00730141$ & $0.00230891$ & $0.000767309$ \\
		$300.-350.$ & $0.0257082$ & $0.00878564$ & $0.00277826$ & $0.000771414$ \\
		$350.-400.$ & $0.046506$ & $0.0139431$ & $0.00440919$ & $0.00139526$ \\
		$400.-470.$ & $0.0257966$ & $0.0160086$ & $0.00506237$ & $0.000774092$ \\
		$470.-550.$ & $0.0126944$ & $0.0248429$ & $0.007856$ & $0.000380961$ \\
		$550.-650.$ & $0.00365842$ & $0.0297732$ & $0.00941513$ & $0.000110922$ \\
		$650.-900.$ & $0.0541991$ & $0.0472457$ & $0.0149404$ & $0.00162197$ \\
		$900.-2500.$ & $0.0367986$ & $0.0994578$ & $0.0314513$ & $0.00110695$ \\
	\bottomrule
	\end{tabular}
    \caption{Expected results of measurements of the $A_{2}$ observable at the LHC at $13$ TeV with $300$ fb$^{-1}$ and $3$ ab$^{-1}$. The columns show the bin range, the expected value of $A_{2}$, the statistical uncertainty with $300$ fb$^{-1}$ and $3$ ab$^{-1}$, and a systematic uncertainty of $3\%$.}
    \label{tab:ptangcoefficientsA2}
\end{table}

\begin{table}[h!]
	\centering
	\small
	\begin{tabular}{ccccc}
		\toprule
		$p_{T}^{\ell\ell} \,\,[\text{GeV}]$ & $A_{0}-A_{2}$ & $\delta (A_{0}-A_{2})_{\text{stat}}^{300}$ & $\delta (A_{0}-A_{2})_{\text{stat}}^{3000}$ & $\delta (A_{0}-A_{2})_{\text{syst}}^{3\%}$\\
		\midrule
        $10.-12.$ & $0.0283314$ & $0.00219976$ & $0.000695624$ & $0.0186437$ \\
        $12.-14.$ & $0.0330746$ & $0.002538$ & $0.000802587$ & $0.0182193$ \\
        $14.-16.$ & $0.0419759$ & $0.00274345$ & $0.000867555$ & $0.0177411$ \\
        $16.-18.$ & $0.0415125$ & $0.00280323$ & $0.000886458$ & $0.0172811$ \\
        $18.-20.$ & $0.0490277$ & $0.00218055$ & $0.000689551$ & $0.0167469$ \\
        $20.-22.5$ & $0.0577262$ & $0.00381117$ & $0.0012052$ & $0.0161526$ \\
        $22.5-25.$ & $0.062618$ & $0.00218661$ & $0.000691467$ & $0.0155138$ \\
        $25.-27.5$ & $0.0653363$ & $0.00239468$ & $0.000757264$ & $0.0148491$ \\
        $27.5-30.$ & $0.066978$ & $0.00249069$ & $0.000787625$ & $0.0142801$ \\
        $30.-33.$ & $0.0764202$ & $0.00216215$ & $0.000683731$ & $0.0136772$ \\
        $33.-36.$ & $0.0761219$ & $0.00246108$ & $0.000778261$ & $0.0131836$ \\
        $36.-39.$ & $0.0853067$ & $0.00485269$ & $0.00153455$ & $0.0127001$ \\
        $39.-42.$ & $0.0835377$ & $0.00245159$ & $0.000775261$ & $0.0124663$ \\
        $42.-45.$ & $0.0908412$ & $0.00268472$ & $0.000848984$ & $0.0122209$ \\
        $45.-48.$ & $0.0870325$ & $0.00300801$ & $0.000951216$ & $0.0123383$ \\
        $48.-51.$ & $0.0925727$ & $0.00348033$ & $0.00110058$ & $0.0123622$ \\
        $51.-54.$ & $0.089258$ & $0.00357317$ & $0.00112994$ & $0.0126788$ \\
        $54.-57.$ & $0.0829$ & $0.00435847$ & $0.00137827$ & $0.0131511$ \\
        $57.-61.$ & $0.094012$ & $0.00298472$ & $0.000943851$ & $0.0133462$ \\
        $61.-65.$ & $0.0920686$ & $0.00319159$ & $0.00100927$ & $0.0140083$ \\
        $65.-70.$ & $0.0915626$ & $0.0031811$ & $0.00100595$ & $0.0146901$ \\
        $70.-75.$ & $0.0909227$ & $0.0030153$ & $0.000953521$ & $0.0154339$ \\
        $75.-80.$ & $0.08731$ & $0.00344346$ & $0.00108892$ & $0.0164048$ \\
        $80.-85.$ & $0.0913945$ & $0.00421305$ & $0.00133228$ & $0.0171533$ \\
        $85.-95.$ & $0.0884333$ & $0.00569044$ & $0.00179947$ & $0.0179584$ \\
        $95.-105.$ & $0.0861138$ & $0.00424156$ & $0.0013413$ & $0.0194852$ \\
        $105.-125.$ & $0.0833509$ & $0.00371347$ & $0.0011743$ & $0.0211513$ \\
        $125.-150.$ & $0.0839705$ & $0.00447094$ & $0.00141384$ & $0.022989$ \\
        $150.-175.$ & $0.0834069$ & $0.00632102$ & $0.00199888$ & $0.0242685$ \\
        $175.-200.$ & $0.0896245$ & $0.00785136$ & $0.00248282$ & $0.0252912$ \\
        $200.-250.$ & $0.0792491$ & $0.00902837$ & $0.00285502$ & $0.0262151$ \\
        $250.-300.$ & $0.0690518$ & $0.0147523$ & $0.0046651$ & $0.0277607$ \\
        $300.-350.$ & $0.0671591$ & $0.0205242$ & $0.00649034$ & $0.0277056$ \\
        $350.-400.$ & $0.07426$ & $0.0326327$ & $0.0103194$ & $0.0286537$ \\
        $400.-470.$ & $0.0752987$ & $0.0362228$ & $0.0114547$ & $0.0281729$ \\
        $470.-550.$ & $0.0679409$ & $0.0490551$ & $0.0155126$ & $0.0292874$ \\
        $550.-650.$ & $0.0166015$ & $0.0724604$ & $0.022914$ & $0.0307811$ \\
        $650.-900.$ & $0.142826$ & $0.0871392$ & $0.0275558$ & $0.0267453$ \\
        $900.-2500.$ & $-0.109047$ & $0.201254$ & $0.063642$ & $0.0343444$ \\
	\bottomrule
	\end{tabular}
    \caption{Expected results of measurements of the $A_{0}-A_{2}$ observable at the LHC at $13$ TeV with $300$ fb$^{-1}$ and $3$ ab$^{-1}$. The columns show the bin range, the expected value of $A_{2}$, the statistical uncertainty with $300$ fb$^{-1}$ and $3$ ab$^{-1}$, and a systematic uncertainty of $3\%$.}
    \label{tab:ptangcoefficientsA0mA2}
\end{table}

\begin{table}[h!]
	\centering
	\small
	\begin{tabular}{cccccc}
		\toprule
		$m_{\ell\ell} \,\,[\text{GeV}]$ & $N_{\text{MC}}$ & $\sigma\,\,[\text{pb}]$ & $\delta\sigma_{\text{stat}}^{300}\,\,[\text{pb}]$ & $\delta\sigma_{\text{stat}}^{3000}\,\,[\text{pb}]$ & $\delta\sigma_{\text{syst}}^{3\%} \,\,[\text{pb}]$\\
		\midrule
		$100-105$ & $8204433$ & $18.1485$ & $0.0118372$                 & $0.00374325$          & $0.544455$ \\
		$105-110$ & $8349055$ & $8.82233$ & $0.00550703$                & $0.00174147$          & $0.26467$ \\
		$110-115$ & $7067831$ & $5.25065$ & $0.00322271$                & $0.00101911$          & $0.15752$ \\
		$115-120$ & $7159110$ & $3.50436$ & $0.00213849$                & $0.000676251$         & $0.105131$ \\
		$120-126$ & $5813465$ & $2.94182$ & $0.0020837$                 & $0.000658923$         & $0.0882547$ \\
		$126-133$ & $5797184$ & $2.37394$ & $0.00193485$                & $0.000611855$         & $0.0712182$ \\
		$133-141$ & $3755474$ & $1.95064$ & $0.00176662$                & $0.000558653$         & $0.0585193$ \\
		$141-150$ & $3813026$ & $1.56363$ & $0.00158484$                & $0.00050117$          & $0.046909$ \\
		$150-160$ & $3096496$ & $1.24575$ & $0.00119425$                & $0.000377655$         & $0.0373726$ \\
		$160-171$ & $3143515$ & $0.992093$ & $0.00101408$               & $0.000320682$         & $0.0297628$ \\
		$171-185$ & $1601319$ & $0.882853$ & $0.00124199$               & $0.000392753$         & $0.0264856$ \\
		$185-200$ & $1631284$ & $0.669961$ & $0.000768605$              & $0.000243054$         & $0.0200988$ \\
		$200-220$ & $1662754$ & $0.607422$ & $0.000784724$              & $0.000248151$         & $0.0182227$ \\
		$220-243$ & $1701384$ & $0.454496$ & $0.000570742$              & $0.000180485$         & $0.0136349$ \\
		$243-273$ & $867442$ & $0.368791$ & $0.000635313$               & $0.000200904$         & $0.0110637$ \\
		$273-320$ & $890247$ & $0.3257$ & $0.000636502$                 & $0.000201279$         & $0.00977101$ \\
		$320-380$ & $917527$ & $0.201384$ & $0.000397336$               & $0.000125649$         & $0.00604152$ \\
		$380-440$ & $709781$ & $0.100279$ & $0.000218025$               & $0.0000689457$        & $0.00300837$ \\
		$440-510$ & $361260$ & $0.0603857$ & $0.000172991$              & $0.0000547046$        & $0.00181157$ \\
		$510-600$ & $370583$ & $0.0400292$ & $0.000123339$              & $0.0000390033$        & $0.00120088$ \\
		$600-700$ & $380320$ & $0.0211948$ & $0.0000512655$             & $0.0000162116$        & $0.000635844$ \\
		$700-830$ & $155170$ & $0.0127793$ & $0.0000579316$             & $0.0000183196$        & $0.000383379$ \\
		$830-1000$ & $158533$ & $0.00713393$ & $0.0000317332$           & $0.0000100349$        & $0.000214018$ \\
		$1000-1500$ & $162895$ & $0.00501588$ & $0.0000223981$          & $7.1\cdot 10^{-6}$    & $0.000150477$ \\
		$1500-3000$ & $166879$ & $0.000848871$ & $5.0\cdot 10^{-6}$     & $1.6\cdot 10^{-6}$    & $0.0000254661$ \\
		$3000-10000$ & $2710$ &	$9.5\cdot 10^{-6}$ & $1.1\cdot 10^{-6}$ & $3.4\cdot 10^{-7}$    & $2.9\cdot 10^{-7}$ \\
	\bottomrule
	\end{tabular}
    \caption{Expected results of measurements of the di-lepton invariant mass spectrum at the LHC at $13$ TeV with $300$ fb$^{-1}$ and $3$ ab$^{-1}$. The columns show the bin range, the number of Monte Carlo events, the expected cross-section, the statistical uncertainty with $300$ fb$^{-1}$ and $3$ ab$^{-1}$, and a systematic uncertainty of $3\%$.}
    \label{tab:mllxsdistributions}
\end{table}

\begin{table}[h!]
	\centering
	\small
	\begin{tabular}{ccccc}
		\toprule
		$m_{\ell\ell} \,\,[\text{GeV}]$ & $A_{0}$ & $(\delta A_{0})_{\text{stat}}^{300}$ & $(\delta A_{0})_{\text{stat}}^{3000}$ & $(\delta A_{0})_{\text{syst}}^{3\%} $\\
		\midrule
		$100-105$ & $0.206069$ & $0.00220099$   & $0.000696013$ & $0.0138179$ \\
		$105-110$ & $0.198655$ & $0.00390891$   & $0.0012361$   & $0.0140404$ \\
		$110-115$ & $0.192236$ & $0.0043067$    & $0.0013619$   & $0.0142329$ \\
		$115-120$ & $0.18408$ & $0.00459632$    & $0.00145349$  & $0.0144776$ \\
		$120-126$ & $0.178093$ & $0.00527618$   & $0.00166847$  & $0.0146572$ \\
		$126-133$ & $0.16969$ & $0.00676013$    & $0.00213774$  & $0.0149093$ \\
		$133-141$ & $0.163594$ & $0.00667648$   & $0.00211129$  & $0.0150922$ \\
		$141-150$ & $0.157821$ & $0.00815895$   & $0.00258008$  & $0.0152653$ \\
		$150-160$ & $0.150422$ & $0.0090537$    & $0.00286303$  & $0.0154873$ \\
		$160-171$ & $0.145611$ & $0.0109549$    & $0.00346424$  & $0.0156316$ \\
		$171-185$ & $0.137872$ & $0.00920768$   & $0.00291172$  & $0.0158639$ \\
		$185-200$ & $0.126617$ & $0.0122699$    & $0.00388007$  & $0.0162015$ \\
		$200-220$ & $0.124962$ & $0.0136978$    & $0.00433163$  & $0.0162512$ \\
		$220-243$ & $0.106773$ & $0.0130855$    & $0.00413799$  & $0.0167968$ \\
		$243-273$ & $0.0990207$ & $0.0193292$   & $0.00611242$  & $0.0170294$ \\
		$273-320$ & $0.097449$ & $0.0197648$    & $0.00625017$  & $0.0170767$ \\
		$320-380$ & $0.0809004$ & $0.0236554$   & $0.00748048$  & $0.017573$ \\
		$380-440$ & $0.0768314$ & $0.0331067$   & $0.0104693$   & $0.0176951$ \\
		$440-510$ & $0.0528575$ & $0.0400658$   & $0.0126699$   & $0.0184143$ \\
		$510-600$ & $0.0517298$ & $0.0535121$   & $0.016922$    & $0.018448$ \\
		$600-700$ & $0.0508636$ & $0.0685334$   & $0.0216721$   & $0.0184748$ \\
		$700-830$ & $0.0460793$ & $0.0885335$   & $0.0279967$   & $0.0186168$ \\
		$830-1000$ & $0.0420531$ & $0.117322$   & $0.0371003$   & $0.0187368$ \\
		$1000-1500$ & $0.0289552$ & $0.141289$  & $0.0446794$   & $0.0191276$ \\
		$1500-3000$ & $0.0236164$ & $0.41192$   & $0.130261$    & $0.0192946$ \\
		$3000-10000$ & $-0.673888$ & $6.51047$  & $2.05879$     & $0.0415498$ \\
	\bottomrule
	\end{tabular}
    \caption{Expected results of measurements of the $A_{0}$ observable at the LHC at $13$ TeV with $300$ fb$^{-1}$ and $3$ ab$^{-1}$. The columns show the bin range, the expected value of $A_{0}$, the statistical uncertainty with $300$ fb$^{-1}$ and $3$ ab$^{-1}$, and a systematic uncertainty of $3\%$.}
    \label{tab:mllangcoefficientsA0}
\end{table}

\begin{table}[h!]
	\centering
	\small
	\begin{tabular}{ccccc}
		\toprule
		$m_{\ell\ell} \,\,[\text{GeV}]$ & $A_{2}$ & $(\delta A_{2})_{\text{stat}}^{300}$ & $(\delta A_{2})_{\text{stat}}^{3000}$ & $(\delta A_{2})_{\text{syst}}^{3\%} $\\
		\midrule
		$100-105$ & $0.0422219$ & $0.00227595$      & $0.000719719$     & $0.00126666$ \\
		$105-110$ & $0.0434971$ & $0.00376089$      & $0.0011893$       & $0.00130493$ \\
		$110-115$ & $0.0417116$ & $0.00381904$      & $0.00120769$      & $0.00125134$ \\
		$115-120$ & $0.038442$ & $0.00474009$       & $0.00149895$      & $0.00115326$ \\
		$120-126$ & $0.0355266$ & $0.00459244$      & $0.00145226$      & $0.0010658$ \\
		$126-133$ & $0.0373691$ & $0.00624652$      & $0.00197532$      & $0.00112104$ \\
		$133-141$ & $0.0345992$ & $0.00571451$      & $0.00180709$      & $0.00103799$ \\
		$141-150$ & $0.0364752$ & $0.0083864$       & $0.00265201$      & $0.00109429$ \\
		$150-160$ & $0.0364136$ & $0.00710229$      & $0.00224594$      & $0.00109239$ \\
		$160-171$ & $0.0291341$ & $0.00948309$      & $0.00299882$      & $0.000874032$ \\
		$171-185$ & $0.0356255$ & $0.0085101$       & $0.00269113$      & $0.0010688$ \\
		$185-200$ & $0.019759$ & $0.0103417$        & $0.00327033$      & $0.000592756$ \\
		$200-220$ & $0.0294363$ & $0.0106759$       & $0.00337601$      & $0.000883093$ \\
		$220-243$ & $0.0237177$ & $0.0105273$       & $0.00332902$      & $0.000711509$ \\
		$243-273$ & $0.0167273$ & $0.0144559$       & $0.00457135$      & $0.000501912$ \\
		$273-320$ & $0.0221835$ & $0.0139955$       & $0.00442577$      & $0.000665533$ \\
		$320-380$ & $0.0170735$ & $0.0210878$       & $0.00666854$      & $0.000512299$ \\
		$380-440$ & $0.0149437$ & $0.0288257$       & $0.00911549$      & $0.00044819$ \\
		$440-510$ & $0.00187905$ & $0.0365037$      & $0.0115435$       & $0.0000557831$ \\
		$510-600$ & $-0.00137813$ & $0.0457439$     & $0.0144655$       & $0.0000413211$ \\
		$600-700$ & $-0.00159732$ & $0.0612091$     & $0.019356$        & $0.0000482714$ \\
		$700-830$ & $-0.00557428$ & $0.0672415$     & $0.0212636$       & $0.000167436$ \\
		$830-1000$ & $0.0137169$ & $0.10256$        & $0.0324322$       & $0.000410183$ \\
		$1000-1500$ & $0.0317234$ & $0.122617$      & $0.038775$        & $0.000951331$ \\
		$1500-3000$ & $-0.00716862$ & $0.396067$    & $0.125248$        & $0.000207753$ \\
		$3000-10000$ & $2.86876$ & $20.4434$        & $6.46477$         & $0.0919646$ \\
	\bottomrule
	\end{tabular}
    \caption{Expected results of measurements of the $A_{2}$ observable at the LHC at $13$ TeV with $300$ fb$^{-1}$ and $3$ ab$^{-1}$. The columns show the bin range, the expected value of $A_{2}$, the statistical uncertainty with $300$ fb$^{-1}$ and $3$ ab$^{-1}$, and a systematic uncertainty of $3\%$.}
    \label{tab:mllangcoefficientsA2}
\end{table}

\begin{table}[h!]
	\centering
	\small
	\begin{tabular}{ccccc}
		\toprule
		$m_{\ell\ell} \,\,[\text{GeV}]$ & $A_{0}-A_{2}$ & $\delta (A_{0}-A_{2})_{\text{stat}}^{300}$ & $\delta (A_{0}-A_{2})_{\text{stat}}^{3000}$ & $\delta (A_{0}-A_{2})_{\text{syst}}^{3\%}$\\
		\midrule
		$100-105$ & $0.0489227$ & $0.00379057$      & $0.00119868$      & $0.0146$ \\
		$105-110$ & $0.0479573$ & $0.00693351$      & $0.00219257$      & $0.0147503$ \\
		$110-115$ & $0.0527973$ & $0.00846707$      & $0.00267752$      & $0.0148349$ \\
		$115-120$ & $0.0466155$ & $0.00778599$      & $0.00246215$      & $0.0150535$ \\
		$120-126$ & $0.0443988$ & $0.0108978$       & $0.0034462$       & $0.0151961$ \\
		$126-133$ & $0.0483033$ & $0.0123165$       & $0.00389482$      & $0.0153476$ \\
		$133-141$ & $0.0355806$ & $0.0115509$       & $0.00365273$      & $0.0155731$ \\
		$141-150$ & $0.0366276$ & $0.0141871$       & $0.00448637$      & $0.0156924$ \\
		$150-160$ & $0.0365152$ & $0.0167127$       & $0.00528503$      & $0.0158598$ \\
		$160-171$ & $0.0364686$ & $0.0246929$       & $0.00780859$      & $0.0159708$ \\
		$171-185$ & $0.0487894$ & $0.0189455$       & $0.00599109$      & $0.0160874$ \\
		$185-200$ & $0.0310485$ & $0.0205385$       & $0.00649484$      & $0.0164532$ \\
		$200-220$ & $0.0324652$ & $0.0236061$       & $0.00746492$      & $0.0164864$ \\
		$220-243$ & $0.0265552$ & $0.0210694$       & $0.00666273$      & $0.0169683$ \\
		$243-273$ & $0.0127414$ & $0.0364795$       & $0.0115358$       & $0.0172249$ \\
		$273-320$ & $0.039359$ & $0.0307471$        & $0.00972308$      & $0.0171653$ \\
		$320-380$ & $0.0332408$ & $0.0434818$       & $0.0137501$       & $0.0176311$ \\
		$380-440$ & $0.0179427$ & $0.0558835$       & $0.0176719$       & $0.017783$ \\
		$440-510$ & $-0.0132197$ & $0.080222$       & $0.0253684$       & $0.0185206$ \\
		$510-600$ & $0.0147696$ & $0.1048$          & $0.0331407$       & $0.0184813$ \\
		$600-700$ & $0.0223771$ & $0.134543$        & $0.0425464$       & $0.0184946$ \\
		$700-830$ & $-0.0145346$ & $0.174328$       & $0.0551275$       & $0.0187056$ \\
		$830-1000$ & $0.0364093$ & $0.226636$       & $0.0716687$       & $0.0187375$ \\
		$1000-1500$ & $-0.0000212533$ & $0.269824$  & $0.0853259$       & $0.0191475$ \\
		$1500-3000$ & $0.0905146$ & $0.904677$      & $0.286084$        & $0.0193981$ \\
		$3000-10000$ & $3.38865$ & $37.2917$        & $11.7927$         & $0.146803$ \\
	\bottomrule
	\end{tabular}
	\caption{Expected results of measurements of the $A_{0}-A_{2}$ observable at the LHC at $13$ TeV with $300$ fb$^{-1}$ and $3$ ab$^{-1}$. The columns show the bin range, the expected value of $A_{0}-A_{2}$, the statistical uncertainty with $300$ fb$^{-1}$ and $3$ ab$^{-1}$, and a systematic uncertainty of $3\%$.}
	\label{tab:mllangcoefficientsA0mA2}
\end{table}

\clearpage

\section{UV models sensitive to our analysis}\label{app::UVmodels}
One of the goals of the SMEFT program is that of indirectly exploring the lanscape of possible UV completions of the SM. Generally, this is done by constraining, through precision measurements, the Wilson coefficients of classes of SMEFT operators, and then match them to a specific BSM model, or, even better, to a class of BMS models. Indeed, it is essential to keep in mind that the SMEFT approach, being fully model-independent, gives little information on the UV, unless it is supplemented by additional hypotheses. For instance, constraints on electroweak precision observables give very different information on the UV if one assumes strongly interacting dynamics at the TeV scale, or weakly coupled extensions of the SM. In other words, the SMEFT gains all its power only when supplemented by reasonable hypotheses at least on the BSM framework that can give rise to the observables at hand.

An interesting question that one could ask on the present paper is which information on the UV one could gather from studying the operators that we consider.
First of all, the fact that dipole and scalar/tensor four-fermion operators break chiral symmetry, rules out the possibility of leveraging our analysis to constrain minimal flavor violating (MFV) UV models. Indeed, if MFV is assumed, then selection rules imply that the Wilson coefficients of chirality-breaking operators are suppressed by (powers of) Yukawa couplings, making our bounds too weak to draw any relevant conclusion. Parity is the other selection rule that makes all constraints on parity-odd operators, both four-fermion and dipole operators, the latter corresponding to Electric Dipole Moments (EDMs) from high energy measurements essentially irrelevant. This is the reason why we considered only parity conserving operators. Concerning parity conserving dipole operatorsa, corresponding to Magnetic Dipole Moments (MDMs), the impact of UV physics is quite well understood, since they often arise from one loop diagrams that contribute to the U(1)$_{Y}$, SU(2)$_L$m and SU(3)$_{c}$ vertex correction, to which there is usually a limited set of contributing Feynman diagrams. The general structure of these operators is therefore relatively easy to understand. For instance,  Ref.~\cite{Valori:2025hlp} studies the one-loop contributions to dipole moments in several UV theories, such as two-Higgs-doublet models, the minimal supersymmetric extension of the SM (MSSM), and scenarios with extra U(1) gauge symmetries, and performes the matching with the SMEFT operators that we consider: the general result is that, within the framework of the aforementioned BSM theories, suitable parameter choices can lead to sizable Wilson coefficients for the CP-even dipole operators. These parameter choices are often fine-tuned, which is expected for sizeable chirality breaking contributions, but are still possible. In this respect, our bounds can be used to constrain such parameter choices, and to limit the possible fine-tuning available in some of these BSM scenarios.

Let us now briefly discuss four-fermion operators. As discussed, for instance, in Refs.~\cite{deBlas:2017xtg,Dorsner:2016wpm}, they can be generated at tree level in BSM models which include leptoquark states. In particular the operators $Q_{\ell equ}^{(1)}$ and $Q_{\ell equ}^{(3)}$ can be generated by integrating out at tree level scalar leptoquarks, while $Q_{\ell edq}$ can be generated by integrating out at tree-level vector leptoquarks. Notice that, because of their chiral structure, the operators $Q_{\ell edq}$ and $Q_{\ell equ}^{(1)}$ introduce terms in the pion decay width $\Gamma \left( \pi \to \ell  \nu_{\ell}  \right) $ that are proportional to the pion mass \cite{Shanker:1982nd}. In particular, one finds that the ratio between the SM and the NP contribution scales as
    \begin{equation}
       \frac{\Gamma(\pi^+ \to \ell^+ \nu_\ell)^{\text{SM}}}{\Gamma(\pi^+ \to \ell^+ \nu_\ell)^{\text{NP}}}  \propto \frac{m_\pi^2}{m_{\ell}(m_d+m_u)}.
    \end{equation}
This contribution is reasonable for the muon channel, but it is hugely enhanced for the electron channel. As a consequence, one can check that for $\Lambda$ of $ \mathcal{O}(1 \text{ TeV})$, the present experimental measurement of the ratio \cite{ParticleDataGroup:2024cfk} 
    \begin{equation}
       R \equiv \frac{\Gamma \left( \pi^+ \to e^+  \nu_e  \right)}{\Gamma \left( \pi^+ \to \mu^+  \nu_\mu  \right)} = (1.230 \pm 0.004)\times 10^{-4},
    \end{equation}
implies, for the electron channel, $c_{ledq}+ c_{lequ}^{(1)}\sim \mathcal{O}(10^{-4})$.\footnote{This was also pointed out in Refs.~\cite{Leurer:1993em,Leurer:1993qx}.} For this reason, as already mentioned in the main text, our bounds on the $Q_{\ell e d q }$ and $Q_{\ell e q u}^{(1)}$ operators are only relevant for the second family.

\newpage
\bibliographystyle{mine}
\bibliography{references}

\end{document}